\documentclass[aps,prc,reprint,groupedaddress,onecolumn,floatfix]{revtex4}
\usepackage{amsmath}
\usepackage{amssymb}
\usepackage{amsfonts}
\usepackage{upgreek}
\usepackage{graphicx}
\usepackage{color}
\usepackage{hyperref}
\hypersetup{colorlinks=true, citecolor=blue, urlcolor=blue}
\def\beq{\begin{equation}}
\def\eeq{\end{equation}}

\def\nhat{\hat{\mathbf{n}}}
\def\lp{\tilde{\mathbf{p}}}

\def\pval{\int\mspace{-19mu}-}
\begin{document}
\title{Simple relativistic quark models}

\author{Sirajudheen Kuthini Kunhammed}
\affiliation{Department of Physics and Astronomy, The University of
Iowa, Iowa City, IA 52242, USA}

\author{W.~N.~Polyzou}
\affiliation{Department of Physics and Astronomy, The University of
Iowa, Iowa City, IA 52242, USA}

\date{\today}


\begin{abstract}
  
   A class of phenomenological relativistic models of hadronic systems
   motivated by QCD that have dual representations as models of mesons
   and nucleons or quarks and gluons is investigated.  These models
   are designed to provided qualitative insight into the role of sea
   quarks in hadronic structure and reactions.  The model assumption
   is that the Hamiltonian can be divided into two parts; one that
   involves degrees of freedom in the same connected local and global
   color singlet and the remaining interactions that allow the
   connected local and global color singlets to interact.  The first
   class of interactions results in infinite towers of bare
   ``particles'' with hadronic quantum numbers.  All but a finite
   number of these remain stable when the second class of interactions
   is included.  The model interactions are expressed in terms of
   sub-hadronic degrees of freedom, which determine the bare hadronic
   spectrum and the interactions involving the bare hadrons in terms
   of a small number of sub-hadronic model parameters.  As a first
   test, this paper considers the simplest case of mesons that
   interact via a string-breaking interaction.  One virtue of this
   model is that all of the bare meson masses and eigenfunctions can
   be computed analytically.  In addition, the string breaking
   interaction leads to production vertices that can also be computed
   analytically.  The relativistic wave functions have a light-front
   kinematic symmetry.  The goal is to find a simple relativistic
   quantum mechanical model based on sub-hadronic degrees of freedom
   that can provide an efficient, qualitatively consistent description of
   hadronic masses, lifetimes, cross sections, sea quark effects, and
   electromagnetic properties.  The simplicity of the model makes it a
   potentially useful tool to study the impact of sea quarks on
   hadronic structure and reactions.

\end{abstract}
\maketitle

\section{Introduction}

It is currently accepted that QCD \cite{Fritzsch:1973pi} is the theory
of the strong interaction.  Lattice calculations
\cite{Wilson:1974sk}\cite{Wilson:1975id} support this belief.
At the same time, models based on baryons exchanging mesons
\cite{GLOCKLE1996107}\cite{Kievsky:1997bg} often provide an
efficient, realistic and quantitatively accurate description of the
structure and dynamics of light nuclei.

While lattice QCD is the most reliable method currently available for
testing the viability of QCD as the theory of nuclear structure and
reactions, it is a finite discretization of a theory with an infinite
number of degrees of freedom.  What makes it compelling is that it
does not make any assumptions about the dominant degrees of freedom;
instead it simply retains all degrees of freedom relevant to a given
volume and resolution.  Because it retains both important and
unimportant degrees of freedom it is not an efficient computational
method, which makes it difficult to compete with meson-exchange models
as a practical computational tool for hadronic reactions.  The
discretization destroys many continuous symmetries of the exact theory
in order to emphasize the role of local gauge invariance in
confinement.  In addition, most calculations use a Euclidean
formulation based on imaginary time, which often requires innovative
methods \cite{Luscher}\cite{Luscher2}\cite{Briceno2}\cite{Briceno1}
to extract physical observables.

A reasonable expectation for the future is that lattice calculations
will provide a reliable means to justify and refine more efficient
realistic models of hadrons by identifying the most important degrees
of freedom and the structure of the most important interactions
between these degrees of freedom \cite{PhysRevD.92.114512}\cite{Hatsuda}.

Observables that are proposed to be measured at the Electron Ion
Collider \cite{Deshpande} and JLAB \cite{Smith} are expressed in terms
of matrix elements of operators in hadronic states.  These observables
include form factors, distribution functions, and Wigner functions
\cite{ji-2004}.  The purpose of these measurements is to understand
the structure of hadronic states at sub-hadronic resolutions.  The
relevant theoretical input needed to compute these observables at this
sensitivity are models of the strong current and the initial and final
hadronic states in different Lorentz frames expressed in terms of the
charge carrying QCD degrees of freedom rather than hadronic degrees of
freedom.
  
These considerations suggest studying fully relativistic models that
are motivated by QCD.  It is desirable to use model degrees of freedom
and interactions that can be directly constrained by lattice
calculations.  Since the physical degrees of freedom in lattice
calculations are locally and globally gauge invariant, this suggests
examining models based on gauge invariant degrees of freedom.  While
there are many gauge invariant degrees of freedom, a flexible class of
models could help identify the dominant gauge invariant degrees of
freedom.  In addition, since the parameters of QCD are quark masses
and one coupling constant, a model motivated by QCD should have all
scales of the model determined by one coupling constant and a
collection of quark masses.

The purpose of this work is to formulate models that are simple enough
to investigate problems where valence and sea quarks interact.  The
goal is to find a class of models that can provide a qualitatively
consistent picture of mass spectra, lifetimes, cross sections and
electromagnetic observables that also include the effects of sea
quarks.  Since the parameters of QCD are quark masses and one coupling
constant, a model of QCD should have all scales determined by the quark masses
and one coupling constant.  A second important goal is that it should
be simple enough to compute observables involving sea quarks that are
qualitatively consistent with experiment.  The expectation is that
refinements can be treated perturbatively.  In order to keep the
dynamics as simple as possible, flavor fine-structure interactions are
not considered in this initial investigation.  These will need to be
added for realistic applications.

The model degrees of freedom in this work are taken to be non-local
objects that are both local and global color singlets.  The three main
QCD motivations for using models based on these structures are
Wilson's Erice lectures of 1976 \cite{Wilson:1975id}
\cite{Wilson:1974sk}, Kogut and Susskind's Hamiltonian formulation of
Lattice QCD \cite{Kogut:1974ag} and Seiler's \cite{Seiler:1982pw}
effort to formulate axioms of QCD based on non-local color singlets.
The choice of model degrees of freedom is most directly motivated by
the Kogut-Susskind Hamiltonian.  In that case the gauge invariant
degrees of freedom are connected networks of quarks, anti-quarks, and
links (gluons).  These states are eigenstates of the part of the
Hamiltonian involving the quark masses and the color electric
interaction, which assigns a mass to each quark and an unperturbed
energy to each link.  The local gauge invariance requires that the
quarks and anti-quarks in the same connected color singlets are
connected by a network of links.  Confinement follows since separating
the quarks requires more links in order to maintain the local gauge
invariance.

This represents a set of gauge invariant states that span a Hilbert
space of gauge invariant degrees of freedom.  The gauge covariant
derivative and color magnetic interactions are operators on this
representation of the Hilbert space that allow these degrees of
freedom to interact, preserving the local gauge symmetry.

A continuum example of a non-local locally gauge invariant
degree of freedom is constructed by applying
operators of the form 
\[
\int \bar{\psi(x)} P e^{\int_x^y A^{\mu}_c \lambda_c d\gamma_{\mu}}
\psi(y) F[\gamma] D[\gamma] 
\]
to the vacuum, 
where $F[\gamma]$ is a functional of paths between $x$ and $y$.

The model assumption for this work is that the strong interaction
Hamiltonian has a decomposition $H=H_1+H_2$, analogous to the Kogut-Susskind
Hamiltonian,  where the eigenstates of $H_1$ are systems of mutually
non-interacting confined connected local and global color singlets and
$H_2$ has interactions that allows the confined connected color
singlets to interact.

If such a decomposition exists then the expectation is that a complete
set of eigenstates of the first part of the Hamiltonian will be a Fock
space of mutually non-interacting bare confined color singlets.  These
quantities have hadronic quantum numbers, with no explicit color
degrees of freedom.  The second part would allow these bare singlets
to interact.  It includes interactions that break bare confined
singlets into pairs of bare confined singlets as well as many-singlet
interactions between bare singlets.

The appeal of this framework is that it only deals with states in the
physical Hilbert space, there are no issues with gauge choices, and
the degrees of freedom have hadronic quantum numbers. They are more
directly related to quantities that are naturally computed using
lattice methods.  Another feature can be understood by considering the
Kogut-Susskind Hamiltonian.  Given two quarks and two anti-quarks in
the same state, they can be made into pairs of connected singlets in
many different ways.  In the inner product of any distinct pairs of
these states, the quark degrees of freedom will disappear, and what
remains will look like the overlap of a connected gauge invariant set
of links with the vacuum which must vanish since they are eigenstates
of the same Hamiltonian with different energies.  The implication is
that, due to the gluonic degrees of freedom, quarks in the same state
in different connected singlets can be treated as distinguishable
particles.  This eliminates Van der Waals forces, which facilitates a
consistent treatment of scattering and bound states
\cite{greenberg_79}\cite{Robson_1}\cite{Robson_2}.

The challenge of working with gauge invariant degrees of freedom is
the large number of non-local degrees of freedom \cite{Seiler:1982pw}.
However, experiment and phenomenology suggest that the dynamics is
dominated a smaller number of hadronic states that interact to first
approximation by meson exchange.  The attitude of this work is to
start with the simplest degrees of freedom and add new degrees of
freedom as needed.  Connected confined color singlets containing a
quark and anti-quark are modeled by treating the quark and anti-quark
as interacting via a confining interaction.  Additional degrees of
freedom with the same quantum numbers corresponding to different
``excited'' confining interactions are anticipated, but will not be
considered.  In a lattice picture the excited interactions correspond
to gluon configurations that make the energy of a quark-anti-quark
system stationary, but not minimal, for fixed quark and anti-quark
positions.  These excited gluon configurations are of interest in
meson searches \cite{Ketzer_2020}.

The simplest interaction between quark-anti-quark-singlets is based on
a ``string-breaking'' vertex at the quark level.  This interaction
generates an infinite number of vertices coupling one bare singlet to
two bare singlets.  Since QCD has only one coupling constant this
vertex should use the same strength parameter, up to a dimensionless
constant of order unity, that is used in the confining interaction.
These parameters, along with the constituent quark masses, fix all of
the production vertices relating two bare confined singlets to one
bare confined singlet.  The model has the appealing feature that all
of production vertices in the meson representation can be computed
analytically.  This facilitates computations involving sea quarks.
The main question that this work addresses is whether such a simple
picture, with all scales fixed by one parameter, can provide a
qualitatively consistent treatment of scattering, resonances, spectral
and electromagnetic properties.

This picture is not new; it has motivated many related quark models
\cite{Micu}
\cite{Yaouanc1}
\cite{Yaouanc2}
\cite{Yaouanc3}
\cite{Yaouanc4}
\cite{ISGUR1983247}
\cite{PhysRevD.35.907w}
\cite{Seiler:1982pw}\cite{miller:1988}\cite{Dubin:1993fk}\cite{Isgur:1984bm}\cite{Dubin:1993fk}
\cite{PhysRevD.41.1595}
\cite{Geiger}
\cite{Ackleh}
\cite{Capstick:1999qq}.
\cite{Blundell}
\cite{Barnes}
\cite{Page}
\cite{Zhou}
\cite{Guo}
\cite{Deng}
\cite{Simonov:2011cm}
\cite{PhysRevC.86.055205}.
\cite{Ketzer_2020}
It is closely related to flux tube models in interpretation.  The
treatment of bare mesons is mathematically a constituent quark model.
The novel feature of this work is that all of the meson vertices can
be computed analytically, making it possible to efficiently
investigate the role of sea quarks, and high-lying states.  There are
explicit formulas for wave functions, which facilitate calculations.
The model is constructed to be fully relativistic so it is applicable
to systems of light quarks and scattering with large momentum
transfers.  In addition, the relativistic dynamics is formulated with
a light-front kinematic symmetry which is useful for hadronic
structure studies.  The framework also provides a dual description of
strongly interacting particles both in terms of QCD and hadronic
degrees of freedom.  The dual description has the advantage that
refinements to the model can be constrained by both lattice QCD and
hadronic phenomenology.

While this paper focuses on the meson sector, which is the simplest
type of hadronic system, there is still significant current interest
in light mesons \cite{Ketzer_2020}, which require a relativistic
treatment.  In addition the model discussed in this paper, with some
minor modifications, can be applied to treat baryons and exotic
mesons.  In the strong coupling limit, the lowest energy states are
the ones with the minimal string length.  This suggests that in the
valence sector baryons can be modeled as bound states of a quark and
diquark, exotic mesons could be modeled as bound states of a diquark
and an anti-diquark.  These mathematically look similar to the meson
models, with all of the computational advantages.  Schwinger-Dyson
studies \cite{Eichmann2019} have demonstrated that diquark models of
baryons compare favorably to full three-body calculations.  Exotic
molecules can be also directly modeled using methods discussed in this
paper.  The relativistic nature of these models are particularly
relevant for hadrons composed of light quarks and for computations of
parton distribution functions \cite{ji-2004}.  Realistic applications
require flavor dependent interactions, which are not considered in
this initial model.

Methods for constructing exactly Poincar\'e invariant models are
discussed in the next section.  It contains a description of
irreducible representations of the Poincar\'e group in a light-front
basis.  It also has explicit expressions that relate the tensor
product of two light-front irreducible representations to a direct
integral of light-front irreducible representations.  These are used
in the construction of a fully relativistic dynamics in the subsequent
sections.  Section three discusses an exactly Poincar\'e-invariant
model of confined quark-anti-quark glue degrees of freedom.
Properties of the confining interaction are discussed in section 4. A
string-breaking vertex is introduced in section 5.  Matrix elements of
the string-breaking vertex with all of the confined bare meson states
are computed analytically.  A minimal Poincar\'e invariant dynamical
model coupling sea and valence quarks is given in section 6.  The
coupled channel bound state problem is discussed in section 7.
Meson-meson scattering is discussed in section 8.  This includes a
discussion of how to formulate scattering with bound states in the
continuum.  The treatment of resonances is discussed in section 9.
Electromagnetic observables are discussed in section 10. Conclusions are
presented in section 11.
The analytic expression for 
matrix elements of the string breaking interaction is derived
in Appendix I and II.
Appendix III
discusses the conditions for the
existence of scattering wave operators in models with
an infinite number of coupled channels.  Appendix IV discusses details of the
scattering calculations.

\section{Kinematic Considerations - Light Front Basics}

In this section the notation for the kinematic variables that are used
in this work is introduced.  A relativistic treatment is utilized
because the kinetic energies of confined light quarks are normally
comparable or much larger than the quark masses.  In addition
relativistic energies and momentum transfers are needed in order to be
sensitive to degrees of freedom inside of a hadron.  Light-front
representations \cite{Dirac:1949cp}
\cite{Leutwyler:1977vy}\cite{Keister:1991sb}\cite{Coester:1992}\cite{Brodsky:1998}
are particularly convenient for these purposes because of the
kinematic nature of the subgroup of light-front-preserving Lorentz
boosts.

A relativistic dynamics is defined by specifying operators
representing the Casimir operators for the Poincar\'e group.  These
are the mass and square of the spin.  The mass operator plays the same
role in relativistic quantum models as the center of mass Hamiltonian
in non-relativistic quantum theories.  The spin must commute with the
mass.  When this happens the Hilbert space can be decomposed into a
direct integral of irreducible representation spaces.  The dynamical
unitary representation of the Poincar\'e group leaves these
irreducible subspaces invariant.  In addition, the transformation
properties on these irreducible subspaces are identical to the
transformation properties of a free particle with the mass and spin
associated with the subspace.  The complication is the
requirement that the unitary representation of the Poincar\'e group
clusters into tensor products implies that the spin is
dynamical.  The reason for this is that the relative orbital angular momentum
depends on subsystem masses, which are dynamical for subsystems of 
two or more particles.  The problem of constructing a relativistic dynamics
involves simultaneously diagonalizing the two dynamical operators.
For the applications in this work mass operators are constructed to
commute with the non-interacting spin, resulting in a dynamical model
that is similar in complexity to a non-relativistic model.  A
non-interacting spin does not necessarily mean that the angular
momentum is kinematic.  The price paid for using representations with
a non-interacting spin is that there will be violations of cluster
properties for systems of more than two hadrons. The states in this
work are truncated to two hadron states where there are no violations
of cluster properties.  There are methods to repair cluster properties
\cite{Sokolov:1977}\cite{Coester:1982vt}\cite{Keister:1991sb} for more
complicated systems.  These methods generate a dynamical spin operator
and many-body interactions.

The kinematic problems are then to describe relativistic one and
two-particle states.  The Hilbert space for a particle of mass $m$ and
spin $j$ is the mass $m$ spin $j$ irreducible representation space of
the Poincar\'e group.  The irreducible representation space is the
space of square integrable functions of a complete set of commuting
observables, that include the mass and spin, and are functions
of the infinitesimal generators of the Poincar\'e group.
 
For a quark-anti-quark pair confined to a local and global color
singlet, there is no experimental way to separate the
quark kinetic energy from the
confining interaction.  This means that there is no experimental
identification that relates the description of a quark or anti-quark to
an irreducible representation of the Poincar\'e group.  To illustrate
this note that a quark-anti-quark mass operator (rest energy) can be
expressed as the sum of a relative kinetic energy and an interaction
in many equivalent ways
\beq
M = K_1 +V_1 = K_2 +V_2 \qquad V_2 = V_1 +K_1-K_2
\label{kc:1}
\eeq
where both $V_1$ and $V_2$ are confining interactions associated with
different choices of the quark relative kinetic energy.  Nevertheless,
it is still convenient to assign a mass to the quark and anti-quark and
treat them as particles that transform irreducibly with respect to the
Poincar\'e group.  This means that their transformation properties are
characterized by a ``mass'' and spin like ordinary particles.  These
quantum numbers label positive-mass positive-energy irreducible
representation spaces of the Poincar\'e group.  Since the single-quark
representation is not the dynamical representation of the Poincar\'e
group, it has no physical consequences, but it is helpful input for
building representations of the Poincar\'e group when the quarks
interact.

In what follows a light-front \cite{Leutwyler:1977vy} basis is used to
label vectors in each irreducible representation space.  A light-front
is a hyperplane of space-time points satisfying
$x^+ = x^0+\nhat \cdot \mathbf{x} =0$ for an arbitrary but fixed
space-like unit vector $\nhat$.  The light-front components of the
four momentum are 
\beq
p^{\pm} = p^0 \pm \nhat \cdot \mathbf{p}
\label{kc:2}\eeq
and
\beq
\mathbf{p}_{\perp} = \mathbf{p}- \nhat (\nhat\cdot \mathbf{p}).
\label{kc:3}
\eeq
While any representation can be used to describe free particles, when
interactions are included a light-front dynamics \cite{Dirac:1949cp}
has advantages
\cite{Leutwyler:1977vy}\cite{Keister:1991sb}\cite{Coester:1992}\cite{Brodsky:1998}
for treating reactions involving electroweak probes, where the initial
and final states are in different reference frames.  This is because
in a light-front dynamics the boosts relating the initial and final
frames are kinematic (independent of interactions) and the magnetic
quantum numbers are invariant under these boosts.  In addition, in a
light-front dynamics the light-front momentum transferred to the
constituents in the impulse approximation is the same as the
light-front momentum transferred to the hadron in all frames related by
light-front boosts.  It is the only one of
Dirac's three forms of dynamics  \cite{Dirac:1949cp} with both of these
properties.

It is useful to express the light-front components of the four momentum
as $2 \times 2$ Hermitian matrices
\beq
P:=
\left (
\begin{array}{cc}
p^+ & p^1 -i p^2 \\
p^1 +i p^2 & p^- 
\end{array}  
\right ) = p^{\mu} \sigma_{\mu} 
\eeq
where $\sigma_{\mu}$ are the Pauli matrices and the identity.
The determinant of $P$ is the invariant $(p^0)^2-\mathbf{p}^2 = -p^2$.
Real Lorentz transformations connected to the identity can be
expressed in this notation as
\beq
P\to P'= APA^{\dagger} \qquad \mbox{det} (A)=1.
\eeq
The group of $2\times 2$ matrices with determinant 1 is
$SL(2,\mathbb{C})$.  It is a double cover of the Lorentz group since
$A$ and $-A$ result in the same Lorentz transformation.

Vectors in the irreducible representation space ${\cal H}_{mj}$ can be
taken as square integrable functions of the light-front three-momentum
components of $p$, $\lp = (p^+,\mathbf{p}_{\perp})$ and the
eigenvalues $\tilde{\mu}$ of the $\nhat$ component of the light-front
spin, $\mathbf{j}_f\cdot \nhat$,
\beq
\langle (m,j ) \lp ,\tilde{\mu} \vert \psi \rangle ,
\qquad
\int d^2 \mathbf{p}_{\perp} \int_0^\infty dp^+ \sum_{\mu=-j}^j
\vert \langle (m,j ) \lp ,\tilde{\mu} \vert \psi \rangle\vert^2 < \infty
. 
\label{kc:4}
\eeq
The $\tilde{}$ is used to denote light-front $3$-vectors and
light-front spins.  The commutation relation determine the
eigenvalue spectrum of these operators.
The spectrum of $p^+$ is $(0,\infty)$ while the
spectrum of $\mathbf{p}_{\perp}$ is $(-\infty,\infty)$.  The four
momenta are on shell and satisfy the light-front dispersion relation
\beq
p^- := {m^2 + \mathbf{p}_{\perp}^2 \over p^+} .
\label{kc:5}
\eeq
Spins in different frames can be compared by boosting to the rest
frame with a specific choice of boost.  This procedure is not unique because
any boost multiplied on the right by a momentum-dependent rotation is
also a boost.  Thus, there are different types of spin observables
that are associated with different choices of boosts.  All of the spin operators satisfy $SU(2)$ commutation relations,  which fixes $j$ to be integer or
half integer, and $\tilde{\mu}$ to vary from $-j$ to $j$ in integer steps.
In the
light-front representation the natural boosts used to define the spin are
the subgroup of light-front preserving boosts,  which are defined
below.

In the basis (\ref{kc:4}) $p^+=0$ corresponds to infinite momentum in the
-$\hat{\mathbf{n}}$ direction, so wave functions representing
normalizable vectors vanish for both large $p^+$ and for $p^+$ near 0.
The rest four vector has light-front components $p^+=p^-=m$.

For quarks and anti-quarks $j=1/2$ and there are additional flavor and
color quantum numbers.  A non-covariant delta-function normalization
is chosen for these irreducible plane-wave basis states
\beq 
\langle (m,{1 \over 2},f',c' ) \lp' ,\tilde{\mu}' \vert
(m,{1 \over 2},f,c ) \lp ,\tilde{\mu} \rangle =
\delta (\lp'-\lp)  
\delta_{\tilde{\mu}' \tilde{\mu}}\delta_{f'f}\delta_{c'c}
\label{kc:6}
\eeq
where
\beq
\delta (\lp'-\lp):=
\delta (p^{+\prime} - p^+) \delta (\mathbf{p}_{\perp}' -\mathbf{p}_{\perp} ).
\label{kc:7}
\eeq
A relativistic quantum mechanics is defined by a unitary
representation, $U(\Lambda, a)$, of the subgroup of the Poincar\'e
group continuously connected to the identity that is consistent with
the dynamics \cite{Wigner:1939cj}.  This ensures that probabilities,
expectation values and ensemble averages are independent of inertial
frame.

The construction of $U(\Lambda,a)$ starts by considering
representations of rotations of a zero-momentum eigenstate.
Since rotations leave the rest four momentum vector unchanged,
the rotation can only affect the magnetic quantum numbers:
\beq
U(R,0) \vert (m,j) \tilde{0},\tilde{\mu} \rangle =
\sum_{\tilde{\nu}-=j}^j \vert (m,j) \tilde{0},\tilde{\nu} \rangle
D^j_{\tilde{\nu}\tilde{\mu}}(R)
\label{kc:8}
\eeq
where $D^j_{\tilde{\nu}\tilde{\mu}}(R):=\langle j,\tilde{\nu} \vert U(R,0)
\vert j,\tilde{\mu} \rangle$
is the $2j$-dimensional unitary representation of 
$SU(2)$:
\beq
D^{j}_{\tilde{\nu}, \tilde{\mu}}[R] =
\sum_{k=0}^{j+\tilde{\mu}}
{\sqrt{(j+\tilde{\nu})!(j-\tilde{\nu})!(j+\tilde{\mu})!(j-\tilde{\mu})!} \over 
k! (j+\tilde{\nu} -k)! (j+\tilde{\mu}-k)! (k -\tilde{\mu} -\tilde{\nu})!}
R_{++}^k R_{+-}^{j+\tilde{\nu}-k}R_{-+}^{j+\tilde{\mu}-k}R_{--}^{k-\tilde{\mu}-
  \tilde{\nu}}
\label{kc:9}
\eeq
in the $\vert j,\tilde{\mu} \rangle$  basis 
and
\beq
R:= 
\left (
\begin{array}{cc}
R_{++} & R_{+-} \\
R_{-+}  & R_{--}\\ 
\end{array}
\right ) = \cos (\theta/2) +i \hat{\pmb{\theta}} \cdot \pmb{\sigma}
\sin (\theta/2) = e^{i{\pmb{\theta} \over 2}\cdot \pmb{\sigma}}
\label{kc:10}
\eeq
is a $SU(2)$ rotation matrix.  
Space-time translations of rest four momentum eigenstates are defined by
\beq
U(I,a) \vert (m,j) \tilde{0},\tilde{\mu} \rangle =
e^{-i m(a^+ + a^-)/2 }
\vert (m,j) \tilde{0},\tilde{\mu} \rangle . 
\label{kc:11}
\eeq
where the light-front components of the rest momentum are
$\tilde{0}:=(m,0,0)$.  The light-front spin is defined so that it does
not change when the system is boosted from the rest momentum to a
specified momentum, $p$, with a light-front preserving, $(x^+=0)$,
boost:
\beq
U(B_f(p/m),0) \vert (m,j) \tilde{0},\tilde{\mu} \rangle =
\vert (m,j) \tilde{p},\tilde{\mu} \rangle \sqrt{{p^+ \over m}} .
\label{kc:12}
\eeq
The square root factors make this unitary for basis
states with a delta-function normalization (\ref{kc:6}) in a
Hilbert space with norm (\ref{kc:4}).
$U(\Lambda,a)$ for any finite Poincar\'e transformation
on any basis state can be expressed as a product of these
elementary unitary transformations.  The result is that the mass $m$ spin
$j$ irreducible unitary representation of the Poincar\'e group
in a light front basis
(\ref{kc:4}) is
\beq
U(\Lambda,a) \vert (m,j),\lp ,\tilde{\mu} \rangle 
=
\sum_{\tilde{\mu}'=-j}^j
\vert (m,j),\lp' ,\tilde{\mu}' \rangle
e^{i a \cdot p'}
\sqrt{{p^{\prime +} \over p^+}}
D^{j}_{\tilde{\mu}' \tilde{\mu}}[B_f^{-1}(p'/m)\Lambda B_f(p/m) ].
\label{kc:13}
\eeq
The $SL(2,C)$ representation of the light-front boosts, $B_f(p/m)$,
that transform $(m,\mathbf{0})$ to $p$
are needed to construct the light-front Wigner rotations,
$B_f^{-1}(p'/m)\Lambda B_f(p/m)$.
They are
\beq
B_f (q) = \pm \left (
\begin{array}{cc}
\sqrt{q^+} & 0 \\
{\mathbf{q}_{\perp}\over \sqrt{q^+}} &{1 \over \sqrt{q^+}} \\ 
\end{array}
\right ) 
\qquad
B_f^{-1} (q) = \pm
\left (
\begin{array}{cc}
{1 \over \sqrt{q^+}}  & 0 \\
-{\mathbf{q}_{\perp}\over \sqrt{q^+}}  & \sqrt{q^+}\\ 
\end{array}
\right ). 
\label{kc:17}
\eeq
These $2\times 2$ matrices $B_f(q)$ are related to the four-vector components of
the $4\times 4$ matrices
$B_f^{\tilde{\mu}}{}_{\tilde{\nu}}(q)$ 
by
\beq
B_f^{\tilde{\mu}}{}_{\tilde{\nu}}(q) = {1 \over 2} \mbox{Tr}(\sigma_{\tilde{\mu}} B_f(q)
\sigma_{\tilde{\nu}} B_f^{\dagger}(q)).
\label{kc:18}
\eeq
The four-vector representation of the light-front boosts are
simply expressed in terms of how they act on the light-front
components of a four vector
\beq
a^+\to a^{+\prime} = q^+ a^+
\qquad \mathbf{a}_{\perp} \to \mathbf{a}_{\perp}' = \mathbf{a}_{\perp}
+\mathbf{q}_{\perp}a^+
\label{kc:18a}
\eeq
with $a^{-\prime}$ determined by the invariance of the proper length of $a$.
The inverse
light-front preserving boost is
\beq
a^+\to a^{+\prime} = {1 \over q^+} a^+
\qquad \mathbf{a}_{\perp} \to \mathbf{a}_{\perp}' = \mathbf{a}_{\perp}
-{\mathbf{q}_{\perp}\over q^+} a^+ .
\label{kc:18b}
\eeq
The $4\times 4 $ matrix representation of the light-front-preserving 
boost, $B_f(q)$ is
\beq
B_f (q)^{\mu}{}_{\nu} =
\left (
\begin{array}{cccc}
{q^{+2}+\mathbf{q}_{\perp}^2+1 \over 2 q^+} & {q^1 \over q^+} & {q^2 \over q^+}
& {q^{+2}+\mathbf{q}_{\perp}^2-1 \over 2 q^+} \\
q^1  &1 & 0  & q^1 \\ 
q^2  &0 & 1  & q^2 \\ 
{q^{+2}-\mathbf{q}_{\perp}^2-1 \over 2 q^+} & -{q^1 \over q^+} & -{q^2 \over q^+} &  {q^{+2}-\mathbf{q}_{\perp}^2+1 \over 2 q^+}\\
\end{array}
\right ). 
\label{kc:14b}
\eeq

Light-front boosts have the distinguishing property that they form a
subgroup of the Poincar\'e group.  This is easy to see from the
matrices (\ref{kc:17}), which are lower triangular with
real entries on the diagonal.  For a general Lorentz
transformation, $\Lambda$, $R_f(\Lambda,p):= B_f^{-1}(\Lambda
p)\Lambda B_f(p)$ is a light-front Wigner rotation.  Because of the
subgroup property this Wigner rotation is the identity when $\Lambda$
is a light-front boost.  This means that light-front boosts leave the
light-front magnetic quantum numbers unchanged.  The light-front
magnetic quantum numbers can be identified with the magnetic quantum
number measured in the particle's rest frame when boosted to the rest
frame with the light-front-preserving boost $B_f^{-1} (p/m)$.

It is typical when defining relativistic spins to choose all types of
spin (light front, canonical, helicity) to be identical in the
particle's or system's rest frame.  Different types of spins are
distinguished by how spins in other frames are related to spins in the
rest frame.  This convention will be followed in constructing
dynamical models.

Because $\hat{\mathbf{n}}$ defines a preferred direction, the
light-front Wigner rotation of a rotation, 
\beq
R \not= B_f^{-1}(Rp/m) R B_f(p/m),  
\label{kc:19}
\eeq
is not the rotation.
Instead it is a conjugate representation of the original rotation that
depends on the direction of $\hat{\mathbf{p}}$ and
the orientation, $\hat{\mathbf{n}}$, of the light front.  This representation
has the structure: 
\beq
B_f^{-1}(R_2 R_1 p) R_2 B_f(R_1p) \times  B_f^{-1}(R_1p) R_1 B_f(p)
=B_f^{-1}(R_2R_1 ) R_2R_1  B_f(p) .
\label{kc:20}
\eeq
In order to add angular momenta with $SU(2)$ Clebsch-Gordan coefficients,
the angular momenta need to be transformed to a representation where they all
undergo the same rotation, independent of the particle's momentum.
This is important for constructing many-body eigenstates of angular momenta.

The canonical boost is the unique boost that has this property.
In the $2 \times 2$ matrix representation the polar
decomposition of a general boost has the form
$B(p)= P(p) R(p)$ where $P(p)$ is a positive Hermitian matrix and
$R(p)$ is an $SU(2)$ rotation. The positive matrix $P(p)$ is the
$SL(2,\mathbb{C})$ matrix
representation of the corresponding canonical boost. Both $P(p)$ and
$R(p)$ are functions of a general boost $B(p)$ defined by:
\beq
P(p) = \left (B(p)B^{\dagger}(p) \right )^{1/2}
\qquad
R(p) = \left (B(p)B^{\dagger}(p) \right )^{-1/2}B(p).
\eeq
When $B(p)$ is a light-front boost the rotation $R(p)$ that relates
the canonical and light-front boosts is called a Melosh
rotation \cite{Melosh:1974cu}.
The way that the light-front spins are added is to start with a
tensor product representation.  For a basis vector representing a
two-body {\it rest} state, the single-particle light-front spins are
converted to canonical spins, which can then be added to the relative
orbital angular momenta using $SU(2)$ Clebsch-Gordan coefficients.
The resulting two-body rest state, which has the same rotational covariance
properties as (\ref{kc:8}), is boosted to the final total
momentum with a light-front boost.

For a non-interacting quark-anti-quark pair the total non-interacting 
four momentum is
\beq
P = p_q+p_{\bar{q}}.
\label{kc:21}
\eeq
The kinematic invariant mass is
\beq
M_0= \sqrt{-P^2}.
\label{kc:22}
\eeq
The boosts are parameterized by the total four velocity
\beq
Q=P/M_0.
\label{kc:23}
\eeq
The momentum of the quark in the rest frame of the
non-interacting quark-anti-quark pair is defined by transforming to the
rest frame with a light-front preserving boost:
\beq
k=k_{f} = k_q= B_f^{-1}(Q) p_q \qquad k_{f\bar{q}} = B_f^{-1}(Q) p_{\bar{q}}. 
\label{kc:24}
\eeq
Note that ${k}$ is not a 4-vector; it is
invariant with respect to light-front boosts.
To show this note that under a general Lorentz transformation, $\Lambda$,  $k$ transforms
like:
\beq
k'= B_f^{-1}(\Lambda Q) \Lambda p_q =
B_f^{-1}(\Lambda Q) \Lambda B_f ( Q) B_f^{-1}(Q) p_q =
B_f^{-1}(\Lambda Q) \Lambda B_f (Q) k.
\label{kc:25}
\eeq
If $\Lambda$ is a light-front boost
the rotation acting on $k$ is the light-front Wigner rotation of a light-front
boost which is the identity.  This gives $k'=k$.

The variables $\tilde{P}$ and $\tilde{\mathbf{k}}$ or 
$\tilde{P}$ and ${\mathbf{k}}$
can be taken as independent variables. 

The light-front components, $\tilde{\mathbf{k}}$, of $k$ are
\beq
\mathbf{k}_{\perp}=
\mathbf{k}_{q\perp} = 
\mathbf{p}_{q\perp} - {p_q^+ \over Q^+} \mathbf{Q}_{\perp}
\qquad 
k^+ = { p_q^+ \over Q^+} , 
\label{kc:26}
\eeq
which follows from (\ref{kc:18b}).
Light-front momentum fractions are defined by
\beq
\xi= {k^+ \over M_0}= {p_q^+\over P^+} \qquad 1-\xi = {p_{\bar{q}}^+ \over P^+}
\label{kc:27}
\eeq
which can be used to express (\ref{kc:26}) as
\beq
\mathbf{k}_{\perp}= \mathbf{p}_{q \perp}- \xi \mathbf{P}_{\perp}
\qquad  k^+ = M_0 \xi .
\label{kc:28}
\eeq

The $\hat{\mathbf{n}}$ component of $k$ is 
\beq
\nhat \cdot k = {1 \over 2} (k^+ - {\mathbf{k}_{\perp}^2 + m_q^2 \over k^+})
\label{kc:29}
\eeq
The dynamical component of the light-front four momentum is
\beq
P^-:= {\mathbf{P}_{\perp}^2 +M_0^2 \over P^+} .
\label{kc:30}
\eeq
The tensor product of the quark and anti-quark Hilbert spaces can be
decomposed into a direct integral of irreducible representation spaces
of the Poincar\'e group
in a light-front basis \cite{Keister:1991sb}.
In what follows the color and
flavor indices are suppressed.
The 
tensor product and two-particle Poincar\'e irreducible light-front bases
is related to the direct integral of irreducible representations by 
\[
\vert (\kappa ,j)\tilde{\mathbf{P}},\tilde{\mu} (l,s) \rangle =
\]
\[
\sum_{\tilde{\nu}_{{q}}, \tilde{\nu}_{\bar{q}},
\mu_q, \mu_{\bar{q}}, m, \mu_s} \int d\hat{\mathbf{k}}_q
\vert (m_q,j_q) \tilde{\mathbf{p}}_q,\tilde{\nu}_q \rangle
\otimes
\vert (m_{\bar{q}},j_{\bar{q}}) \tilde{\mathbf{p}}_{\bar{q}},\tilde{\nu}_{\bar{q}}
\rangle   
D^{j_q}_{\tilde{\nu}_{{q}} \mu_q}[R_{fc}(\mathbf{k}_q/m_q)]
D^{j_{\bar{q}}}_{\tilde{\nu}_{\bar{q}} \mu_{\bar{q}}}[R_{fc}(\mathbf{k}_{\bar{q}}/m_{\bar{q}})]
\times
\]
\[
\langle j_q, \mu_q, j_{\bar{q}}, \mu_{\bar{q}} \vert s, \mu_s \rangle 
Y_l^{m} (\hat{\mathbf{k}})
\langle s, \mu_s, l, m \vert j, \tilde{\mu} \rangle
\times
\]
\beq
\sqrt{{p_q^+ p_{\bar{q}}^+ (\omega_q({\kappa}_q) + \omega_{\bar{q}}({\kappa}_{\bar{q}}) )  \over \omega_q({\kappa}_q)  \omega_{\bar{q}}({\kappa}_{\bar{q}})  (p_q^+ + p_{\bar{q}}^+)}}.
\label{kc:31}
\eeq
The magnetic quantum numbers without the tildes are 
canonical spin labels.
The coefficients relating the tensor product to the irreducible two-body
states are Clebsch-Gordan coefficients for the Poincar\'e group in
the light-front basis.
In (\ref{kc:31}) $\tilde{\mathbf{p}}_{{q}}$ and 
$\tilde{\mathbf{p}}_{\bar{q}}$ on the right are functions of
$k_q$ implicitly defined by (\ref{kc:24}).
$\kappa := \vert \mathbf{k}_q\vert =
\mathbf{k}_{\bar{q}}$ and $\omega_q(\kappa)= \sqrt{m_q^2+\kappa^2}$.
The rotation
$R_{fc}(\mathbf{k}_q/m_q):= B_f^{-1}(\mathbf{k}_q/m_q)B_c(\mathbf{k}_q/m_q)$,
where $B_c(\mathbf{k}_q/m_q)$ is a rotationless (canonical) boost, 
transforms the canonical spins to light-front spins.
$R_{fc}(\mathbf{k}_q/m_q)$ is the
Melosh \cite{Melosh:1974cu} rotation discussed above.
The $SL(2,\mathbb{C})$  representation the rotationless boost
(the canonical boost) is  
\beq
B_c(p/m) = e^{{\pmb{\rho}\cdot \pmb{\sigma}\over 2}} =
\cosh (\rho/2) + \pmb{\hat{\rho}}\cdot \pmb{\sigma} \sinh (\rho/2) =
\sqrt{p^0 +m  \over 2m}I+ \hat{\mathbf{p}} \cdot \pmb{\sigma}
\sqrt{{p^0-m \over 2m}}
\label{kc:32}
\eeq
where $\rho$ is the rapidity of the Lorentz transformation.
The inverse is
\beq
B^{-1}_c(p/m) = e^{-{\pmb{\rho}\cdot \pmb{\sigma}\over 2}} =
\cosh (\rho/2) - \pmb{\hat{\rho}}\cdot \pmb{\sigma} \sinh (\rho/2) =
\sqrt{p^0 +m  \over 2m}I- \hat{\mathbf{p}} \cdot \pmb{\sigma}
\sqrt{p^0-m \over 2m}.
\label{kc:33}
\eeq
The canonical spins and orbital angular momenta rotate together,
independent of the quark and anti-quark momenta, so they can be
coupled with $SU(2)$ Clebsch-Gordan coefficients.

The Poincar\'e group Clebsch-Gordan coefficients can be applied to any
product of positive mass irreducible representations to construct a
two-body positive mass irreducible representation.


The two-particle irreducible basis and the bases in terms of momentum
fractions and transverse momenta are related by
\[
\vert \tilde{\mathbf{P}},\mathbf{k}_{q\perp},\xi_q ,
\tilde{\nu_1},\tilde{\nu}_2 \rangle =
\]
\[
\sum_{\mu_q, \mu_{\bar{q}}, m, \mu_s}
\vert (\kappa,j)\tilde{\mathbf{P}},\tilde{\mu} (l,s)\rangle
\langle j, \mu \vert s, \mu_s, l, \mu_l \rangle  
Y^{*l}_{m}(\hat{\mathbf{q}}) \langle s,\mu_s \vert 
j_q, \mu_q, j_{\bar{q}}, \mu_{\bar{q}} \rangle \times
\]
\beq
D^{j_q}_{\mu_q \nu_q}[R_{cf}(\mathbf{k}_q/m_q)]
D^{j_{\bar{q}}}_{\mu_{\bar{q}} \nu_{\bar{q}}}[R_{cf}(\mathbf{k}_{\bar{q}}/m_{\bar{q}})]
\sqrt{{\omega_{{q}}(\kappa)\omega_{\bar{q}}(\kappa)\over
(\omega_q(\kappa)+\omega_{\bar{q}}(\kappa))
\xi (1-\xi))}}.
\label{kc:35}
\eeq
These can be combined to get
\beq
\vert \tilde{\mathbf{P}},\mathbf{k}_{q\perp},\xi_q ,
\tilde{\nu_1},\tilde{\nu}_2 \rangle =
\vert (m_q,j_q) \tilde{\mathbf{p}}_q,\tilde{\nu}_q \rangle
\otimes
\vert (m_{\bar{q}},j_{\bar{q}}) \tilde{\mathbf{p}}_{\bar{q}},\tilde{\nu}_{\bar{q}}
\rangle   \sqrt{ {p^+_q p^+_{\bar{q}}\over
(p^+_q + p^+_{\bar{q}} )
\xi (1-\xi)}}.
\label{kc:36}
\eeq
Change of basis matrix elements can be read off from these matrix
elements.  These were formulated for  quark-anti-quark
states, but analogous relations hold for two hadron states.

Finally, for the purpose of constructing interacting light-front models
it is useful work in a basis where the angular momentum couplings are
removed from (\ref{kc:31}).  The advantage of this representation is
that rotationally covariant interactions are easily included
in this basis.
The result is a representation where
spins and single particle 3 momenta in the two particle rest frame transform 
together:
\[
\vert \tilde{\mathbf{P}}, \pmb{\kappa},  {\mu}_q,
{\mu}_{\bar{q}} \rangle :=
\]
\[
\sum_{\tilde{\nu}_{{q}}, \tilde{\nu}_{\bar{q}}}
\vert (m_q,j_q) \tilde{\mathbf{p}}_q (\tilde{\mathbf{P}}, \pmb{\kappa} ),\tilde{\nu}_q \rangle
\otimes
\vert (m_{\bar{q}},j_{\bar{q}}) \tilde{\mathbf{p}}_{\bar{q}}(\tilde{\mathbf{P}}, \pmb{\kappa} ),\tilde{\nu}_{\bar{q}}
\rangle   
D^{j_q}_{\tilde{\nu}_{{q}} \mu_q}[R_{fc}(\mathbf{k}_q/m_q)]
D^{j_{\bar{q}}}_{\tilde{\nu}_{\bar{q}} \mu_{\bar{q}}}[R_{fc}(\mathbf{k}_{\bar{q}}/m_{\bar{q}})]
\times
\]
\beq
\sqrt{{p_q^+ p_{\bar{q}}^+ (\omega_q({\kappa}_q) + \omega_{\bar{q}}({\kappa}_{\bar{q}}) )  \over \omega_q({\kappa}_q)  \omega_{\bar{q}}({\kappa}_{\bar{q}})  (p_q^+ + p_{\bar{q}}^+)}}.
\label{kc:37}
\eeq
In this representation the spins are not subsystem light-front spins, but
when they are added to the orbital angular momentum using
$SU(2)$ Clebsch-Gordan coefficients the resulting
spin is the light-front spin of the irreducible two-particle system.

\section{Confined Color Singlets - Bare Mesons}

The model degrees of freedom are local and global confined color
singlet quark-anti-quark-gluon states.  These are modeled by a
quark-anti-quark pair interacting via a confining interaction.  While
it is possible to use a color-dependent confining interaction, the
color indices are summed so the final degrees of freedom are local and
global color singlets.  Thus, in what follows the color degrees of freedom
of both the quarks and the interactions do not appear.

The model mass operator for a bare confined 
quark-anti-quark-glue singlet has the form
\beq
M_c = M_0 + U_c .
\label{cq:1}
\eeq
The confining interaction $U_c$ is chosen to have the form
\beq
U_c = 
\sqrt{\kappa^2 + V_c + m_q^2} + \sqrt{\kappa^2 + V_c + m_{\bar{q}}^2}
- M_0
\label{cq:2}
\eeq
where
\beq
V_c = -{\lambda^2 \over 4} \nabla^2_k + V_0
\label{cq:3}
\eeq
and $V_0$ and $\lambda$ are constants and $\kappa^2=\mathbf{k}^2$.
With this choice the
confined singlet quark-anti-quark invariant mass operator, $M_c$, is
\beq
M_c= \sqrt{\kappa^2 + V_c + m_q^2} + \sqrt{\kappa^2 + V_c + m_{\bar{q}}^2}.
\label{cq:4}
\eeq
For equal mass quarks and anti-quarks  this interaction is an addition to the 
square of the mass operator: $M_c^2 = M_0^2+ 4V_c$.  The form (\ref{cq:4})
is a generalization that can be used to treat unequal quark masses, which
can be used to model mesons with heavier quarks.
The only flavor dependence is assumed to be 
in the quark masses. 

This mass operator is the rest energy operator - which is 
the relativistic analog of the center-of-mass Hamiltonian.  
The mass operator (\ref{cq:4}) is a function of the
operator
\beq
\kappa^2 + V_c = 
\kappa^2   -{\lambda^2 \over 4} \nabla^2_{\pmb{\kappa}} + V_0
\label{cq:5}
\eeq
which is, up to constants, the Hamiltonian for a quantum mechanical
harmonic oscillator.  It follows that the wave functions of
(\ref{cq:1}) are harmonic oscillator wave functions, however the
spectrum is different; the eigenvalues of $M_c$ are obtained by
replacing $\kappa^2 + V_c$ in (\ref{cq:4}) by the harmonic oscillator
eigenvalues.  The resulting bare meson mass eigenvalues are
\beq
M_{nl} \to \sqrt{m_q^2 + \lambda (2n+l+{3 \over 2}) + V_0}  +
\sqrt{m_{\bar{q}}^2 + \lambda (2n+l+{3 \over 2}) + V_0}.  
\label{cq:6}
\eeq

While additional flavor dependent interactions would be needed for a
quantitatively more realistic model, that is not the goal of this work.  On
the other hand, as a few-body model, this model does not have
dynamical chiral symmetry breaking.  However a light pion (Goldstone
boson) is essential for a hadronic dynamics with the correct range.
This can be realized in this model by adding a spin-spin interaction 
that acts in
the $l=0$ states and gives the physical pion mass and the physical
$\pi$-$\rho$ mass splitting:
\beq
V_{ss} := (a + b \mathbf{s}_q \cdot \mathbf{s}_{\bar{q}})\delta_{l0} .
\label{cq:7}
\eeq
This interaction is easy to include in the model; it has the
advantage that is does not change the radial wave functions.

The mass operator obtained by adding this interaction to the confining
interaction is
\beq
M_c= \sqrt{\kappa^2 + V_c + m_q^2} + \sqrt{\kappa^2 + V_c + m_{\bar{q}}^2} 
+ V_{ss} .
\label{cq:8}
\eeq
The eigenvalue spectrum becomes
\beq
M_{nls} \to \sqrt{m_q^2 + \lambda (2n+l+{3 \over 2})+V_0}  +
\sqrt{m_{\bar{q}}^2 + \lambda (2n+l+{3 \over 2})+ V_0}  
+ \delta_{l0} ( a + {b\over 2}(s(s+1)))).
\label{cq:9}
\eeq
Solving for $a$ and $b$ to get the $\pi$ and $\rho$ masses
and splittings gives
\beq
a = m_{\pi} - 
\sqrt{m_q^2 + \lambda {3 \over 2}+ V_0}  -
\sqrt{m_{\bar{q}}^2 + \lambda {3 \over 2}+V_0}  
\label{cq:10}
\eeq
\beq
b= m_{\rho}-m_{\pi}.
\label{cq:11}
\eeq
The $\pi$ and $\rho$ masses are bare meson masses; In this and the
next section the constants $a$ and $b$ are chosen so the bare $\pi$ and
$\rho$ masses are the physical masses.  When the string-breaking
interaction is introduced the bare $\pi$ mass will be renormalized and
the $\rho$ will become unstable with a resonant peak shifted relative to
the bare value.  After the string breaking vertex is introduced the
constants $a$ and $b$ will be chosen so the bare $\pi$ and $\rho$ masses
become $m_{\pi-bare}=.16GeV$ and $m_{\rho-bare}=.882GeV$.
With these choices the string breaking interaction brings the
masses closer to their experimental values. 

The harmonic oscillator wave functions are 
known,
\beq
R_{nl} (r) = (-)^n ({\lambda \over 2})^{3/4}
\sqrt{{2 \Gamma (n+1) \over \Gamma (n+l+{3 \over 2})}}
(\sqrt{{\lambda \over 2}r})^l
L_n^{l+{1\over 2}}({\lambda \over 2}r^2)e^{-{\lambda \over 4} r^2}. 
\label{cq:11_r}
\eeq
In this expression the variable $r$ is the Fourier-Bessel transform
of the light-front invariant variable $\kappa$.
In momentum space they have the same form with
\beq
\tilde{R}_{nl} (\kappa) = (-)^n ({2 \over \lambda })^{3/4}
\sqrt{{2 \Gamma (n+1) \over \Gamma (n+l+{3 \over 2})}}
(\sqrt{{2 \over \lambda}\kappa})^l
L_n^{l+{1\over 2}}({2 \over \lambda }\kappa^2)e^{-{1 \over \lambda} \kappa^2}. 
\label{cq:11_k}
\eeq
In the non-interacting two-particle irreducible bases the coordinate
and momentum-space mass/spin eigenfunctions have the form
\beq
\langle \tilde{\mathbf{P}},j,\tilde{\mu},r,l,s\vert 
\tilde{\mathbf{P}}',j',\tilde{\mu}',n',l',s' \rangle =
\delta (\tilde{\mathbf{P}}-\tilde{\mathbf{P}}') 
\delta_{\tilde{\mu} \tilde{\mu}'} \delta_{j'j}
\delta_{s's} \delta_{l'l}  
R_{n'l'} (r)
\label{cq:17}
\eeq
and
\beq
\langle \tilde{\mathbf{P}},j,\tilde{\mu},\kappa,l,s\vert 
\tilde{\mathbf{P}}',j',\tilde{\mu}',n',l',s' \rangle =
\delta (\tilde{\mathbf{P}}-\tilde{\mathbf{P}}') 
\delta_{\tilde{\mu} \tilde{\mu}'} \delta_{j'j}
\delta_{s's} \delta_{l'l}  
\tilde{R}_{n'l'} (\kappa)
\label{cq:18}
\eeq
respectively.  In these expressions the r-space basis
functions are
obtained from the corresponding momentum-space basis functions
using a Fourier-Bessel transform on the momentum-space functions.
The total light-front momentum is not transformed.

A unitary irreducible representation of the Poincar\'e group consistent
with this dynamics can be defined on the mass-spin basis states,
$\vert \tilde{\mathbf{P}},j,\tilde{\mu},n,l,s\rangle$.  These states span
a subspace ${\cal H}_{j,n,l,s}$.  This subspace is invariant
under the following light-front unitary representation of the Poincar\'e
group
\beq
U_{njls}(\Lambda , a) \vert \tilde{\mathbf{P}},j,\tilde{\mu},n,l,s\rangle =
e^{-i a\cdot \Lambda P_{nls}}\sum_{\tilde{\nu}=-j}^j
\vert \tilde{\pmb{\Lambda}}P_{nls},j,\tilde{\nu},n,l,s\rangle
\sqrt{{(\Lambda P_{nls})^+ \over P^+}}
D^j_{\tilde{\nu} \tilde{\mu}}[B_f^{-1} (\Lambda P_{nls})\Lambda B_f(P_{nls})] 
\label{cq:19}
\eeq
where the dynamics enters in $P^-$: 
\beq
P_{nls} = (P^+,\mathbf{P}_{\perp} ,P^-_{nls})=
(P^+,\mathbf{P}_{\perp} , {\mathbf{P}_{\perp}^2 + M^2_{nls}
\over P^+}).
\label{cq:20}
\eeq
Note that the mass in this expression is the eigenvalue of the 
confining mass operator.  What makes this a light-front dynamics 
is that the mass eigenvalues do not appear on the right hand 
side of equation  (\ref{cq:19}) 
for Poincar\'e transformations $\Lambda$ that leave the light front invariant. 

This dynamical representation differs from the non-interacting
two-particle irreducible representation by the replacement $M_0\to
M_{nls}$.  Since the model has additional string breaking
interactions, this is not the physical unitary representation of the
Poincar\'e group.  Note that in this light-front representation
rotational invariance is exactly preserved, although it is a dynamical
transformation since it depends on the mass eigenvalues.

The Hadronic representation of the Hilbert space is the infinite direct sum of
these irreducible subspaces over the confined (bare) mass channels
\beq
{\cal H}_H := \oplus {\cal H}_{njls}.
\label{cq:21}
\eeq 
The corresponding dynamical unitary representation of the Poincar\'e
group on ${\cal H}_H$ is
\beq
U_H(\Lambda ,a ) = \sum_{njls} U_{njls}(\Lambda,a)
\label{cq:22}
\eeq
where each $U_{njls}(\Lambda,a): 
{\cal H}_{njls} \to{\cal H}_{njls}$.
The wave functions 
$\langle \tilde{\mathbf{P}}, \mathbf{k}, \mu_{{q}}, \mu_{\bar{q}}
\vert \tilde{\mathbf{P}}', n, j, l, s, \mu \rangle 
$
define a unitary mapping $\Phi_{njls}$ from the irreducible 
representation space ${\cal H}_{njls}$ to the two-free-quark
space, ${\cal H}_Q$.  The sum
\beq
\Phi := \sum_{njls} \Phi_{njls}
\label{cq:23}
\eeq
defines a unitary map from the hadronic Hilbert space ${\cal H}_H$
to the quark-anti-quark Hilbert space ${\cal H}_Q$.  The
unitary representation of Poincar\'e group on ${\cal H}_{Q}$
in the quark representation is
defined by
\beq
U_Q(\Lambda ,a) = \Phi U_H(\Lambda ,a) \Phi^{\dagger}.
\label{cq:24}
\eeq
The quark representation is needed to compute current matrix
elements involving electrically charged quarks.
\section{Properties of the Confining Interaction}

The spectrum of the confined singlet mass operator  is 
\beq
M_{nls} \to \sqrt{m_q^2 + \lambda (2n+l+{3 \over 2})+V_0}  +
\sqrt{m_{\bar{q}}^2 + \lambda (2n+l+{3 \over 2})+ V_0}  
+ \delta_{l0} ( a + {b\over 2}(s(s+1))).
\label{pci:1}
\eeq
The RMS relative displacement and momentum of the quark and anti-quark
in each of the oscillator states is
\beq
\langle r^2_{nls} \rangle^{1/2} = \sqrt{{2 \over \lambda}( 
2n +l   + {3 \over 2})}
\qquad
\langle k^2_{nls} \rangle^{1/2} = \sqrt{{\lambda\over 2}(
2n +l   + {3 \over 2})}.
\label{pci:2}
\eeq
The following scaling relations emerge from these expressions in
the limit that the oscillator quantum number, $2n+l$, gets
large
\beq
M_{nls}  \approx \sqrt{2} \lambda \langle r_{nls}^2 \rangle^{1\over 2}  
\label{pci:3}
\eeq
This shows an asymptotically linear confinement with respect to the
light-front invariant variable $r$.
In addition as $l$ gets large
\beq
l \approx {1 \over 4 \lambda} M_{nls}^2 = \alpha M_{nls}^2
\label{pci:4}
\eeq
which shows Regge behavior and as $n$ gets large.  
These observations suggest that in spite of its simplicity, this model
has some properties that are qualitatively consistent with
observations.  Note that models with a similar structure, where the
mass square eigenstates are harmonic oscillator wave functions, also
emerge from Holographic QCD \cite{deTeramond:2016}.

The parameters of this model are the quark masses, the oscillator
coupling strength $\lambda$, the constant $V_0$, and the parameters
$a$ and $b$ in the spin-spin interaction.  The quark masses and $V_0$
appear in the spectrum in Hamiltonian in the combination,
$m_q^2 + V_0$, so they are essentially the same parameter.  The difference
is only relevant for matrix elements of currents. 

In the absence of the string-breaking interaction, which will be
introduced in the next section, the parameters $a$ and $b$ can be
chosen to get the experimental pion mass and $\pi-\rho$ mass splitting.
The quark masses and coupling constant can be chosen to get the correct
Regge slope and intercept.  While there is not a single Regge slope or
intercept, all of the slopes are similar.

In what follows the Regge parameters from the family containing the 
$\rho$ meson ($\rho$ and $a$ mesons), which has \cite{faustov} $\alpha=.887 (GeV)^{-2}$ and $\alpha_0= .456$, are used to 
fix $\lambda=.282$.  The intercept fixes the value of
\beq
m_q^2 + {3 \over 2} \lambda + V_0
\label{pci:6}
\eeq
$m_q^2$ can be given any value provided it is compensated with $V_0$.
Setting $V_0$ to zero requires  $m_q^2$ be negative to get the
Regge intercept.  For the choice ${3 \over 2} \lambda + V_0 =0$,
the Regge trajectory can be approximately realized with $m_q=m_\rho/2$.
This work initially uses $m_q=.385 GeV$.

With this choice of $\lambda$ the pion $RMS$ radius is
\beq
\sqrt{{3 \over \lambda}} \approx 3.3 (GeV)^{-1}=.64 fm 
\label{pci:7}
\eeq
Note that in the light-front representation, this is not the same as
the charge radius due to the additional momentum dependence in the
Melosh rotations that are needed to couple the quark spins to a
rotationally invariant eigenstate.  

The mean quark momentum in the pion is 
\beq
\sqrt{{3 \lambda \over 4}} \approx .46 (GeV)
\label{pci:8}
\eeq
which is still relativistic even for the relatively heavy 
quark masses.

The calculated (bare) and measured masses in
the $\rho$ meson's Regge trajectory are given in table I.
\begin{table}[t]
\caption{\label{table.1}Regge trajectories, $J=L+1, S=1$ $m_q=.385$, $\lambda=.282$}
\begin{tabular}{lllllll}
\hline
\hline
meson &l & exp. mass & exp. (mass)$^2$& j & calc. mass  & calc. (mass)$^2$  \\
\hline
$\rho$   & 0 &0.770 & 0.593 &1&  0.770 & 0.593   \\   
$a_2$    & 1 &1.320 & 1.742 &2&  1.311 & 1.719   \\
$\rho_3$ & 2 &1.690 & 2.856 &3&  1.687 & 2.846   \\
$a_4$    & 3 &2.040 & 4.162 &4&  1.994 & 3.976   \\
$\rho_5$ & 4 &2.350 & 5.522 &5&  2.259 & 5.103   \\
$a_6$    & 5 &2.450 & 6.000 &6&  2.497 & 6.335   \\
\hline
\end{tabular}
\label{omega}
\end{table}
The table exhibits qualitative agreement with the data
from the particle data book.
The general Regge behavior and linear confinement are
illustrated in figures 1 and 2.  The first figure shows
the Regge trajectories for different values of the principal
quantum number.  The first figure is a plot the square of the masses
in table 1 as a function of $l$.  The second figure shows   
approximately linear confinement by plotting the bare masses 
against the RMS quark-anti-quark separation.  The numerical values are
in table II.
\begin{figure}
\centering
\includegraphics[width=1\linewidth]{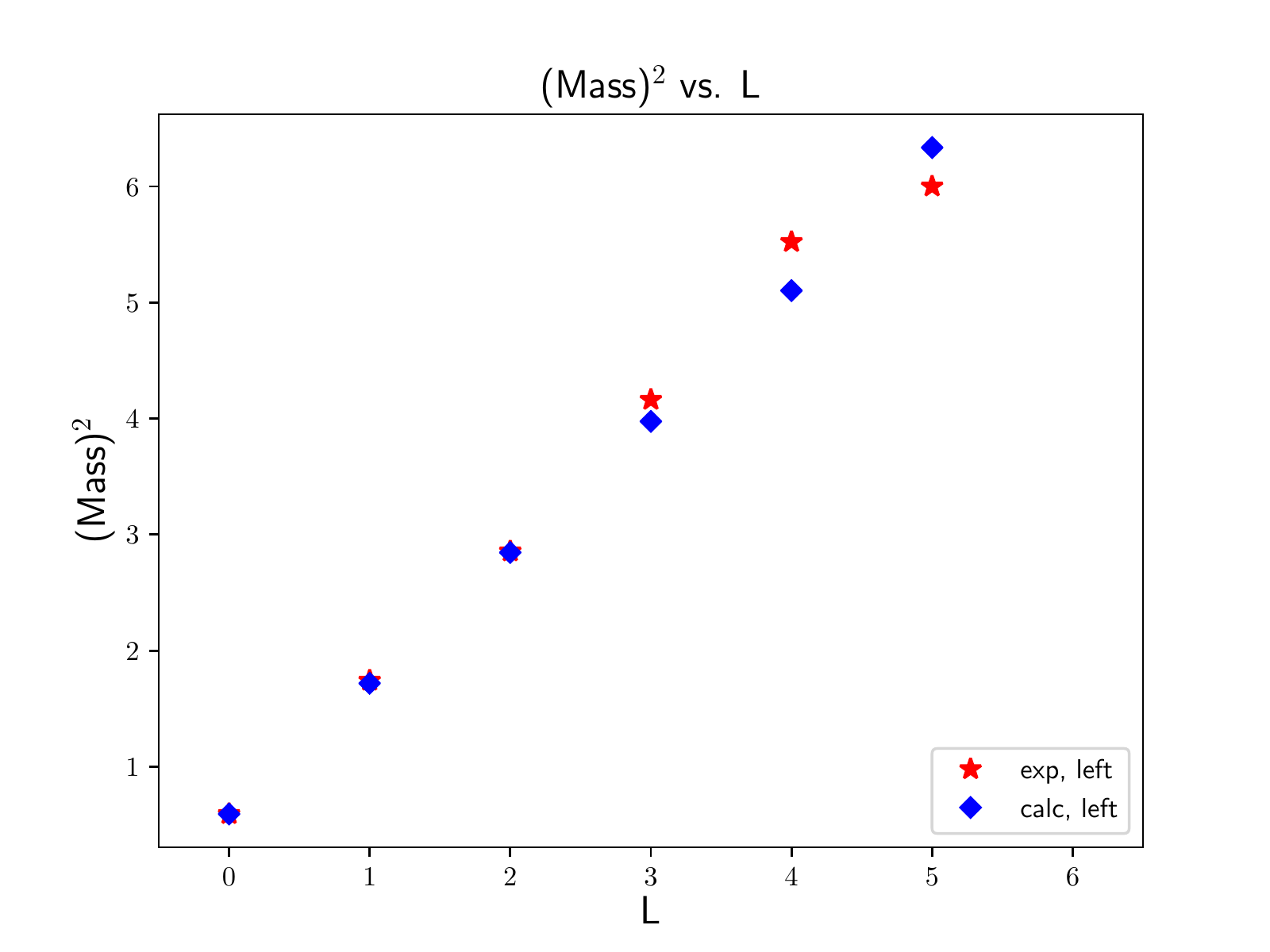}
\caption{\label{fig.1}Regge slope}
\label{fig:1}
\end{figure}

\begin{figure}
\centering
\includegraphics[width=1\linewidth]{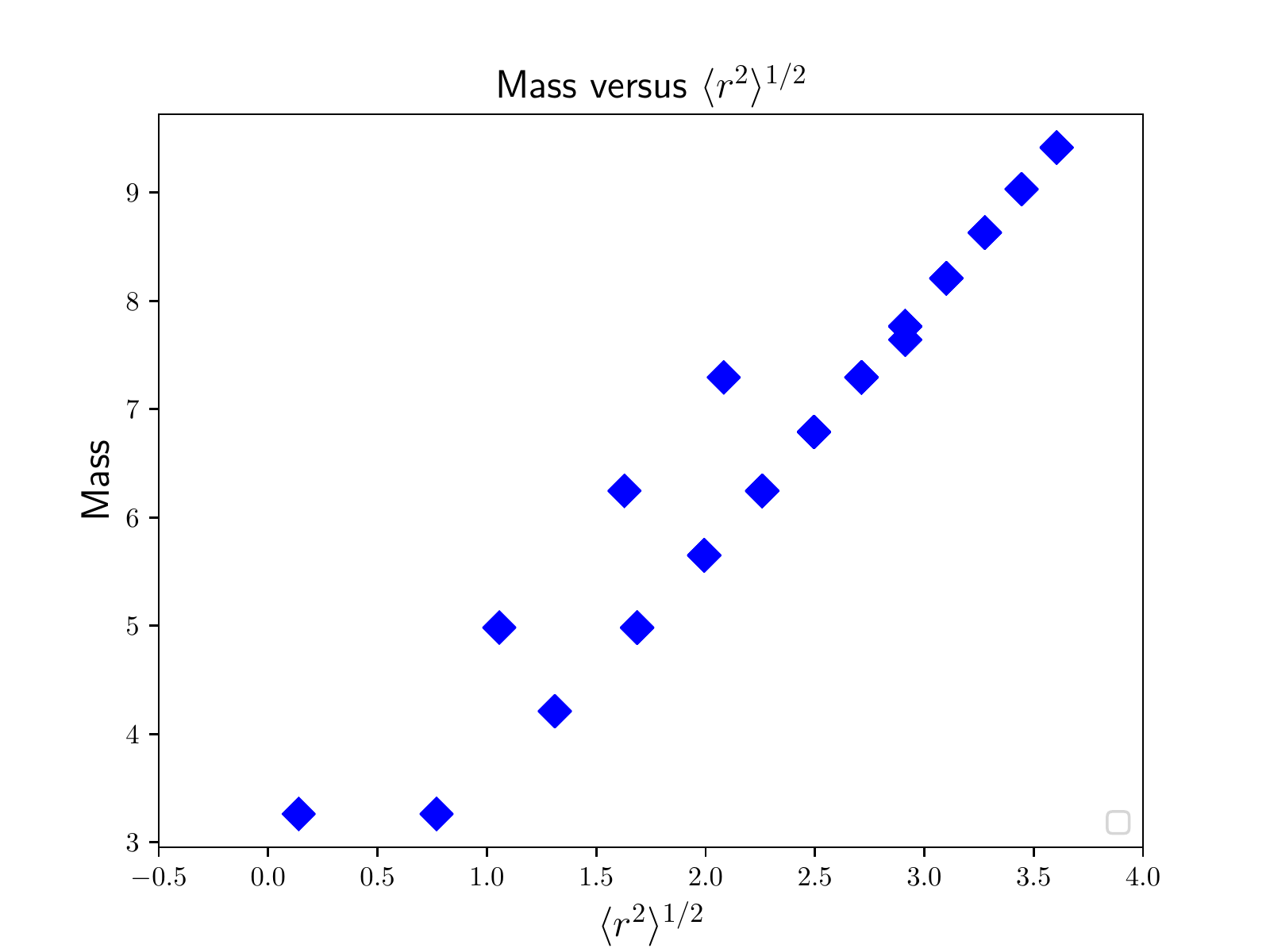}
  \caption{\label{fig.2}Mass versus r}
  \label{fig:2}
\end{figure}

\begin{table}[t]
\caption{\label{table.2}Meson mass versus RMS radius}
\begin{tabular}{llllllllll}
\hline  
\hline		
$n$ & $l$ & $s$ & mass(GeV) & radius(GeV)$^{-1}$ & $n$ & $l$ & $s$ & mass(GeV) & radius(GeV)$^{-1}$\\ 
\hline
 0 & 0&  0& 0.140 & 3.261 & 2 & 0&  0& 1.629 & 6.245\\ 
 0 & 0&  1& 0.770 & 3.261 & 2 & 0&  1& 2.259 & 6.245\\
 0 & 1&  0& 1.311 & 4.210 & 2 & 1&  0& 2.496 & 6.789\\
 0 & 1&  1& 1.311 & 4.210 & 2 & 1&  1& 2.496 & 6.789\\
 0 & 2&  0& 1.687 & 4.982 & 2 & 2&  0& 2.713 & 7.293\\
 0 & 2&  1& 1.687 & 4.982 & 2 & 2&  1& 2.713 & 7.293\\
 0 & 3&  0& 1.994 & 5.649 & 2 & 3&  0& 2.913 & 7.764\\
 0 & 3&  1& 1.994 & 5.649 & 2 & 3&  1& 2.913 & 7.764\\
 0 & 4&  0& 2.259 & 6.245 & 2 & 4&  0& 3.101 & 8.208\\
 0 & 4&  1& 2.259 & 6.245 & 2 & 4&  1& 3.101 & 8.208\\
 0 & 5&  0& 2.496 & 6.789 & 2 & 5&  0& 3.277 & 8.629\\
 0 & 5&  1& 2.496 & 6.789 & 2 & 5&  1& 3.277 & 8.629\\
 1 & 0&  0& 1.057 & 4.982 & 3 & 0&  0& 2.083 & 7.293\\
 1 & 0&  1& 1.687 & 4.982 & 3 & 0&  1& 2.713 & 7.293\\
 1 & 1&  0& 1.994 & 5.649 & 3 & 1&  0& 2.913 & 7.640\\
 1 & 1&  1& 1.994 & 5.649 & 3 & 1&  1& 2.913 & 7.764\\
 1 & 2&  0& 2.259 & 6.245 & 3 & 2&  0& 3.101 & 8.208\\
 1 & 2&  1& 2.259 & 6.245 & 3 & 2&  1& 3.101 & 8.208\\
 1 & 3&  0& 2.496 & 6.789 & 3 & 3&  0& 3.277 & 8.629\\
 1 & 3&  1& 2.496 & 6.789 & 3 & 3&  1& 3.277 & 8.629\\
 1 & 4&  0& 2.713 & 7.293 & 3 & 4&  0& 3.445 & 9.031\\
 1 & 4&  1& 2.713 & 7.293 & 3 & 4&  1& 3.445 & 9.031\\
 1 & 5&  0& 2.913 & 7.764 & 3 & 5&  0& 3.605 & 9.415\\
 1 & 5&  1& 2.913 & 7.764 & 3 & 5&  1& 3.605 & 9.415\\
 \hline
\end{tabular}
\label{tablex}
\end{table}

\section{String Breaking}

The second class of interactions allow the confined singlets to
interact.  The interaction is assumed to be a string-breaking
interaction that assumes in the first approximation that a
quark-anti-quark pair is produced with equal probability at any point
on a line between the quark and anti-quark, causing it to break up into
a pair of confined singlets.  The interaction is taken to be local in
the sense that the quark anti-quark pair is produced at a point.  This is
consistent with the assumption that string breaking is generated by
the local covariant derivative operator.
Since
this naive interaction is singular, the delta functions that keep the
produced pair on the line between the original quark and anti-quark are
replaced by Gaussian approximations of delta functions where the width
of the Gaussian is the same as the oscillator ground state.  This has
the effect of fattening the string to a flux tube with a width consistent
with the size of the ground state.  Given that QCD has only one
coupling constant, it seems reasonable that the scale that determines
the bare meson sizes should also determine the bare hadronic vertex
form factors.  Subsequent calculations using this vertex indicate that
in order to get a consistent description of the Regge trajectories and
the $\rho$ lifetime, it is necessary to use a unitary scale
transformation to reduce the width of the flux tube by a factor of two
(see equations (\ref{sb:37}-\ref{sb:38}).  This transformation is applied after
the vertex is computed.  The resulting interaction is still consistent
with the scale set by the confining interaction.

An additional virtue of this string breaking vertex is that it is
possible to analytically perform the nine dimensional integrals that
smear the vertex with one initial and two final confined quark
anti-quark singlets.  This property simplifies calculations with sea
quarks.

The light-front translationally invariant, spin-independent part 
of the string-breaking vertex is taken to have a $r$-space 
kernel of the form
\beq
\langle \mathbf{r}_1,\mathbf{r}_2,\mathbf{r}_{12} \vert v_{2:1} \vert \mathbf{r}
\rangle := 
\gamma  \delta (\mathbf{r}-2\mathbf{r}_{12}) \int_0^1 d\eta
\delta_{\sqrt{{\lambda \over 2}}} (\mathbf{r}_1 - \eta \mathbf{r})
\delta_{\sqrt{{\lambda \over 2}}} (\mathbf{r}_2 - (1-\eta) \mathbf{r})
\label{sb:1}
\eeq
where the Gaussian approximate delta function is 
\beq
\delta_{\sqrt{{\lambda \over 2}}} (\mathbf{r}):=
({\lambda \over 4 \pi})^{3/2}
e^{-\lambda r^2 \over 4}
\qquad 
\int \delta_{\sqrt{{\lambda \over 2}}} (\mathbf{r})
d\mathbf{r} = 1 .
\label{sb:2}
\eeq

The coordinates are defined as Fourier transforms of the light-front
invariant
momentum variable
$\mathbf{k}$.
In order for matrix elements of this vertex 
to have dimensions of energy the matrix element
\beq
\langle \psi \vert v_{1:2} v_{2:1} \vert \psi \rangle
\qquad v_{1:2} :=v_{2:1}^{\dagger} 
\label{sb:3}
\eeq
should have dimension (energy)$^2$.  It follows that 
$\gamma$ should have dimensions of (energy)$^{-1}$.
Since $\lambda$ has dimension (energy)$^2$,
it can be replaced by
\beq
\gamma \to g/\sqrt{\lambda} 
\label{sb:4}
\eeq
which makes $g$ a dimensionless factor.  Since the model assumption is
that one parameter should fix all scales, the dimensionless parameter
$g$ should be of order unity.  In the calculations that follow $g$ was
taken to be $5.44$.   This choice gives a qualitatively consistent picture
of the $\rho$ lifetime and the $\pi -\pi$ scattering cross section.

The variables $\mathbf{r}_1$ and $\mathbf{r}_2$ are the displacement
variables of the quark-anti-quark pairs in the produced confined
singlets.  The variable $\mathbf{r}$ is the displacement
variable for the quark-anti-quark pair in the initial
confined singlet.  The variable $\mathbf{r}_{12}$ represents the
displacement between the centers of the two produced singlets.
The assumption that the quark anti-quark pair is produced at a point
leads to $\mathbf{r} = 2\mathbf{r}_{12}$.  The general structure is illustrated in figure 3.

\begin {figure}
\centering
\includegraphics[width=0.7\linewidth]{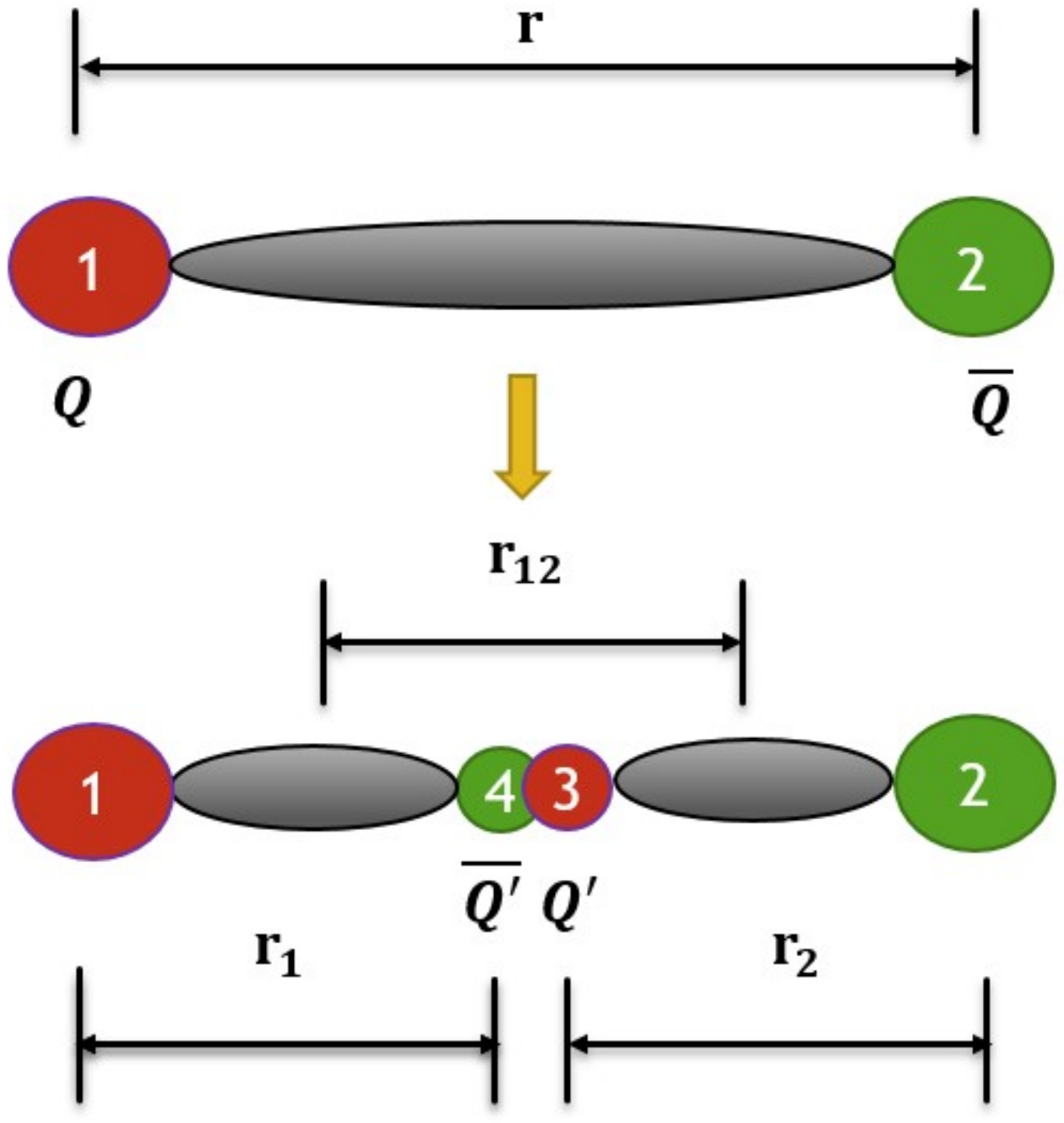}
\caption{\label{fig.3}String breaking vertex}
\label{fig:3}
\end{figure}

In the limit that the Gaussians become delta functions the interaction
makes the string break with equal probability at any point along the
line between the initial quark and anti-quark.

The hadronic vertices, which are defined by the overlap of this
string-breaking vertex with three harmonic oscillator states can be
computed analytically.  The result, which is derived in appendix I and II
is 
\[
\langle n_1,l_1,m_1,n_2,l_2,m_2,\mathbf{r}_{12} \vert v_{2:1} \vert
n,l,m \rangle =
\]
\[
\int 
\psi_{n_1l_1m_1}^*(\mathbf{r}_1)
\psi_{n_2l_2m_2}^*(\mathbf{r}_2)
v(\mathbf{r}_1,\mathbf{r}_2,\mathbf{r}_{12};\mathbf{r})
\psi_{nlm}(\mathbf{r})
d\mathbf{r}_1 d \mathbf{r}_2 d \mathbf{r} =
\]
\[
{g \over \sqrt{\lambda}}
R_{nl}(2 {r}_{12})
({2\lambda})^{3/2}
{(\sqrt{{\lambda \over 2}}{r}_{12})^{2n_1+l_1+2n_2+l_2} 
\over
\sqrt{2 n_1! \Gamma (n_1+l_1+{3 \over 2}})
\sqrt{2 n_2! \Gamma (n_2+l_2+{3 \over 2}})
}\times
\]
\[
e^{-{\lambda \over 4} r_{12}^2} \sum_{k_1+k_2=2r} 
{(l_1+2n_1)! (l_2+2n_2)! \over
k_1! k_2! (l_1+2n_1-k_1)! (l_2+2 n_2 -k_2)!}(-)^{k_2}
({1 \over 2})^{l_1+2n_1+l_2 +2n_2 } 
\times
\]
\beq
{1 \over 2r+1}
M({1 \over 2}+r ,{3 \over 2}+r, -{\lambda r^2_{12} \over 4})
Y_{lm} (\hat{\mathbf{r}}_{12})
Y_{l_1m_1}^*(\hat{\mathbf{r}}_{12})
Y_{l_2m_2}^*(\hat{\mathbf{r}}_{12}).
\label{sb:22}
\eeq
where $M({1 \over 2}+r ,{3 \over 2}+r, -{\lambda r^2_{12} \over 4})$
is the confluent hypergeometric function.

This matrix element does not include any spin dependence.  The model
assumption is that the string breaking creates a quark anti-quark pair
out of the vacuum at the point where the string breaks.  In what
follows the initial quark-anti-quark is labeled $1-2$, the pair created
out of the vacuum is labeled $3-4$ and the final quark anti-quark pairs
are labeled $1-4$ and $2-3$ (see figure 3).  Since the quark and anti-quark have
opposite parity, if they are created out of the vacuum the pair must
be in an odd l state.  Since the spin of the pair can be 0 or 1, to
get $j=0$ and $l$ odd the quark and anti-quark must be created with
$s=l=1$.  The following spin-dependent addition is motivated by
assuming that the string-breaking operator is oriented 
parallel to the original quark anti-quark pair
\beq 
Y_{1 m}(\hat{\mathbf{r}}_{12}) 
\langle s_3, \mu_3, s_4, \mu_4 
\vert 1, \mu_s \rangle \langle 1 , m_l, 1, \mu_s \vert 0, 0\rangle . 
\label{sb:25}
\eeq
This operator is included as a dimensionless multiplicative factor.
The directional dependence is motivated by the string breaking model, where the
breaking is caused by a quark-anti-quark-link oriented parallel to the
line between the quark and anti-quark.  
The $Y^{1}_{m_1}
(\hat{\mathbf{r}}_{12})$ gives the spherical components of the unit
vector along the line between original the quark and anti-quark.  The
full-spin dependence is obtained by multiplying this factor by
an additional spin factor that couples the above to the spins of the 
other quarks and anti-quarks
\beq
\langle s_{23}, \mu_{s23} \vert s_2, \mu_2, s_3, \mu_3 \rangle  
\langle s_{14}, \mu_{s14} \vert s_1, \mu_1, s_4 ,\mu_4 \rangle  
\langle s_1, \mu_1, s_2, \mu_2 \vert s_{12}, \mu_{s12} \rangle
\label{sb:26}
\eeq
and summing over the single quark spins.  

The spin-dependent vertex is obtained by multiplying
the spin-independent vertex by these two spin-dependent factors
to get:
\[
\langle n_{23}, l_{23},m_{23},s_{23},\mu_{s23},n_{14},l_{14},m_{14},
s_{14},\mu_{s14},
\mathbf{r}_{12} \vert v_{2:1} \vert
n_{12},l_{12},m_{12},s_{12},\mu_{s12}  \rangle  
\]
\[
\sum
\langle n_{23}, l_{23},m_{23} ,n_{14},l_{14},m_{14},
\mathbf{r}_{12} \vert v_{2:1} \vert
n_{12},l_{12},m_{12} \rangle
Y_{1, m_{1}}(\hat{\mathbf{r}}_{12}) \times
\]
\[
\langle {1 \over 2}, \mu_3, {1 \over 2}, \mu_4 
\vert 1, \mu_s \rangle \langle 1 ,m_1, 1, \mu_s \vert 0, 0\rangle  
\times
\]
\beq
\langle s_{23}, \mu_{s23} \vert s_2, \mu_2, s_3, \mu_3 \rangle  
\langle s_{14}, \mu_{s14} \vert s_1, \mu_1, s_4 ,\mu_4 \rangle  
\langle s_1, \mu_1, s_2, \mu_2 \vert s_{12}, \mu_{s12} \rangle .
\label{sb:27}
\eeq
The radial dependence can be projected on angular momentum
states by multiplying by the  spherical harmonic $\langle 
l_v, m_v \vert \hat{\mathbf{r}}_{12} \rangle$ and integrating over angles. 
The resulting angular integral is over a product of five spherical harmonics
\[
\langle 1_{14},m_{14} \vert \hat{\mathbf{r}}_{12} \rangle 
\langle 1_{23},m_{23} \vert \hat{\mathbf{r}}_{12} \rangle 
\langle l_{v},m_{v} \vert \hat{\mathbf{r}}_{12} \rangle
\langle \hat{\mathbf{r}}_{12} \vert l_{12} m_{12} \rangle
\langle \hat{\mathbf{r}}_{12} \vert 1,m_{l34} \rangle =
\]
\beq
Y^*_{l_{14} m_{14}} (\hat{\mathbf{r}}_{12})
Y^*_{l_{23} m_{23}} (\hat{\mathbf{r}}_{12}) 
Y^*_{l_{v} m_{v}} (\hat{\mathbf{r}}_{12}) 
Y_{l_{12} m_{12}}(\hat{\mathbf{r}}_{12})
Y_{1 m_{1}} (\hat{\mathbf{r}}_{12}).
\label{sb:28}
\eeq
This integral can be expressed as sums of
products of Clebsch-Gordan coefficients:
\[
\int d\hat{\mathbf{r}}_{12}
Y^*_{l_{14} m_{14}} (\hat{\mathbf{r}}_{12})
Y^*_{l_{23} m_{23}} (\hat{\mathbf{r}}_{12}) 
Y^*_{l_{v} m_{v}} (\hat{\mathbf{r}}_{12}) 
Y_{l_{12} m_{12}}(\hat{\mathbf{r}}_{12})
Y_{1 m_{34}} (\hat{\mathbf{r}}_{12})=
\]
\beq
\sum_{l_t,l_{12:34}}
\langle l_{14}, m_{14}, l_{23}, m_{23} \vert l_{14:23}, m_{l_{14:23}} \rangle
\langle l_{14:23}, m_{l_{14:23}}, l_{v}, m_{v} \vert l_t, m_t \rangle 
\langle l_t, m_t \vert l_{12}, m_{12}, 1, m_{34} \rangle .
\label{sb:29}
\eeq
Since the bare meson states are eigenstates states of total 
angular momentum, it is useful to couple the spins and
orbital angular momenta of each bare meson state. 
This involves multiplying by three more Clebsch-Gordan coefficients.
The result for the vertex is 
\[
\langle n_{23},j_{23}, l_{23},s_{23},\mu_{23},n_{14},j_{14},l_{14},s_{14},
\mu_{14}
{r}, l_v, m_v  \vert v_{2:1} \vert
n_{12},j_{12},l_{12},s_{12}, \mu_{12}  \rangle =
\]
\beq
\langle n_{23}, l_{23}, n_{14}, l_{14},
{r} \vert v_{2:1} \vert
n_{12},l_{12} \rangle \times
C(j_{23}, l_{23},s_{23},\mu_{23},j_{14},l_{14},s_{14},
\mu_{14}, l_v, m_v ; 
j_{12},l_{12},s_{12}, \mu_{12})
\label{sb:30}
\eeq
where
\[
C(j_{23}, l_{23},s_{23},\mu_{23},j_{14},l_{14},s_{14},
\mu_{14}, l_v, m_v ; 
j_{12},l_{12},s_{12}, \mu_{12}) :=
\]
\[
\sum
\langle j_{14},\mu_{14} \vert l_{14}, m_{14}, s_{14} ,\mu_{s14} \rangle 
\langle j_{23},\mu_{23} \vert l_{23}, m_{23}, s_{23} ,\mu_{s23} \rangle 
\times
\]
\[
\langle l_{14}, m_{14}, l_{23}, m_{23} \vert l_{14:23}, m_{l_{14:23}} \rangle
\langle l_{14:23}, m_{l_{14:23}}, l_{v}, m_{v} \vert l_t, m_t \rangle 
\langle l_t, m_t \vert l_{12}, m_{12}, 1, m_{34} \rangle 
\times
\]
\[
\langle s_{23}, \mu_{s23} \vert s_2, \mu_2, s_3, \mu_3 \rangle  
\langle s_{14}, \mu_{s14} \vert s_1, \mu_1, s_4 ,\mu_4 \rangle  
\langle s_1, \mu_1, s_2, \mu_2 \vert s_{12}, \mu_{s12} \rangle
\times
\]
\beq
\langle s_3, \mu_3, s_4, \mu_4 
\vert 1, \mu_s \rangle \langle 1 ,m_l, 1, \mu_s \vert 0, 0\rangle  
\langle l_{12}, m_{12}, s_{12}, \nu_{s12} \vert j_{12}, \nu_{12} \rangle .
\label{sb:31}
\eeq
This has the form of the product of the
spin-independent vertex (\ref{sb:24}) multiplied by a spin-dependent
coefficient (\ref{sb:31}).

The more useful form is the momentum-space version of these 
vertices.  They can be obtained by performing a Fourier-Bessel transform
of the vertex (\ref{sb:30}).  Because the spin-independent vertex factors out
of this expression it is
enough to replace the spin-independent coefficient
\beq
\langle n_{23}, l_{23}, n_{14}, l_{14},
{r} \vert v_{2:1} \vert
n_{12},l_{12} \rangle
\label{sb:31b}
\eeq
in the above expression by its Fourier-Bessel transform
\beq
\langle n_{23}, l_{23}, n_{14}, l_{14},
{\kappa} ,l_v \vert v_{2:1} \vert
n_{12},l_{12} \rangle
:=
\sqrt{{2 \over \pi}} \int_0^\infty r^2 dr j_{l_v} (\kappa r) 
\langle n_{23}, l_{23}, n_{14}, l_{14},
{r} \vert v_{2:1} \vert
n_{12},l_{12} \rangle
\label{sb:32}
\eeq
so the momentum-space vertex becomes
\[
\langle n_{23},j_{23}, l_{23},s_{23},\mu_{23},n_{14},j_{14},l_{14},s_{14},
\mu_{14},
\kappa , l_v, m_v  \vert v_{2:1} \vert
n_{12},j_{12},l_{12},s_{12}, \mu_{12}  \rangle =
\]
\beq
\langle n_{23}, l_{23}, n_{14}, l_{14},
\kappa ,l_v \vert v_{2:1} \vert
n_{12},l_{12} \rangle \times
C(j_{23}, l_{23},s_{23},\mu_{23},j_{14},l_{14},s_{14},
\mu_{14}, l_v, m_v ; 
j_{12},l_{12},s_{12}, \mu_{12}) .
\label{sb:33}
\eeq

Both the strength and size scale of the vertex is determined by the
same parameter that is responsible for the confinement.  This
elementary string breaking vertex at the quark level leads to analytic
expressions for all vertices relating one bare confined meson
eigenstate to two bare confined meson eigenstates as a function of the
initial relative quark-anti-quark displacement.

The only integral that needs to be computed numerically is the
one-dimensional radial integral in the Fourier-Bessel transform.
While the model is too crude to expect that all of these vertices lead
to accurate results, the hope is that they provide a rough
characterization of the size of the contributions from the higher
lying states assuming that the physics is largely determined by the
string-breaking mechanism.

This vertex has the property that the kernel of this 
operator is rotationally covariant:
\[
\sum
D^{j_{23}}_{\mu_{23} \nu_{23}}[R]
D^{j_{14}}_{\mu_{14} \nu_{14}}[R]
D^{l_v}_{m_v m_v'}[R]
\langle n_{23},j_{23}, l_{23},s_{23},\nu_{23},n_{14},j_{14},l_{14},s_{14},
\nu_{14}
\kappa , l_v, m_v'  \vert v_{2:1} \vert
n_{12},j_{12},l_{12},s_{12}, \mu_{12}  \rangle =
\]
\beq
\sum
\langle n_{23},j_{23}, l_{23},s_{23},\mu_{23},n_{14},j_{14},l_{14},s_{14},
\mu_{14},\kappa , l_v, m_v  \vert v_{2:1} \vert
n_{12},j_{12},l_{12},s_{12}, \nu_{12}  \rangle
D^{j_{12}}_{\nu_{12} \mu_{12}}[R] .
\label{sb:34}
\eeq
This is also true for the $r$-space vertex.  This means that the
operator is rotationally invariant.  This property will be important
in making a Poincar\'e invariant dynamics that includes the
vertex. This will be discussed in the next section.

For the purpose of using this vertex in a relativistic light-front
dynamical model it is useful decompose this vertex into invariant
kernels for different partial waves.  This is achieved by coupling the
final meson spins and relative orbital angular momenta
\[
\langle (\kappa,j) n_{23}, j_{23}, l_{23},s_{23},n_{14},j_{14},l_{14},s_{14},
l_v, s, \tilde{\mu}  \vert v^j_{2:1} \vert
n_{12},j_{12},l_{12},s_{12},\tilde{\mu}' \rangle =
\]
\beq
\delta_{jj_{12}} \delta_{\tilde{\mu}  \tilde{\mu}'}
\langle n_{23}, l_{23}, n_{14}, l_{14},
\kappa ,l_v \vert v_{2:1} \vert
n_{12},l_{12} \rangle \times
G^j( j_{23}, l_{23},s_{23},j_{14},l_{14},s_{14},
l_v, s ; l_{12},s_{12} )
\label{sb:35}
\eeq
where the rotationally invariant coefficient can be computed by
using rotational covariance and integrating over the $SU(2)$ Haar measure, which is
equivalent to averaging over the magnetic quantum numbers
\[
G^j( j_{23}, l_{23},s_{23},j_{14},l_{14},s_{14},
l_v, s ; l_{12},s_{12} )
:=
\]
\beq
{1 \over 2j +1}
\sum_{\mu_{12}=-j}^{j}
C(j_{23}, l_{23},s_{23},\mu_{23},j_{14},l_{14},s_{14},
\mu_{14}, l_v, m_v ; 
j ,l_{12},s_{12}, \mu_{12})
\langle j , \mu_{12} \vert l_v, m_v, s, \mu_s \rangle
\langle s, \mu_s \vert j_{23}, \mu_{23}, j_{14}, \mu_{14}  
\rangle .
\label{sb:36}
\eeq
The spins in (\ref{sb:35}) are the system light-front spins.

Qualitative agreement with the $\rho$ lifetime was used to determine the
dimensionless strength, $g$ of the string breaking interaction.  In order to achieve
the desired agreement, in addition to using the freedom to make order of unity
adjustments to the strength, it was also necessary to make the following
scale transformation on the vertex
on the spin-independent
part of the string breaking vertex
\[
\langle n_{23}, l_{23}, n_{14},l_{14},r \vert v_{2:1} \vert
n_{12},l_{12} \rangle \to
\langle n_{23}, l_{23}, n_{14},l_{14},r \vert v'_{2:1} \vert
n_{12},l_{12} \rangle :=
\]
\beq
(2)^{3/2} \langle n_{23}, l_{23}, n_{14},l_{14},2r \vert v_{2:1} \vert
n_{12},l_{12} \rangle
\label{sb:37}
\eeq
and 
\[
\langle n_{23}, l_{23}, n_{14},l_{14},\kappa,l_v \vert v_{2:1} \vert
n_{12},l_{12} \rangle \to
\langle n_{23}, l_{23}, n_{14},l_{14},\kappa,l_v \vert v'_{2:1} \vert
n_{12},l_{12} \rangle :=
\]
\beq
(2)^{-3/2} \langle n_{23}, l_{23}, n_{14},l_{14},{\kappa \over 2} ,l_v \vert v_{2:1} \vert
n_{12},l_{12} \rangle . 
\label{sb:38}
\eeq
This scale transformation has the effect of reducing the width of the
string breaking vertex by a factor of two, which is still a
modification of order unity of the original string breaking
vertex. The calculations that follow use the re-scaled vertex unless
otherwise specified.

Figure 4 shows the strength of the spin-independent part
(\ref{sb:31b}) 
of the re-scaled vertices for $l_{23}=l_{14}=l_{12}=0$ and different
values of $n_{23}=n_{14}=n_{12}=n$.  The lower $n$ contributions
dominate at lower energies.  The contribution for each $n$
is significant in a given energy range and falls off rapidly outside of
that range.  We note that most of the computational complexity is due
to the spin coupling.
\begin {figure}
\centering
\includegraphics[width=.7\linewidth]{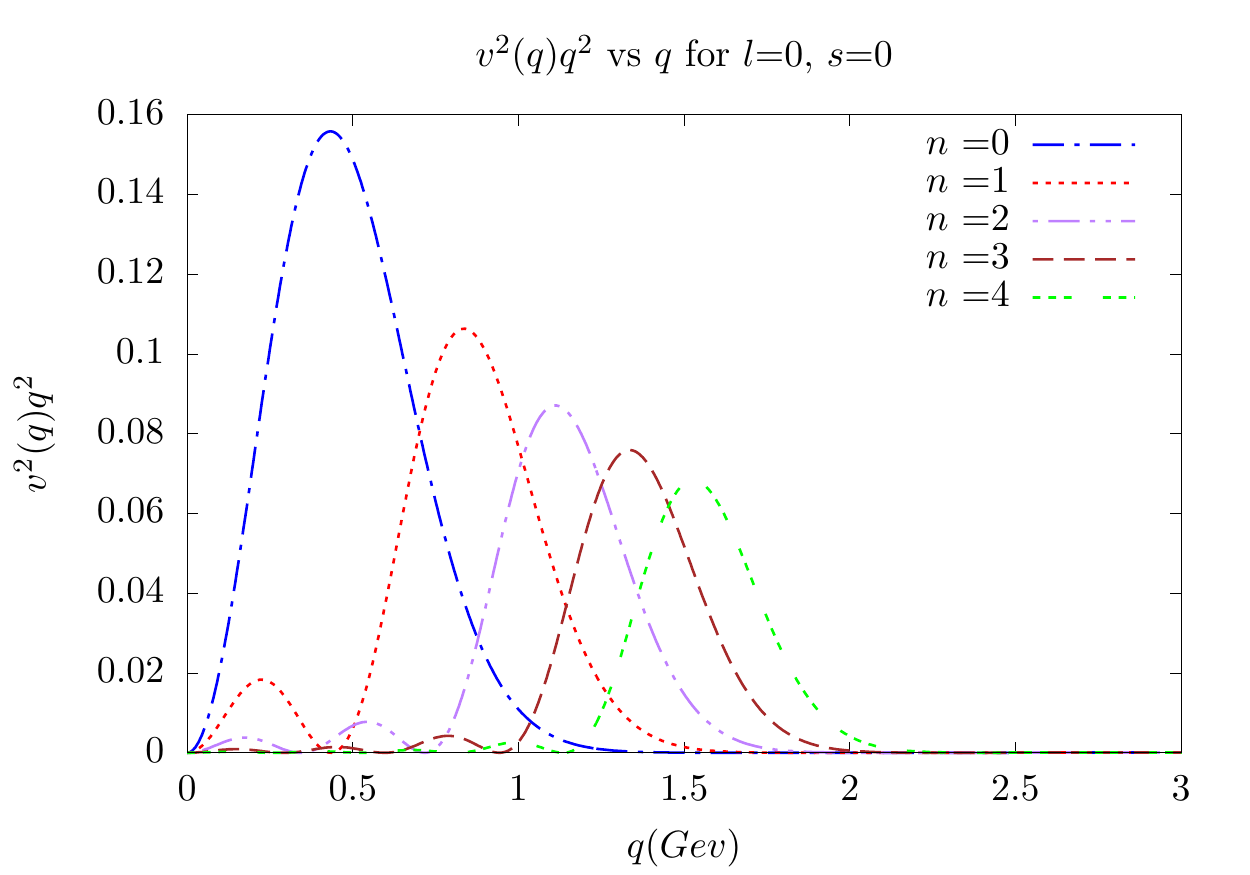}
\caption{\label{fig.4}Strength of vertex as value of n changes for l=0 s=0}
\label{fig:4}
\end{figure}

\section{Sea Quarks - Relativistic 1$+$2 Model}

The simplest extension of the bare confined singlet model allows
the bare confined
singlets to interact with the string breaking vertex to produce pairs
of bare confined singlets.  The model Hilbert space is the orthogonal
direct sum of the one-singlet Hilbert space with the tensor product of
two copies of the one-singlet Hilbert space
\beq
{\cal H} = {\cal H}_H \oplus ({\cal H}_H\otimes {\cal H}_H).
\label{sq:1}
\eeq
This is unitarily equivalent to the corresponding representation
in terms of confined quark-anti-quark-gluon singlet pairs  
\beq
{\cal H} = {\cal H}_Q \oplus ({\cal H}_Q\otimes {\cal H}_Q).
\label{sq:2}
\eeq
This equivalence exhibits the duality between representations in terms
of one and two quark-anti-quark-gluon singlet pairs and one in terms of
infinite towers of one or two bare mesons.  In this case the
two-singlet or two-bare meson sectors represent sea quark degrees of
freedom.

In section 3 a unitary representation of the Poincar\'e group
on the single-meson states (\ref{cq:19}), was constructed in
both the hadronic (\ref{cq:22}) and
quark (\ref{cq:24}) representations
\beq
U(\Lambda ,a): {\cal H}_H \to {\cal H}_H
\label{sq:3}
\eeq
\beq
U(\Lambda ,a): {\cal H}_Q \to {\cal H}_Q .
\label{sq:4}
\eeq
The next step is to construct a relativistic dynamics on the Hilbert
space (\ref{sq:1}-\ref{sq:2}).
The first step is to use 
(\ref{sq:3}-\ref{sq:4}) to
define the {\it free} dynamics on the space (\ref{sq:1})
\beq
U_0 (\Lambda ,a) =
\left (
\begin{array}{cc}
U(\Lambda ,a) & 0 \\
0 & U(\Lambda ,a)\otimes U(\Lambda ,a) \\
\end{array}  
\right ).
\label{sq:5}
\eeq
Here free means free bare mesons, not free quarks.  This
representation treats the bare mesons as if they were elementary
particles.  This representation has to be modified to include the
dynamics defined by the string-breaking vertex.

The next step in constructing a relativistic dynamics is to decompose
the operator $U_0(\Lambda ,a)$ into a direct integral of irreducible
representations.  This can be done by decomposing the direct product
of single-particle irreducible basis states into a direct integral of
irreducible basis states using the Poincar\'e group Clebsch-Gordan
coefficients in the light front basis: (\ref{kc:31})
\[
\vert (\kappa,j) \tilde{\mathbf{p}}, \tilde{\mu} ;
n_1, j_1, l_1, s_1, n_2, l_2, s_2, s_{12}, l_{12} \rangle =
\]
\[
\sum \vert (M_{n_1, l_1, s_1}, j_1 ) \tilde{\mathbf{p}}_1,\mu_1 \rangle
\vert (M_{n_2, l_2, s_2}, j_2 ) \tilde{\mathbf{p}}_2,\mu_2 \rangle 
D^{j_1}_{\mu_1, \nu_1} [R_{fc}(k_1/M_{n_1,j_1,l_1,s_1})]
D^{j_2}_{\mu_2, \nu_2} [R_{fc}(k_2/M_{n_2,j_2,l_2,s_2})] \times
\]
\beq
Y_{l_{12}m_{12}}(\hat{\mathbf{q}}_1) \langle j_1 ,\nu_1, j_2 ,\nu_2 \vert s_{12} \nu_{12} \rangle \langle l_{12}, m_{12}, s_{12}, \nu_{12} \vert j, \tilde{\nu} \rangle
\sqrt{{p_1^+ p_2^+ (\omega_1 (\kappa) + \omega_2 (\kappa )) \over
\omega_1(\kappa ) \omega_2 (\kappa ) (p_1^++ p_2^+)}} 
\label{sq:6}
\eeq
where in this representation the two-body kinematic mass is
the multiplication operator
\beq
M_{02} := \omega_1(\kappa ) + \omega_2 (\kappa )=
\sqrt{\kappa^2 + M_{n_1,j_1,l_1,s_1}^2} +
\sqrt{\kappa^2 + M_{n_2,j_2,l_2,s_2}^2}.
\label{sq:7}
\eeq

The states 
\beq
\left (
\begin{array}{c}
\vert (M_{nls}, j ) \tilde{\mathbf{p}},\tilde{\mu} ; l,s \rangle \\
0 \\
\end{array}
\right ) 
\label{sq:8}
\eeq
and 
\beq
\left (
\begin{array}{c}
0 \\
\vert (\kappa,j) \tilde{\mathbf{p}}, \tilde{\mu} ;
n_1, j_1, l_1, s_1, n_2, l_2, s_2, s_{12}, l_{12} \rangle
\end{array}
\right ). 
\label{sq:9}
\eeq
are a basis that
transforms irreducibly under $U_0(\Lambda ,a)$:
\[
U_0(\Lambda ,a)
\left (
\begin{array}{c}
\vert (M_{nls}, j ) \tilde{\mathbf{p}},\tilde{\mu} ; l,s \rangle \\
0 \\
\end{array}
\right ) =
\]
\beq
\left (
\begin{array}{c}
\vert (M_{nls}, j ) \tilde{\pmb{\Lambda}}p,\tilde{\nu} ; l,s \rangle \\
0 \\
\end{array}
\right )
e^{-i \Lambda p \cdot a}D^j_{\tilde{\nu}\tilde{\mu}} [B_f^{-1}(\Lambda p/M_{nls})
\Lambda B_f(p/M_{nls})]
\sqrt{{(\Lambda p)^+ \over p^+}} .
\label{sq:8a}
\eeq
\[
U_0(\Lambda ,a)
\left (
\begin{array}{c}
0 \\
\vert (\kappa,j) \tilde{\mathbf{p}}, \tilde{\mu} ;
n_1, j_1, l_1, s_1, n_2, l_2, s_2, s_{12}, l_{12} \rangle
\end{array}
\right )=
\]
\beq
\left (
\begin{array}{c}
0 \\
\vert (\kappa,j) \tilde{\pmb{\Lambda}}p,\tilde{\nu} ;
n_1, j_1, l_1, s_1, n_2, l_2, s_2, s_{12}, l_{12} \rangle
\end{array}
\right ) 
e^{-i \Lambda p \cdot a}D^j_{\tilde{\nu}\tilde{\mu}} [B_f^{-1}(\Lambda p/M_{02})
\Lambda B_f(p/M_{02})]
\sqrt{{(\Lambda p)^+ \over p^+}} .
\label{sq:9a}
\eeq

The next step is to express the kernel of the vertex operator 
in this basis:
\[
\langle  (\kappa,j) \tilde{\mathbf{p}}, \tilde{\mu} ;
n_1, j_1, l_1, s_1, n_2, l_2, s_2, s_{12}, l_{12} \vert 
v_{2:1} \vert   (n , j' ) \tilde{\mathbf{p}}',\tilde{\mu}' ; l,s
\rangle
= 
\]
\beq
\delta_{jj'} \delta_{\tilde{\mu} \tilde{\mu}'}\delta (\tilde{\mathbf{p}}-
\tilde{\mathbf{p}}') 
\langle n_{1}, l_{1}, n_{2}, l_{2},
\kappa ,l_{12} \vert v_{2:1} \vert
n_{},l_{} \rangle \times
G^j( j_{1}, l_{1},s_{1},j_{2},l_{2},s_{2},
l_{12}, s_{12} ; l ,s )
\label{sq:10}
\eeq
where $G^j( j_{1}, l_{1},s_{1},j_{2},l_{2},s_{2},
l_{12}, s_{12} ; l ,s )$ is the spin coefficient in (\ref{sb:36}).
Using this vertex the dynamical mass operator is defined by 
\beq
M  = M_0 +V =
\underbrace{
\left (
\begin{array}{cc}
M_{c}  & 0 \\
0 & \sqrt{ M_{c1}^2+{\kappa}^2}+ \sqrt{ M_{c2}^2+{\kappa}^2}   \\
\end{array}  
\right )}_{M_0} +
\underbrace{\left (
\begin{array}{cc}
0  & v_{1:2} \\
v_{2:1} & 0 \\
\end{array}  
\right )}_{V},
\label{sq:11}
\eeq
where $M_0$ is the mass Casimir operator of $U_0(\Lambda,a)$.
With this definition $M$
commutes with $\tilde{\mathbf{p}}$ and $\mathbf{j}$ but not with $M_0$
because $[M_0,V] \not=0$.  This means that it is possible to
simultaneously diagonalize $M$, $\tilde{\mathbf{p}}$, $j^2$ and $j_z$.
If this is done in the basis of simultaneous eigenstates of $M_0$,
$\tilde{\mathbf{p}}$, $j^2$ and $j_z$ then it is only necessary to
diagonalize $M$.  The resulting eigenfunctions have the form
\[
\langle j, \tilde{p} ,\tilde{\mu}, \cdots \vert (m,j')\tilde{p}' ,\tilde{\mu}' ; X \rangle =
\left (
\begin{array}{c}
\langle (M_{nls}, j ) \tilde{\mathbf{p}},\tilde{\mu} ; l,s  \vert
(m,j')\tilde{\mathbf{p}}' ,\tilde{\mu}' ; X \rangle\\
\langle (\kappa,j) \tilde{\mathbf{p}}, \tilde{\mu} ;
n_1, j_1, l_1, s_1, n_2, l_2, s_2, s_{12}, l_{12}
\vert (m,j')\tilde{\mathbf{p}}' ,\tilde{\mu}' ; X \rangle
\\
\end{array}  
\right ) =
\]
\beq
\delta (\tilde{\mathbf{p}}' -  \tilde{\mathbf{p}})
\delta_{j'j} \delta_{\tilde{\mu}'\tilde{\mu}} \times 
\left (
\begin{array}{c}
\langle (M_{nls}, j ) l,s  \vert
(m,j)  ; X \rangle\\
\langle (\kappa,j) ;
n_1, j_1, l_1, s_1, n_2, l_2, s_2, s_{12}, l_{12}
\vert (m,j); X \rangle
\\
\end{array}  
\right )
\label{sq:12}
\eeq
where $X$ represents possible degeneracy parameters.

The rotational invariance of (\ref{sq:11}) means that the rest eigenstates
in this representation transform under the $2j+1$ dimensional
representation of the $SU(2)$.

A dynamical unitary representation of the Poincar\'e group with a light-front 
kinematic symmetry is defined in this basis by 
\beq
U(\Lambda ,a)\vert (m,j) \tilde{\mathbf{p}},\tilde{\mu} , X \rangle := 
\vert (m,j) 
\tilde{\pmb{\Lambda}}p,\tilde{\nu} , X \rangle
e^{-i \Lambda p \cdot a}D^j_{\tilde{\nu}\tilde{\mu}} [B_f^{-1}(\Lambda p/m) \Lambda B_f(p/m)]
\sqrt{{(\Lambda p)^+ \over p^+}} .
\label{sq:13}
\eeq
This reduces the dynamical problem to finding the 
eigenvalues of the mass operator in the kinematic irreducible basis.
The representation (\ref{sq:13}) has the property that 
$e^{i \Lambda p \cdot a}$, $D^j_{\nu\mu} [B_f^{-1}(\Lambda p/m) \Lambda B_f(p/m)]$
and $\sqrt{{(\Lambda p)^+ \over p^+}}$ are all independent of $m$
for all Poincar\'e transformations that leave the light front invariant.
Since the interaction only changes the mass in this representation,
it follows that the light-front-preserving subgroup is kinematic.
This means that for these light-front-preserving Lorentz transformations
\beq
\left (
\begin{array}{c}
\langle (M_{nls}, j ) \tilde{\mathbf{p}},\tilde{\mu} ; l,s  \vert
U(\Lambda ,a) \vert
(m,j')\tilde{p}' ,\tilde{\mu}' ; X \rangle\\
\langle (\kappa,j) \tilde{\mathbf{p}}, \tilde{\mu} ;
n_1, j_1, l_1, s_1, n_2, l_2, s_2, s_{12}, l_{12}
\vert U(\Lambda ,a) \vert (m,j')\tilde{p}' ,\tilde{\mu}' ; X \rangle
\\
\end{array}  
\right ) ,
\label{sq:14}
\eeq
can be evaluated by acting to the right on the irreducible eigenstate or
 to the left with the adjoint on the non-interacting basis state.  
 
Equation (\ref{sq:13}) defines $U(\Lambda ,a)$ on a complete set of
irreducible states.  This defines a relativistic light-front dynamics
on this space that is consistent with the dynamical mass operator.
This model has the advantage that the mass operator is exactly rotationally
invariant.  Given these eigenstates they can be transformed to
single-quark bases or any other convenient basis.

In what follows calculations will used to investigate to what extent
this simple model provides a qualitatively consistent description 
of the following observables:

\begin{itemize}
\item[1.] Probability of finding sea quarks in bound states.

\item[2.] Cross section for scattering of bare mesons.

\item[3.] Lifetimes and energy shifts of unstable bare mesons.

\item[4.] Convergence as the number of bare meson channels are
increased.

\item[5.] Form factor calculations.
\end{itemize}

\section{Bound States}

The first question of interest is how do the sea quark contributions
generated by the string-breaking vertex impact the mass spectrum and
the sea quark content of the mesons.  Clearly the eigenvalues of the
lowest-mass bare mesons will shift and it will make the high-lying
bare mesons unstable.

In the hadronic representation the bound state problem requires
solving a system of equations with an infinite number of coupled
channels.

Bound-states are eignevectors of $M= (M_{0} + V)$
with eigenvalues $\eta$ in the point spectrum:
\beq
(M_{0} + V) \vert \Psi \rangle = \eta \vert \Psi \rangle .
\label{bs:1}
\eeq

Truncating to the one $+$ two-bare meson sectors the 
mass eigenvalue equation has the operator form
\beq
\left (
\begin{array}{cc}
M_c & v_{1:2} \\
v_{2:1} & \sqrt{ M_{c1}^2+{\kappa}^2}+ \sqrt{ M_{c2}^2+{\kappa}^2}
\end{array}
\right)
\left (
\begin{array}{c}
\vert \psi_1 \rangle  \\
\vert \psi_2 \rangle 
\end{array}
\right)=
\eta \left (
\begin{array}{c}
\vert \psi_1 \rangle  \\
\vert \psi_2 \rangle 
\end{array}
\right) 
\label{bs:2}
\eeq
where $M_c$ is the bare quark-anti-quark mass operator, (\ref{cq:8}).
These equations are equivalent to the pair of coupled equations
\beq
\vert \Psi_1 \rangle = {1 \over \eta- M_c}v_{1:2} \vert \Psi_2 \rangle 
\label{bs:3}
\eeq
\beq
\vert \Psi_2 \rangle = {1 \over \eta - 
\sqrt{ M_{c1}^2+{\kappa }^2}- \sqrt{ M_{c2}^2+{\kappa}^2}}
v_{2:1} \vert \Psi_1 \rangle . 
\label{bs:4}
\eeq
Using the second equation in the first gives a single equation for
the valence component,
$\vert \Psi_1 \rangle$:
\beq
\left (I -{1 \over \eta - M_{c} }
v_{1:2} 
{1 \over \eta - 
\sqrt{ M_{c1}^2+\kappa^2}+ \sqrt{ M_{c2}^2+\kappa^2}  
} 
v_{2:1} \right ) \vert \Psi_1 \rangle =0 .
\label{bs:5}
\eeq
The exact meson mass eigenvalues are the real values of $\eta$ 
between zero and the two-bare meson threshold,
where (\ref{bs:5}) has non-zero solutions.  

The short-hand notation is used
\beq
m \leftrightarrow \{n_1,j_1,l_1,s_1 \}
\label{bs:6}
\eeq
\beq
n_1, n_2, \kappa \leftrightarrow
\{s_{12}, l_q, n_1, j_1, l_1, s_1, n_2, j_2, l_2, s_2 \}
\label{bs:7}
\eeq
to label the one and two singlet channels, and
\beq
M_n := M_{nls} \qquad
M_{n_1,n_2}(\kappa) := \sqrt{\kappa^2 + M^2_{n_1,l_1,s_1}} +
\sqrt{\kappa^2 + M^2_{n_2,l_2,s_2}} 
\label{bs:8}
\eeq
to label the mass eigenvalues of the bare meson systems.
In this notation the mass eigenvalue problem (\ref{bs:5}) 
has the form
\beq
\sum_m \left (\delta_{nm} - \sum_{n_1,n_2} \int_0^\infty 
{1 \over \eta -M_n} \langle n \vert v_{1:2} \vert
n_1, n_2, \kappa \rangle 
{\kappa^2 d\kappa \over \eta -M_{n_1, n_2}(\kappa)} \langle n_1, n_2,
\kappa \vert v_{2:1} \vert
m \rangle  \right ) \langle  m \vert \Psi_1 \rangle =0  .
\label{bs:9}
\eeq
Even this truncated problem is a coupled channel problem with an
infinite number of channels.  It can be solved by
an additional truncation to a finite number of channels.
The ability to analytically calculate vertices that couple
any combination of channels can be used to estimate the
size of the  error  due to eliminated channels.

Equation (\ref{bs:9}) has the form
\beq
\sum_m (\delta_{nm}- A(\eta)_{nm}) \langle m \vert \Psi_1 \rangle =0 .
\label{bs:10}
\eeq
The eigenvalues are the real zeros, $\eta_k$, of the determinant of the
matrix, $\delta_{nm}- A(\eta)_{nm}$,  considered a function of $\eta$:
\beq
f(\eta)= \mbox{det} (I-A(\eta)) =0 .
\label{bs:11}
\eeq
The eigenvalues, which are the zeroes of the determinant, can be determined
by plotting the
determinant, $f(\eta)$, for
values of $\eta$ between 0 and the two meson threshold.
For this range of values of $\eta$,
the determinant is a real valued function of $\eta$ and the momentum
integrals in (\ref{bs:9}) are non-singular. 

To determine the wave function for the $k^{th}$ eigenvalue,
$\eta_k$, consider the ordinary eigenvalue
problem for $\chi$:
\beq
\sum_m (\chi \delta_{nm} - A_{nm} (\eta_k) )\xi_m^k=0
\label{bs:12a}
\eeq
By construction $\chi=1$ is the eigenvalue of this equation
for $\eta = \eta_k$.  The valence wave function in the hadronic representation is
\beq
\langle n \vert \Psi_1 \rangle = N_k \xi^k_n
\eeq
where $\xi^1_n$ is the eigenvector associated with $\chi=1$ of
(\ref{bs:12a}) and 
$N_k$ is a normalization constant.

The sea quark component of the wave function is
\beq
\langle n_1,n_2,\kappa \vert \Psi_2  \rangle =
\sum_m 
{1 \over \eta_k -M_{n_1, n_2}( \kappa)}
\langle n_1, n_2, \kappa \vert v_{2:1} \vert
m \rangle
{1 \over \eta_k -M_m}
N_k \xi^k_m .
\label{bs:15}
\eeq
The normalization constant $N_k$ is chosen so
\beq
1 = \langle \Psi_1 \vert \Psi_1 \rangle + \langle \Psi_2 \vert \Psi_2 \rangle .
\eeq
With this normalization   
$
\langle \Psi_1 \vert \Psi_1 \rangle
$
represents the probability of finding the meson in the valence sector
and $
\langle \Psi_2 \vert \Psi_2 \rangle$ 
represents the probability of finding the meson with sea quarks.

The evaluation of 
$
\langle \Psi_2 \vert \Psi_2 \rangle$ involves a one-dimensional integral and a
sum over two hadron states that couple to the valence sector. 

These ``exact calculations'' can be compared to the perturbative
result to determine the applicability of perturbation theory.  The
perturbative calculation treats the string breaking interaction as a
perturbation.  A formal expansion parameter $\epsilon$ is introduced
to keep track of powers of the string breaking vertex.  In the end it
is set to 1.  The unperturbed state is taken to be one of the low-mass
confined quark-anti-quark states ($n_0= m_0,l_0$):
\beq
\left (
\begin{array}{c}
\vert \Psi_{10} \rangle \\
\vert \Psi_{20} \rangle 
\end{array} 
\right ) =
\left (
\begin{array}{c}
\delta_{nn_0}  \\
0 
\end{array} 
\right ) .
\label{bs:16}
\eeq 
Standard Rayleigh-Schr\"odinger perturbation theory gives the 
leading correction to the unperturbed eigenvalue
$M_{n_0}$ in powers of $\epsilon$:
\beq
M_{n_0} (\epsilon ) = M_{n_0} + \epsilon^2
\sum_{n_1 n_2} \int \kappa^2 d\kappa 
{\vert \langle n_0 \vert v_{12} \vert n_1, n_2, \kappa \rangle \vert^2
\over 
M_{n_0} - M_{n_1,n_2}(\kappa) } + o(\epsilon^4). 
\label{bs:17}
\eeq
The corresponding un-normalized wave function is 
\beq
\left (
\begin{array}{c}
\langle n \vert \Psi_{1} \rangle \\
\langle n_1, n_2, \kappa \vert \Psi_{2} \rangle 
\end{array} 
\right ) =  
\left (
\begin{array}{c}
\delta_{nn_0} \\
\epsilon {\langle n_1, n_2, \kappa \vert v_{21} \vert n_0 \rangle \over 
M_{n_0} - M_{n_1,n_2}(\kappa) }
\end{array} 
\right )  + \cdots
\label{bs:18}
\eeq
In what follows the model calculation uses a bare $\pi$ mass of $m_{\pi
  0}=.16$ GeV, a bare $\rho$ mass of $m_{\rho 0}= .882$ GeV and a
dimensionless vertex coupling constant $g=5.44$.  The correction to
the pion mass and wave function, keeping the 2 quark-2 anti-quark
channels with $n\leq 4$, is computed using both second order
perturbation theory (\ref{bs:17}) and the exact calculation obtained
by finding zeroes of the determinant (\ref{bs:11}).  The perturbative
result gives a pion mass of $.1327$ GeV while the exact method gives a
pion mass of $.1329$ GeV.  While the perturbative and ``exact''
calculated masses are close, they differ from the bare $\pi$ mass of
$.16$ GeV by about $.027$ GeV, which is about a 17\% shift.  The
calculation of the corresponding pion wave function results in a
probability of .82 of measuring the pion to be in the valence sector
and .18 of measuring pion to be in a two bare singlet state.
This indicates that in this model sea quarks make up a non-trivial part of the
pion wave function.

\section{Meson-Meson Scattering}

This model can also be used to treat meson-meson scattering.  Due to
the truncation, the particles that scatter are bare mesons.  The
scattering integral equations have an infinite number of poles in the
continuum.  These are not real, because the vertex makes the
associated mesons unstable.  The treatment of scattering in models with
confined particles is discussed in
\cite{Dashen:2009kj}\cite{PhysRevD.14.2773}.  These results are
applicable to this model.

In this dual quark-hadron setting it is natural to formulate the
scattering problem using a two Hilbert space formulation
\cite{Coester:1982vt}.  In one Hilbert space, the asymptotic space,
the bare mesons are treated like elementary particles.  Their internal
structure is included in a mapping from the asymptotic space to the
dynamical Hilbert space.  The asymptotic Hilbert space is the direct
sum of scattering channel Hilbert spaces.  The $2+1$ model scattering
channels involve pairs of bare mesons labeled by the quantum numbers
$n_1$ and $n_2$.  A normalizable vector in the two bare-meson subspace
of the Hilbert space has the form
\beq
\vert \Psi_{0 n_1n_2}(0) \rangle = \sum_{\tilde{\mu}_1 \tilde{\mu}_2} \int d\tilde{p}_1 ,d\tilde{p}_2
\vert (n_1)\tilde{p}_1 ,\tilde{\mu}_1  \rangle \times
\vert (n_2)\tilde{p}_2 ,\tilde{\mu}_2  \rangle  
f_{n_1} (\tilde{p}_1 ,\tilde{\mu}_1) f_{n_2} (\tilde{p}_2 ,\tilde{\mu}_2)
\label{sc:1}
\eeq
where  $f_{n_1} (\tilde{p}_1 ,\tilde{\mu}_1)$ and $ f_{n_2} (\tilde{p}_2 ,\tilde{\mu}_2)$
are wave packets in the meson momenta and magnetic quantum numbers and
the $n_i$ represent the particle quantum numbers $(n_i,j_i,l_i,s_i)$.
Equation (\ref{sc:1}) can be expressed in the form of a mapping from the
Hilbert space ${\cal H}_{n_1n_2}$ of square integrable
functions of light-front momenta and meson magnetic quantum numbers to the
two bare-mesons subspace of the Hilbert space:
\beq
\vert \Psi_{0 n_1n_2} (0) \rangle := 
\Phi_{0 n_1 n_2}  \vert f_{n_1} f_{n_2} \rangle .
\label{sc:2}
\eeq
For identical mesons this state should be symmetrized:
\beq
\vert \Psi_{0 n_1 n_2} \rangle  \to \vert \Psi_{0,{\cal S},n_1, n_1}  \rangle =
{\cal S} \vert \Psi \rangle 
\label{sc:3}
\eeq
where ${\cal S}$ is the projector on symmetrized states.  Note that in this
model the symmetrization is only with respect to identical mesons, not quarks.
The symmetrizer and normalization factors can be absorbed in the definition of
$\Phi_{0 n_1 n_2}$,
In both cases the free two-meson states should be normalized to unity:
\beq 
\Vert \vert \Psi_{0 n_1n_2} (0) \rangle \Vert =1 .
\label{sc:4}
\eeq

The asymptotic Hilbert space is defined as the orthogonal direct sum
of all of the channel spaces:
\beq
{\cal H}_A := \oplus_{n_1,n_2} {\cal H}_{n_1,n_2}.
\label{sc:5}
\eeq
In this model there are an infinite number of channel subspaces.
This is an artifact of the truncation to the $2+1$ sector, which  eliminates
the interactions in the two-meson subspace that allow unstable bare
mesons to decay.

An injection operator from ${\cal H}_A$ to the two-meson subspace of the
$2+1$ Hilbert space ${\cal H}$ is defined by
\beq
\Phi := \sum_{n_1,n_2} \Phi_{0 n_1 n_2}
\label{sc:6}
\eeq
where each $\Phi_{0 n_1 n_2}$ is understood to act on the subspace
${\cal H}_{n_1,n_2}$ of ${\cal H}_{A}$.  The two Hilbert space
representation separates the quantities (momenta and magnetic quantum
numbers) that are used to prepare initial and final meson wave packets
from the internal structure of each meson in terms of quarks, anti-quarks
and gluons.  The asymptotic space contains
the degrees of freedom that can be ``controlled in an experiment''. 

A light-front asymptotic Hamiltonian and mass operator is
defined on ${\cal H}_{A}$.  On each subspace ${\cal H}_{n_1,n_2}$ 
they are:
\beq
P^-_A = {M_{cn1}^2 + \mathbf{p}_{n_1 \perp}^2 \over p_{n_1}^+} +
{M_{cn2}^2 + \mathbf{p}_{n_2 \perp}^2 \over p_{n_2}^+}
\label{sc:7}
\eeq
and
\beq
M_A = \sqrt{M_{cn1}^2 +\kappa^2} + \sqrt{M_{cn2}^2 +\kappa^2}.
\label{sc:8}
\eeq

In the absence of the string breaking vertex 
the asymptotic states evolve in time like two free particles (mesons)
\beq
\vert \Psi_{0 n_1 n_2}(t) \rangle =
\Phi_{0n_1 n_2} e^{- {i \over 2} (P_A^-+P_A^+) t }\vert f_{n_1} f_{n_2} \rangle   
\label{sc:9}
\eeq
Dynamical solutions of the Schr\"odinger equation $\vert \Psi_{\pm n_1 n_2} (t)\rangle$
that look like these states
long before or long after interacting satisfy the scattering
asymptotic conditions
\beq
\lim_{t\to \pm \infty}
\Vert \vert \Psi_{\pm n_1 n_2} (t)\rangle -  \vert \Psi_{\pm 0 n_1 n_2}(t) \rangle \Vert
=0
\label{sc:10}
\eeq
where $\vert \Psi_{\pm 0 n_1 n_2}(t)\rangle $ are free meson states of the form (\ref{sc:9})
that are approached by the exact scattering states in the asymptotic future
$(+)$ or past $(-)$.

Using the unitarity of the time evolution operator this can be expressed as
\beq
\lim_{t\to \pm \infty} \Vert \vert \Psi_{\pm n_1 n_2} (0)\rangle - e^{i
{1 \over 2} (P^-+P^+) t  }  \Phi_{n_1 n_2} e^{- {i \over 2}
(P_A^-+P_A^+) t }\vert f_{n_1} f_{n_2} \rangle \Vert =0 .
\label{sc:11}
\eeq
Since $P^+$ and $\mathbf{P}_{\perp}$ are kinematic operators (\ref{sc:11})
becomes
\beq
\lim_{t\to \pm \infty} \Vert \vert \Psi_{\pm} (0)\rangle - e^{i
{M^2 \over 2 P^+} t }  \Phi_{0n_1 n_2} e^{-i {M_A^2 \over 2P^+} t }\vert f_{n_1} f_{n_2} \rangle \Vert =0 .
\label{sc:12}
\eeq
Because the spectrum of $P^+$ is positive,  the time limit can be replaced by a
$\tau := {t /2 P^+}$ limit:
\beq
\lim_{\tau\to \pm \infty} \Vert \vert \Psi_{\pm n_1 n_2 } (0)\rangle -
e^{i
  M^2 \tau} \Phi_{0n_1 n_2} e^{-i M_A^2 \tau }
\vert f_{n_1} f_{n_2} \rangle
\Vert =0 .
\label{sc:13}
\eeq
Finally the invariance principle \cite{simon}\cite{Chandler:1976} can be
used to replace $M^2$ by $M$
leading to the following expression for the normalized scattering state
at time $0$:
\beq
\vert \Psi_{\pm n_1 n_2}(0) \rangle := 
\lim_{\tau\to \pm \infty}e^{i M \tau} \Phi_{0n_1 n_2} e^{-i M_A \tau }
\vert f_{n_1} f_{n_2} \rangle  =
\Omega_{\pm n_1 n_2} \vert f_{n_1} f_{n_2} \rangle .
\label{sc:14}
\eeq
In this notation the probability that a solution of the
Schr\"odinger equation that asymptotically looks like
\beq
\vert \Psi_{-0 n_1 n_2}(t) \rangle
\label{sc:15}
\eeq
in the past will be measured to be in a state that looks like
\beq
\vert \Psi_{+0 n_1' n_2'}(t) \rangle
\label{sc:16}
\eeq
in the future is given by the square of the inner product of the
unit normalized vectors (\ref{sc:14}) at any common time (normally taken as $(t=0)$)
\beq
P = \vert \langle \Psi_{+ n_1' n_2'}(0) \vert \Psi_{- n_1 n_2}(0) \rangle \vert^2
=
\vert \langle f'_{n_1'} f'_{n'_2} \vert 
\Omega^{\dagger}_{+ n_1' n_2' } 
\Omega_{-n_1 n_2} \vert f_{n_1} f_{n_2} \rangle
\vert ^2 =
\vert \langle f'_{n_1'} f'_{n'_2} \vert 
S_{n_1'n_2';n_1 n_2 }
\vert f_{n_1} f_{n_2} \rangle
\vert^2 .
\label{sc:17}
\eeq
The operator
\beq
S_{n_1'n_2';n_1 n_2 }:=
\Omega^{\dagger}_{+ n_1' n_2' } 
\Omega_{-n_1 n_2}
\label{sc:18}
\eeq
is the channel scattering operator.

In this model the interaction looks like a separable potential with
an infinite number of terms.  While truncation to a finite number of terms
leads to a short-range interaction, whether the infinite sum also leads
to a short-range interaction depends of properties of the string breaking vertex.  A sufficient condition
for the existence of the limit in (\ref{sc:14}) is given by
the Cook condition \cite{Cook:1957} which expresses this limit as
the integral of a derivative, and bounds the integral by the integral of the 
norm of the integrand.  In this model the Cook condition has the form  
\beq
\int_a^{\infty}
\Vert  
v_{1:2}  \Phi_{0 n_1, n_2} e^{-i M_{n_1,n_2,q} t}\vert f_{n_1}
f_{n_2} \rangle
\Vert dt < \infty 
\label{sc:19}
\eeq
which gives a sufficient condition on the string breaking vertex for
the existence of the limits (\ref{sc:10}).  In appendix III this
condition is shown to hold for the vertex (\ref{sb:1}) for the case
$n_1=n_2$.  This is possible because the sum over channels can be
replaced by an integral that can be bound.  In the appendix it is
shown that the integrand falls off like $t^{-3/2}$ for large $t$, so
this integral is finite.  This is an important result because in the
context of this 2+1 truncation, it means that truncations to a finite
numbers of channels are actually approximations in the sense that they
converge in the infinite channel limit.  

The channel scattering operator can be expressed in terms of
a multi-channel scattering operator, $S$, on ${\cal H}$: 
\beq
S_{n_1'n_2';n_1 n_2 } = \Phi_{0n_1' n_2'}^{\dagger} S \Phi_{0n_1 n_2}.
\label{sc:20}
\eeq
The $S_{n_1'n_2';n_1 n_2 }$ are matrix
elements of $S$ in bare meson mass eigenstates, $\vert m \cdots \rangle $, of
$M_0$.  These matrix elements have the form
\beq
\langle m \cdots \vert S \vert m' \cdots\rangle =
\langle m \cdots \vert I \vert m' \cdots \rangle
-2 \pi i \delta (m-m') \langle m \cdots \vert T (m+i0^+) \vert m'
\cdots \rangle 
\label{sc:21}
\eeq
where
\beq
T(z) = V + V (z-M)^{-1} V 
\label{sc:22}
\eeq
and $V$ is the string breaking interaction.
Note that all of the operators in (\ref{sc:22}) commute with all three
kinematic
components of the total light-front momentum of the system.
The second resolvent identity 
\cite{Hille:1957}, 
\beq
(z-M)^{-1} = (z-M_{c0})^{-1} +
(z-M_{c0})^{-1} V (z-M)^{-1} ,
\label{sc:23}
\eeq
can be used to construct a
Lippmann-Schwinger integral equation for $T(z)$ of the form
\beq
\left (
\begin{array}{cc}
T^{11}(z) & T^{12}(z) \\
T^{21}(z) & T^{22}(z) \\
\end{array}
\right ) = 
\left (
\begin{array}{cc}
0 & v_{1:2} \\
v_{2:1} & 0 \\
\end{array}
\right ) + 
\left (
\begin{array}{cc}
0 & v_{1:2} \\
v_{2:1} & 0 \\
\end{array}
\right )
\left (
\begin{array}{cc}
(z-M_1)^{-1} & 0 \\
0 & (z-M_2)^{-1}) \\
\end{array}
\right )
\left (
\begin{array}{cc}
T^{11}(z) & T^{12}(z) \\
T^{21}(z) & T^{22}(z) \\
\end{array}
\right ).
\label{sc:24}
\eeq
The derivation is justified when $(z-M)^{-1}$ and $(z-M_{c0})^{-1}$
are both defined.  The property of this model that deviates from
conventional scattering theory is that $(z-M_{c0})^{-1}$ has an
infinite number of poles on the positive real axis.  These poles are not
expected to appear in $(z-M)^{-1}$ or (\ref{sc:22}) because the vertex
will cause the
bare bound states in the continuum to decay.  Both resolvents are
analytic for $z$ off of the real axis.  The solution of equations
(\ref{sc:24}) 
has the correct limit at the poles in the continuum,
but the equations are ill-defined at these poles.

To understand this in more detail note that if $(M_{c0}-\eta)\vert B
\rangle =0$ and $z=\eta+i \epsilon$ then (\ref{sc:23}) leads to
\[
\langle B \vert (z-M)^{-1} = \langle B \vert 
(i \epsilon)^{-1} + \langle B \vert 
(i \epsilon )^{-1} V (z-M)^{-1}=
\]
\[
(i \epsilon)^{-1} \langle B \vert 
(1 +  V (z-M)^{-1})=
\]
\[
(i \epsilon)^{-1} \langle B \vert 
(\eta+i \epsilon -M_{co}) (z-M)^{-1})=
\]
\beq
(i \epsilon)^{-1} \langle B \vert 
(i \epsilon (z-M)^{-1})
\label{sc:25}
\eeq
which shows that the singular terms in the resolvent
$(z-M_c)^{-1}$ 
in the driving term cancel with corresponding singular terms in the kernel
of the integral equation when the equation is applied to 
the bare discrete eigenstates.

In what follows the equations will be recast in a form where there
are no poles in the continuum.  The equations ({\ref{sc:24}) break up into
uncoupled pairs of coupled equations.   The component that is
needed to calculate the scattering operator is $T^{22}(e+i0^+)$
which is the solution of the coupled pair of equations
\beq
T^{22}(e+i0^+)= 0 + v_{2:1} (e-M_1 +i 0^+)^{-1} T^{12}(e+i0^+)
\label{sc:26}
\eeq
\beq
T^{12}(e+i0^+)= v_{1:2} + v_{1:2} (e-M_2 +i 0^+)^{-1} T^{22}(e+i0^+) .
\label{sc:27}
\eeq
Using (\ref{sc:26}) in (\ref{sc:27})  gives a single equation for
$T^{12}(e+i0^+)$:
\beq
T^{12}(e+i0^+)= v_{1:2} + v_{1:2} (e-M_2 +i 0^+)^{-1} v_{2:1} 
(e-M_1 +i 0^+)^{-1} T^{12}(e+i0^+).
\label{sc:28}
\eeq
The solution of (\ref{sc:28}) can be used to calculate the transition 
operator for scattering.  
\beq
T^{22}(e+i0^+)=  v_{2:1} (e-M_1 +i 0^+)^{-1} T^{12}(e+i0^+).
\label{sc:29}
\eeq
The kernel of (\ref{sc:28}) has poles in the continuum.
To transform equation (\ref{sc:28})
to a form where the poles in the continuum do not appear 
define:
\beq
\Gamma_{12}(e+i0^+)  := (e-M_1 +i 0^+)^{-1} T^{12}(e+i0^+)
\label{sc:30}
\eeq
Multiplying equation (\ref{sc:28}) by $(e-M_1 +i 0^+)^{-1}$ gives an
equivalent equation for $\Gamma_{12}(e+i0^+)$:
\beq
\Gamma_{12}(e+i0^+)= (e-M_1 +i 0^+)^{-1} v_{1:2} + (e-M_1 +i 0^+)^{-1} v_{1:2} (e-M_2 +i 0^+)^{-1} v_{2:1} 
\Gamma_{12}(e+i0^+) .
\label{sc:31}
\eeq
Multiply both sides of (\ref{sc:31}) by $(e-M_1)$ to get
\beq
(e-M_1 - v_{1:2} (e-M_2 +i 0^+)^{-1} v_{2:1}) \Gamma_{12}(e+i0^+)
=  v_{1:2}.
\label{sc:32}
\eeq
Formally solving for $\Gamma_{12}(e+i0^+)$ gives 
\beq
\Gamma_{12}(e+i0^+) =
(e-M_1 - v_{1:2} (e-M_2 +i 0^+)^{-1} v_{2:1})^{-1}
v_{1:2}
\label{sc:33}
\eeq
which leads to the following expression for $T^{22}(e+i0^+)$
\beq
T^{22}(e+i\epsilon^+) = v_{2:1}\Gamma_{12}(e+i\epsilon^+) =
v_{2:1} {1 \over e-M_1 - v_{1:2} (e-M_2 +i 0^+)^{-1} v_{2:1}}v_{1:2}.
\label{sc:34}
\eeq
The advantage of this form is that there are no poles in the
continuum.   The denominator is complex for
energies above the scattering threshold, because
$v_{1:2} (e-M_2 +i 0^+)^{-1} v_{2:1}$ is complex above
threshold for two-particle scattering.

The scattering problem is to invert the matrix
\beq
K_{mn}(e+i\epsilon^+) = (e-M_n) \delta_{mn} - M_{mn} (e+i0^+)
\label{sc:35}
\eeq
where
\beq
M_{mn} (e+i0^+) = \sum_{m_1m_2}
\int_0^\infty {\langle m \vert v_{1:2} \vert m_{1}, m_{2},\kappa \rangle
\kappa^2 d\kappa
\langle  m_{1}, m_{2} , \kappa \vert
v_{2:1} \vert n \rangle
\over (e-M_{2m_1m_2}(\kappa) +i 0^+)} .
\label{sc:36}
\eeq
Given this solution the transition matrix elements can be expressed
in terms of the vertex and the inverse of $K_{mn}(e+i 0^+)$:
\beq
\langle m_{1}, m_{2},\kappa \vert T^{22}(e+i0^+)
\vert m_{1}', m_{2}',\kappa' \rangle =
\sum
\langle  m_{1}, m_{2} , \kappa \vert
v_{2:1} \vert m \rangle
K_{mn}^{-1}(e+i 0^+) \langle n \vert v_{1:2} \vert  m_{1}', m_{2}' , \kappa' 
\rangle
\label{sc:37}
\eeq

The quantity that needs to be computed is the matrix
$M_{mn} (e+i0^+)$ in equation (\ref{sc:36}) which is used to
compute $K_{mn}(e+i0 ^+)$.
In principle $K_{mn}(e+i 0 ^+)$ is an infinite matrix.  For the purpose of
calculations the number of channels must be truncated to
a finite number.  As mentioned earlier,  this truncation is actually
a controlled approximation in the context of this 2+1 model.
The calculation is reduced to inverting a complex
matrix.  Convergence can be checked by adding more channels.

The scattering computation requires computing the matrix
$M_{mn}(e+i0^+)$,  constructing $K_{mn}(e+i0^+)$, and
inverting $K_{mn}(e+i0^+).$

The details of how this calculation is performed is discussed in
Appendix IV.

Transition matrix elements have to be computed in a basis.  A natural
basis for scattering calculations uses as commuting observables
the discrete quantum numbers of
the bare mesons, the kinematic light front components of the total
momentum, and three momentum, $\pmb{\kappa}$, of one of the mesons
boosted to the two-meson rest frame with a light-front boost.  The
total kinematic light-front momenta are conserved and factor out of
the transition matrix elements.  The following shorthand notation is
used for the reduced matrix elements:
\beq
\langle  n_1', n_2', \pmb{\kappa}_f \vert T^{22}  \vert n_1, n_2, \pmb{\kappa}_i \rangle := 
\langle n_1',j_1',l_1', s_1',\mu_1';n_2',j_2',l_2',s_2',\mu_2', \pmb{\kappa}'
\vert T^{22} \vert
n_1,j_1,l_1, s_1,\mu_1;n_2,j_2,l_2,s_2,\mu_2, \pmb{\kappa} \rangle .
\eeq
The center of mass differential cross section is
\beq
{d\sigma \over d\Omega (\hat{\pmb{\kappa}}_f)} =
{(2 \pi)^4} 
({\omega_{n_1}(\kappa_f)\omega_{n_2}(\kappa_f) \over
\omega_{n_1}(\kappa_f)+\omega_{n_2}(\kappa_f)} 
{\omega_{m_1}(\kappa_i)\omega_{m_2}(\kappa_i) \over
\omega_{m_1}(\kappa_i)+\omega_{m_2}(\kappa_i)})
\vert {\kappa_f \over \kappa_i} 
\vert \langle n_1, n_2, \pmb{\kappa}_f \vert T^{22} \vert m_1, m_2, \pmb{\kappa}_i \rangle\vert^2 
\label{sc:51}
\eeq
where $\omega_m(\kappa) = \sqrt{m^2+ \kappa^2}$.
For identical mesons 
$\vert \langle n_1, n_2, \pmb{\kappa}_f \vert T^{22} \vert m_1, m_2, \pmb{\kappa}_i \rangle\vert^2$  is replaced by
\beq
\vert \langle n_1, n_1, \pmb{\kappa}_f \vert T^{22} \vert m_1, m_1, \pmb{\kappa}_i \rangle +
 \langle n_1, n_1, -\pmb{\kappa}_f \vert T^{22} \vert m_1, m_1, \pmb{\kappa}_i \rangle\vert^2 .
\label{sc:52}
\eeq

Phase shifts can be expressed in terms of the transition matrix elements by
\beq
\langle n_1, n_2, \kappa_f \vert T^{22} \vert m_1, m_2, \kappa_i \rangle =
- 4 \pi^2 {\omega_{m_1}(\kappa_i)\omega_{m_2}(\kappa_i) \over
\omega_{m_1}(\kappa_i)+\omega_{m_1}(\kappa_i)} e^{i \delta_j} \sin (\delta_j)
\eeq

The simplest case of interest is $\pi-\pi$ scattering.  
In this case the transition matrix element for each partial wave
has the form 
\beq
\langle 0, 0, l_\kappa, \kappa \vert T^{22} \vert 0 , 0, l_\kappa, \kappa \rangle =
\sum_n \langle 0, 0, l_\kappa, \kappa \vert v_{2:1} \vert n \rangle 
(e -M_n)^{-1} (\delta_{nm} - M_{nm})^{-1} 
\langle m \vert v_{12} \vert 0,0,l_\kappa, \kappa \rangle .
\label{sc:53}
\eeq
The matrix $M_{mn}(e+i0^+) $ is approximated by 
\beq 
M_{nm} (e+i\epsilon^+) =
\sum_{k,m_1 m_2} \langle n \vert v_{1:2} \vert m_{1}, m_{2},l_\kappa, \chi_{k} \rangle
g_k (e+i\epsilon^+)   
\langle  m_{1}, m_{2}, l_\kappa ,\chi_{k} \vert
v_{2:1} \vert m \rangle
\label{sc:54}
\eeq
where the $\chi_k$ denote the box functions (see eq. (\ref{sc:39}) in Appendix IV)
for a suitably fine mesh and
$g_k(e+i\epsilon)^+$ are the matrix elements (\ref{sc:46}). 
For a channel truncation with one one-body and one two-body
channel this becomes
\beq 
M_{00} (e) =
\sum_n \langle 0 \vert v_{1:2} \vert 0, 0,l_\kappa, \chi_n \rangle
g_n(e+i\epsilon^+)    
\langle  0, 0,l_\kappa , \chi_n \vert
v_{2:1} \vert m \rangle
\label{sc:55}
\eeq
and
\beq
\langle 0, 0, 0, l_\kappa \vert T^{22}(+i\epsilon^+) \vert 0 , 0, l_\kappa, \kappa \rangle =
\langle 0, 0, l_\kappa, \kappa \vert v_{2:1} \vert 0 \rangle 
(e -M_0  - v_{1:2}M_{00}(e+i\epsilon^+)v_{2:1} )^{-1} 
\langle 0 \vert v_{1:2} \vert 0,0,l_\kappa, \kappa \rangle
\label{sc:56}
\eeq

For the simplest truncation, where the vertex only includes
the $\pi^+- \pi^- \leftrightarrow \rho$ channels (note
$\pi^0- \pi^0 \leftrightarrow \rho$ is forbidden)
,  
the transition matrix elements have the from
\beq
T_l(\kappa) = {1 \over \pi \kappa \omega_{\pi}(\kappa)}
{\Gamma (\kappa)/2 \over
2 \omega_{\pi}(\kappa) - m_{\rho}-\Delta (\kappa) + i \Gamma(\kappa)/2}
\label{sc:57}
\eeq
where
\beq
\Gamma (\kappa) = \pi \kappa \omega_{\pi}(\kappa) \vert
\langle \rho \vert v_{12} \vert \pi ,\pi , \kappa \rangle\vert^2
\label{sc:58}
\eeq
\[
\Delta (\kappa) := P\int_0^\infty {\kappa^{\prime 2} d\kappa'
\langle \rho \vert v_{12} \vert \pi , \pi , \kappa' \rangle\vert^2 \over 2 \omega_{\pi} (\kappa) - 2 \omega_{\pi}(\kappa')} \approx
\]
\[
\int_0^\Lambda {d\kappa' \over \kappa^2 -\kappa^{\prime 2}}\left
( {\kappa^{\prime 2}(\kappa^{\prime 2}-\kappa^2) \langle \rho
\vert v_{12} \vert \pi , \pi , \kappa' \rangle\vert^2 \over 2
\omega_{\pi} (\kappa') - 2 \omega_{\pi}(\kappa)} - \kappa^2
\omega_{\pi} (\kappa) \langle \rho \vert v_{12} \vert \pi , \pi , \kappa \rangle\vert^2 \right ) +
\]
\beq
{1 \over 2}
\kappa \omega_{\pi} (\kappa)
\langle \rho \vert v_{12} \vert \pi , \pi , \kappa \rangle\vert^2
\ln ({\Lambda+\kappa \over \Lambda -\kappa})
+
\int_\Lambda^\infty {d\kappa' \over \kappa^2 -\kappa^{\prime 2}}\left
( {\kappa^{\prime 2}(\kappa^{\prime 2}-\kappa^2) \langle \rho
\vert v_{12} \vert \pi , \pi , \kappa' \rangle\vert^2 \over 2
\omega_{\pi} (\kappa') - 2 \omega_{\pi}(\kappa)} \right )
\label{sc:59}
\eeq
where the last term in this expression can be ignored for sufficiently
large cutoff $\Lambda$.  Expression (\ref{sc:57}) has a resonant form, although
because of the momentum dependence it is not necessarily a
Breit-Wigner form.  

In general the scattering calculation is a coupled channel calculation.
The total cross section is 
\beq
\sigma (\kappa)  = (\pi)^3 \omega_{\pi}(\kappa)^2 
\sum_l (2l+1) \vert T_l(\kappa) \vert^2 .
\label{sc:60}
\eeq

Figures 4 and 5 show the real and imaginary parts of the transition
operator for $\pi-\pi$ scattering for the parameters used section VII.
The calculation shown in these
figures is a coupled channel calculation using $\rho$ and $f_2$
intermediate states.  The total $\pi-\pi$ differential cross section
computed using these channels is shown in figure 6, and compared to
data from references \cite{Srinivasab1} \cite{Protopopescu}.

The computed cross section exhibits a qualitative agreement with the
data, but it is quantitatively about 25\% below the observed cross section.
The calculation is limited to $s$-channel exchanges.
\begin{figure}
\centering
\centering
\includegraphics[width=.9\linewidth]{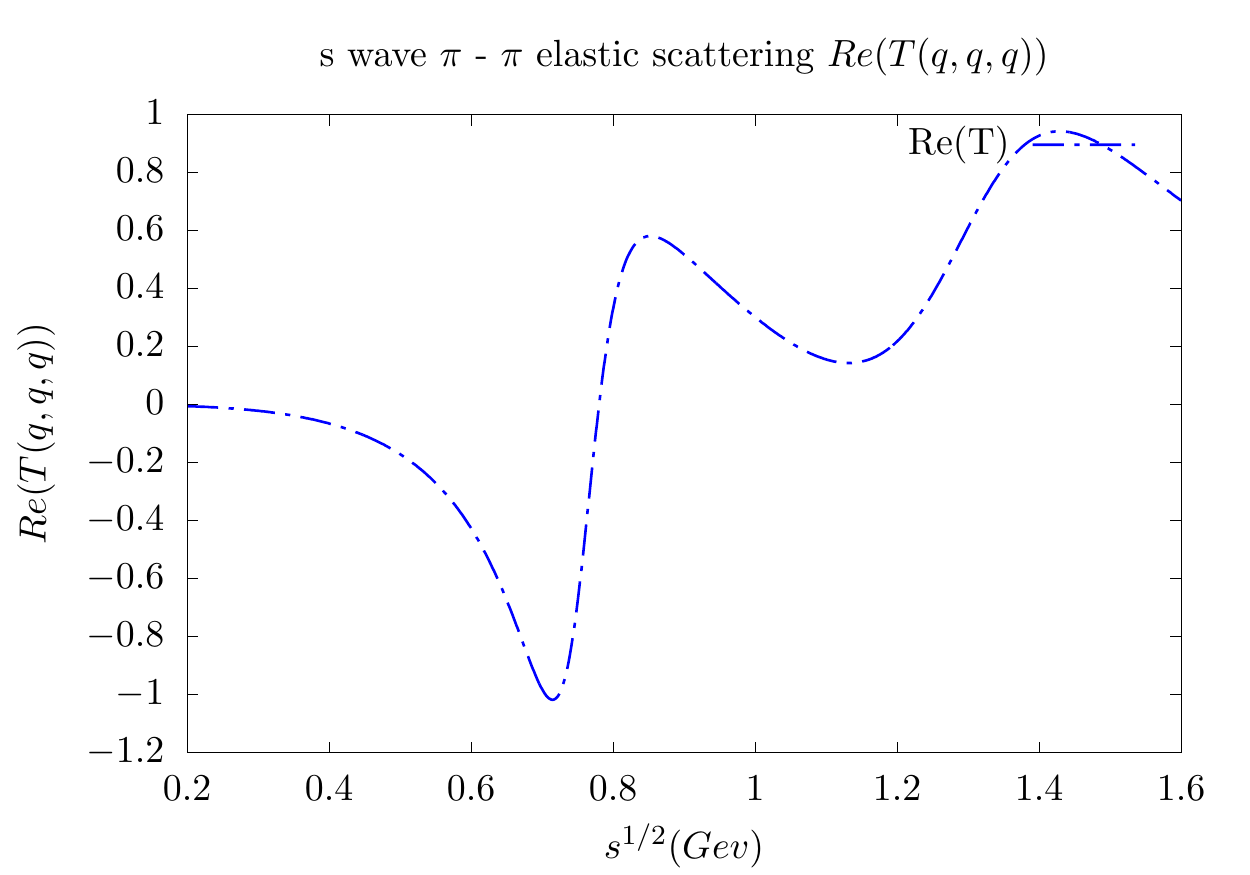}
\caption{\label{fig.5}Real part of T}
\label{fig:5}
\end{figure}
\begin{figure}
\centering
\includegraphics[width=.9\linewidth]{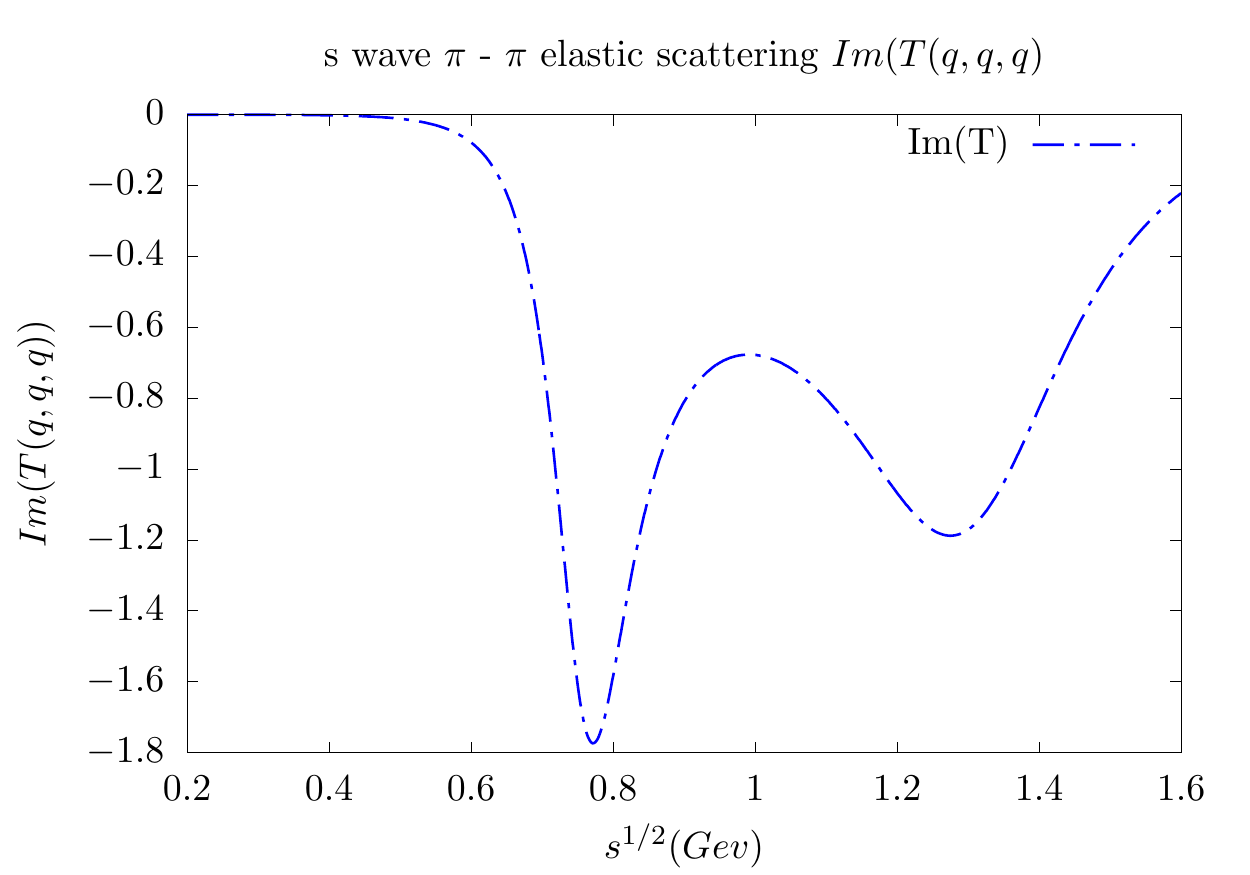}
\caption{\label{fig.6}Imaginary part of T}
\label{fig:6}
\end{figure}

\begin{figure}
\centering
\includegraphics[width=.7\linewidth]{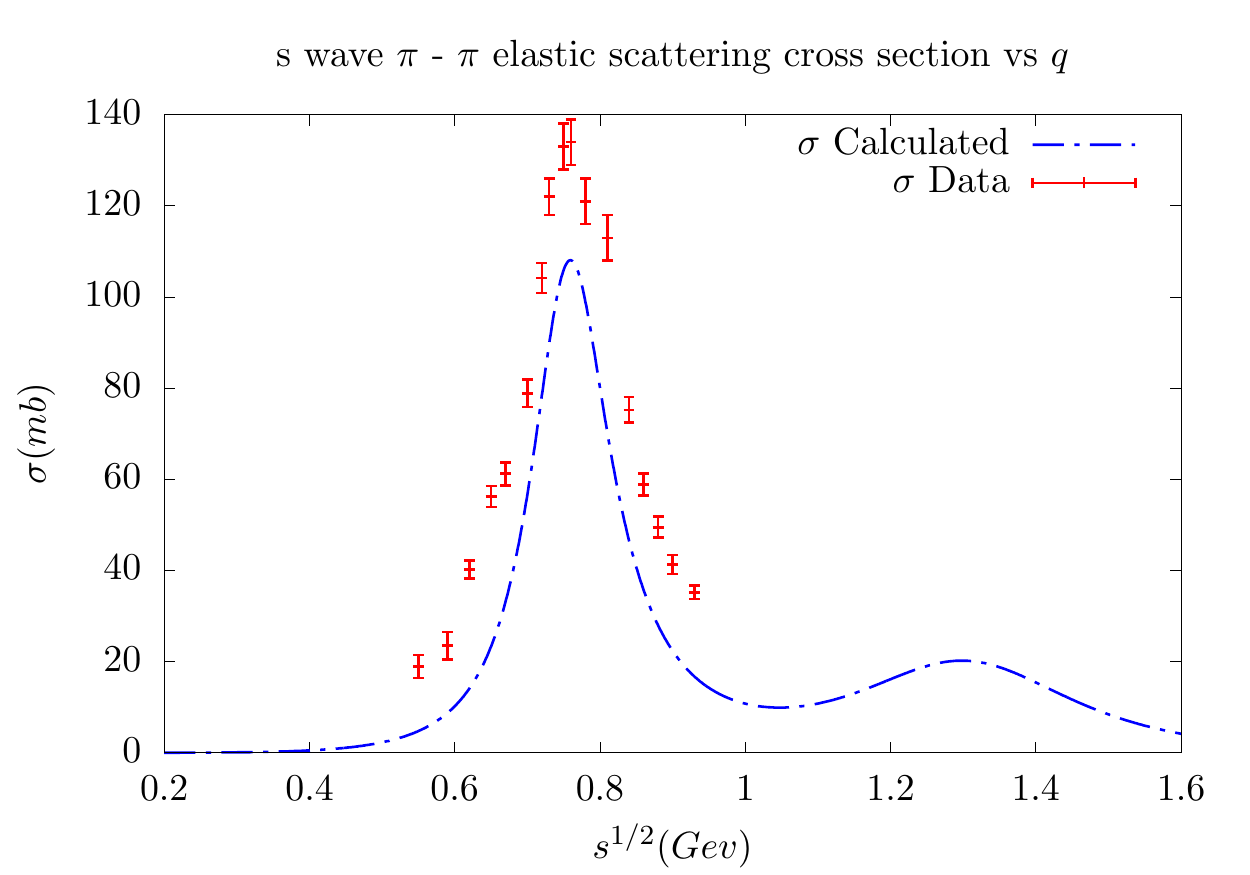}
\caption{\label{fig.7}Cross section}
\label{fig:7}
\end{figure}

\section{Unstable particles}
 
Bound states occur when 
\beq
M_{n_1,n_2,\kappa_{12}} - M_{n_0} >0 
\label{res:1}
\eeq
for all $n_1$ and $n_2$.  When (\ref{res:1}) is negative the denominator in
(\ref{bs:5}) can vanish.  When this happens the bound state becomes
unstable with respect to decay into the two-bare meson channels.  All
but a small number of the bare mesons states will fall into this
category.

The correct way to understand the resonant behavior is the same way that they
are observed experimentally, by looking for peaks in the differential
cross section. Equation (\ref{sc:57}) already has the resonant form
exhibiting a complex shift in the position of the bare meson pole.

When perturbation theory is justified, expressions for the lifetime
and mass corrections are simple: The expression for the second order
correction to the binding energy in (\ref{bs:17}) becomes complex
\[
\pval
\sum_{n_1,n_2}  q_{12}^2 dq_{12} 
{\vert \langle n_0 \Vert v_{1:2} \Vert n_1, n_2, q_{12} 
\rangle\vert ^2
\over M_{n_0} -M_{n_1,n_2,q_{12}} }
\]
\beq
-i \pi \int \delta (M_{n_0} - M_{n_1,n_2, q_{12}} ) q_{12}^2 dq_{12}
\vert \langle n_1, n_2, q_{12} \Vert v_{2:1} \Vert n_0 \rangle \vert^2  
\label{res:3}
\eeq
where $\pval$ indicates a principal value integral.  The principal
value term gives the leading correction to the bare mass.  The
imaginary term arises because the bound state can decay to the
two-particle continuum.  The imaginary part leads to a wave function
with an amplitude that decays like
\beq
e^{-\Gamma t/2}
\label{res:4}
\eeq
where $\tau = 1/\Gamma$ is the lifetime of the  unstable state
in perturbation theory:
\beq
\Gamma = \sum_{\mathbf{n}_1 \mathbf{n}_2}
2 \pi {q_{120} \omega_{n1}(q_{120})\omega_{n2}(q_{120})
\over \omega_{n1}(q_{120})+ \omega_{n2}(q_{120})}
\vert \langle n_1, n_2, q_{120} \vert v_{21} \vert n_0 \rangle  \vert^2
\label{res:5}
\eeq
The sum in (\ref{res:5}) sum is over the open decay channels which are the 
channels where 
\beq
M_{{n}_1,{n}_2,q_{120}}  = M_{{n}_0}
\label{res:6}
\eeq
has solutions for real $q^2_{120}$ that depend on $n_1$ and $n_2$.
This requires
\beq
M_{{n}_1,{n}_2,0} < M_{{n}_0}
\label{res:7}
\eeq
These solutions are
\beq
{q}_{120}^2 = 
{M_{{n}_1}^4 + M_{{n}_2}^4 + M_{{n}_0}^4 -
2 M_{{n}_1}^2 M_{{n}_2}^2 
-2 M_{{n}_1}^2  M_{{n}_0}^2 
-2 M_{{n}_2}^2 M_{{n}_0}^2 \over
4 M_{{n}_0}^2 }
\label{res:8}
\eeq
when the numerator is positive.

Partial widths for decay into the bare 
mesonic states $n_1$ and $n_2$ are given by
\beq
\Gamma_{n_1,n_2} = 
2 \pi {q_{120} \omega_{n1}(q_{120})\omega_{n2}(q_{120})
\over \omega_{n1}(q_{120})+ \omega_{n2}(q_{120})}
\vert \langle n_1, n_2, q_{120} \Vert v_{21} \Vert n_0 \rangle  \vert^2
\label{res:9}
\eeq
Again, $n_1$ and $n_2$ represent several quantum numbers.

This perturbative expression is identical to the $\Gamma$
in (\ref{sc:36}) obtained by looking at the scattering
amplitude near resonance, except in the exact case the shift is a
function of $\kappa$, so the position of the peak has to be determined by
by finding the value of $\kappa$ that makes the real part of denominator of
(\ref{sc:57}) vanish.

This analysis is applied to treat the decay of the $\rho$ meson into a
pair of pions.  Using the parameters given in section VII, the bare $\rho$
mass is shifted down by $.122$ GeV from its value of $.882$ to
the physical value of $.770$ GeV.  The resulting width is $\Gamma_p =
.134\times GeV$ which is qualitatively consistent with the observed
width of $.150$ GeV.  The size of the resonant shift, 12.7\%,  is consistent
with the size of the 17\% correction to the pion mass due to the coupling
to the sea quarks in this model.  Similar calculation could be performed for
higher lying states; these will generally involve more open channels.

\section{form factors}

The last type of observables of interest are electromagnetic observables.
Electron scattering from a hadron includes contributions from  both
valence and sea quarks.

The simplest electron scattering reaction is the scattering of an electron
from a charged pion.  The relevant observable is the pion form
factor.  Because the momentum transfer is space-like the form factor
can always be calculated in a frame where the $+$ component of the
momentum transfer is zero, $Q^+=0$.  The form factor can be expressed
in terms of the $+$ component of the current at $x=0$:
\beq
F_\pi (Q^2) = \langle \pi , \tilde{\mathbf{p}}' \vert I^+(0) \vert
\pi ,\tilde{\mathbf{p}} \rangle  
\eeq
where the pion state (in this model) will in general include both
valence and sea quark contributions.  In this model the pion state vector
has the form
\beq
\left (
\begin{array}{c}    
\vert \pi, \mathbf{\tilde{p}} \rangle_1 \\ 
{1 \over m_\pi - M_{2}}  v_{2:1}  
{1 \over m_\pi - M_{1}} \vert \pi ,\mathbf{\tilde{p}} \rangle_1 \\
\end{array}
\right )
\eeq
where $m_{\pi}$ is the mass eigenvalue.
The current matrix element has the general form
\[
\langle \pi , \mathbf{p}_{\perp}' \vert I^+(0) \vert
\pi ,\mathbf{p}_{\perp} \rangle =
\]
\[
_1\langle \pi ,\mathbf{\tilde{p}}' \vert I^{\mu}(0) 
\vert \pi, \mathbf{\tilde{p}} \rangle_1 + 
\]
\[
_1\langle \pi, \mathbf{\tilde{p}}' \vert I^{\mu}(0)
\vert  {1 \over m_\pi - M_{2}}v_{2:1}
{1 \over m_\pi - M_{1}} \vert \pi ,\mathbf{\tilde{p}} \rangle_1
+
\]
\[
_1\langle
\pi, \mathbf{\tilde{p}} \vert
{1 \over m_\pi - M_1}
v_{12} {1 \over m_\pi - M_{2}}
\vert I^{\mu}(0)
\vert \pi , \mathbf{\tilde{p}} \rangle_1 +
\]
\beq
_1\langle
\pi ,\mathbf{\tilde{p}} \vert
{1 \over m_\pi - M_{1}}
v_{12} {1 \over m_\pi^* - M_{2}}
\vert I^{\mu}(0)
\vert  {1 \over m_\pi^* - M_{2}}v_{2:1}
{1 \over m_\pi - M_{1}} \vert \pi, \mathbf{\tilde{p}}' \rangle_1
\label{cur:1}
\eeq
This expression involves both the wave functions and current
operators.  The evaluation of the current matrix elements is naturally
performed in the quark-anti-quark-gluon representation.  The advantage
of the light front representation is invariance of the single quark
magnetic quantum numbers under light front boosts, which lead to
frame-independent impulse approximations.  In addition the boosts are
kinematic.  On the other hand the quark current operator necessarily
has many-body contributions due to the dynamical nature of rotational
covariance and current conservation.  These relations involve the
string breaking vertex.

The structure of the full current is beyond the scope of this model.
It is nevertheless useful to examine the pion form factor assuming
that the quarks can be treated a point particles in the valence sector
of this model to determine if there is any kind of qualitative
agreement with experiment.
 
Equation (\ref{cur:1}) can be interpreted as the matrix element of a
two-body current in the valence state, however in this case valence
state is not the same as the bare state, and there is an additional
normalization correction that appears in the current.

The simplest calculation is the calculation of the charge form factor
for the bare pion assuming that the single quark and anti-quark are treated
as point charges with no magnetic form factors.  In this case the pion
form factor is given by the first term in (\ref{cur:1}) where the
valence wave function is replaced by the bare wave function.  The
results of the calculation are shown in figures 8 and 9, where they are
compared to low and high energy data from
\cite{amendolia}\cite{volmer}\cite{bebek}.  The solid curves use the
quark masses and oscillator parameters of this model. The dash-dot
curves use an oscillator wave function with $\lambda=.259$ and
quark masses of $m_q=.20$,
which were the values used in \cite{coester2}, and the
dotted curves use $\lambda=.282$ and $m_q=.2$.
\begin{figure}
\centering
  \centering
  \includegraphics[width=1\linewidth]{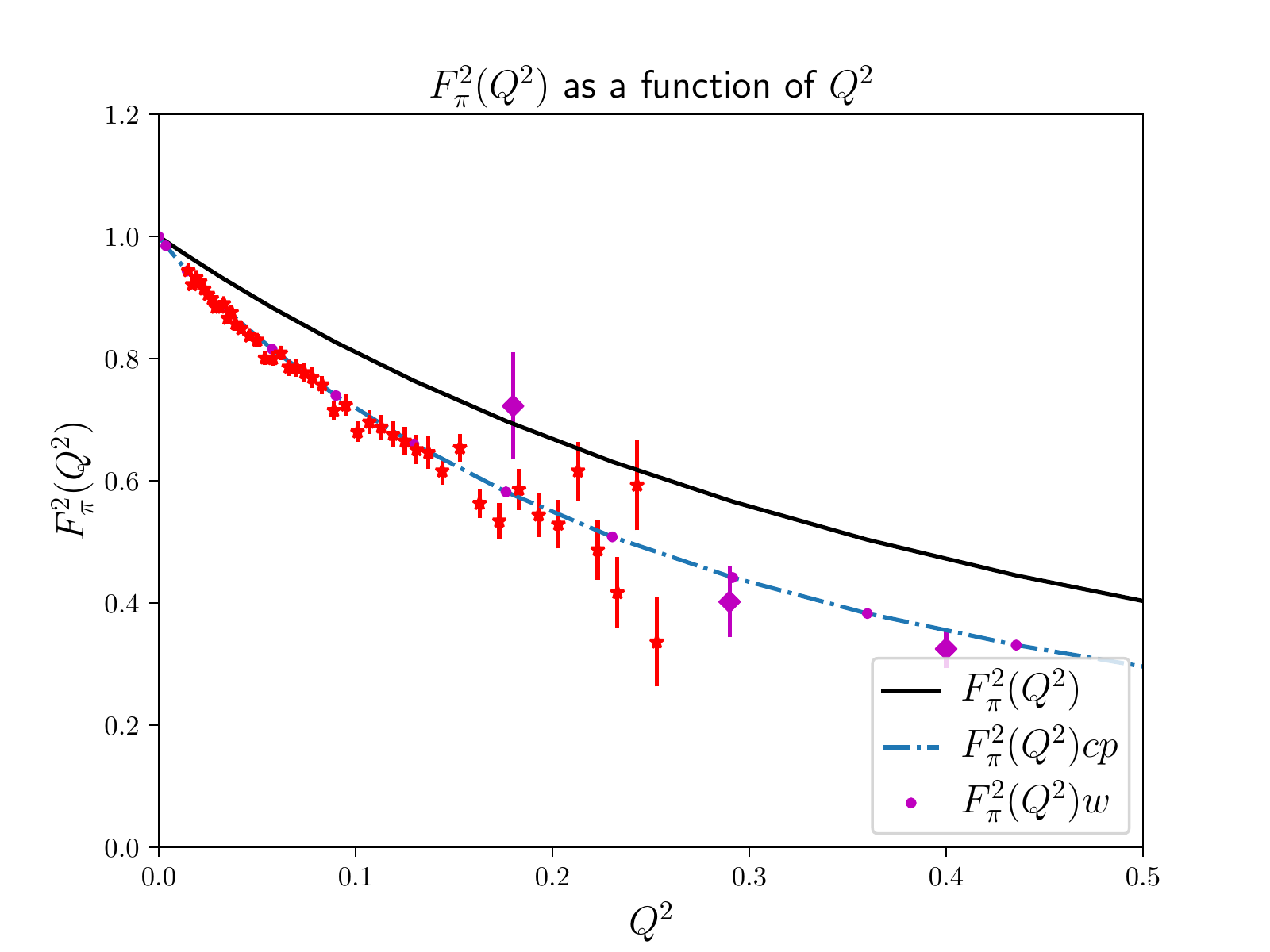}
  \caption{\label{fig.8}$F_\pi(Q^2)$}
  \label{fig:test5}
\end{figure}
\begin{figure} 
  \centering
  \includegraphics[width=1\linewidth]{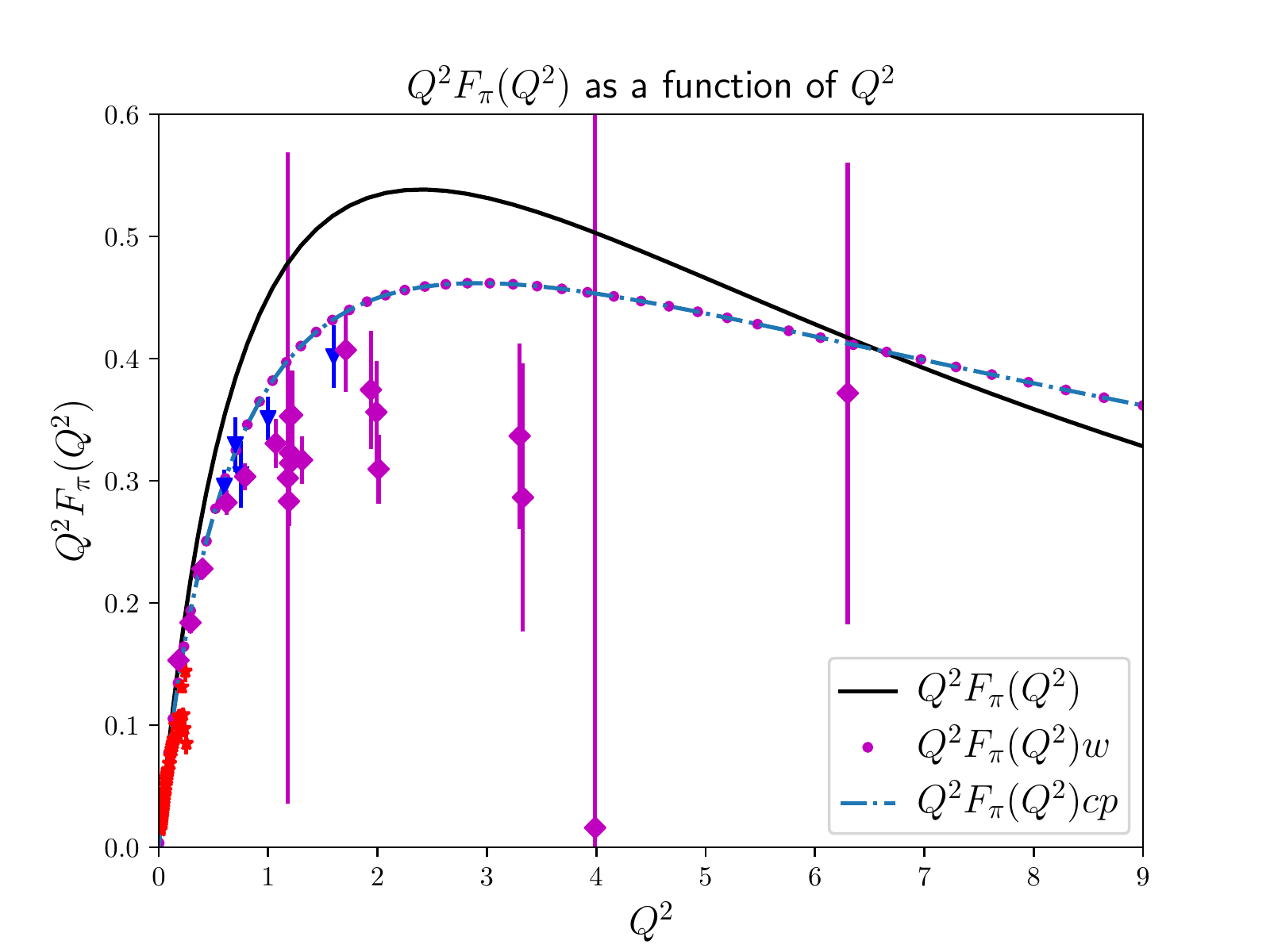}
  \caption{\label{fig.9}$Q^2 F_{\pi}(Q^2)$}
  \label{fig:test9}
\end{figure}
The single quark and anti-quarks are treated as point charges with no
magnetic form factors.  Equation (\ref{sq:6}) is used to express the
valence wave function in terms of the single quark degrees of freedom.
The calculated form factor is too large compared to data for both low
and high $Q^2$, however a rough agreement was achieved in
\cite{coester2} using the same valence wave function with 
$\lambda$ decreased by about 8\% to .259 and a lighter quark mass,
$m=.2$ . Since the quark masses appear in the
Poincar\'e generators in
the in the combination $m_q^2+V_0$, there is an additional freedom to
reduce the quark masses keeping $m_q^2+V_0$ constant.  This freedom
leaves the Poincar\'e generators unchanged.  It leaves the bound state,
scattering, and lifetime calculations unchanged.
Reducing the quark masses
to $m_q=.2$ without changing the coupling constant gives the curves
with the dots, which are indistinguishable from the 
results of \cite{coester2} which use a slightly reduced value of $\lambda$.
Further calculations are needed to determine the corrections due to
sea quarks and two-body currents.

\section{conclusion}

This work examined a simple model of mesons, based on non-local
degrees of freedom that are both globally and locally gauge invariant.
The structure of the model was largely motivated by the strong
coupling limit of Hamiltonian lattice QCD.  Gauge invariant systems of
quarks and anti-quarks connected by links were modeled by quark
anti-quark pairs interacting via a confining interaction.  A key
feature is that quarks in different connected singlets are treated as
distinguishable particles.  The model has dual representations in
terms of hadronic and QCD degrees of freedom.  While there are many
possible interactions involving connected gauge invariant degrees of
freedom, this work examined the consequence of the assumption that the
dominant contribution comes from string breaking, which produces
quark-anti-quark pairs from a given local/global singlet state.  The
model was formulated so all matrix elements of the string breaking
vertex with bare meson states could be computed analytically in terms
of the underlying QCD degrees of freedom.  Calculations were performed
by truncating to the direct sum of a one and two-singlet subspace.
The model is formulated to be fully relativistic and designed so
computations can be performed efficiently.  In order to stay close in
spirit to QCD, one parameter was used to set all of the scales.  In
this application that parameter was fixed by the Regge slope of the
family of excitations of the $\rho$ meson in the valence model, which
fixed a confinement scale.  The same parameter was used to fix the
size of the string breaking vertex, however a scale transformation on
the string breaking vertex was needed to get a qualitatively
consistent treatment of lifetimes and spectral properties.  The vertex
also included a dimensionless coupling constant of order unity.
The model
was used to calculate sea quark contributions to bound states,
spectral properties, cross sections, lifetimes and electromagnetic
properties.  The calculations of spectral properties, lifetimes, and
scattering cross sections generally gave results that were
qualitatively consistent with experiment.  The calculation were
consistent with a non-trivial sea quark component in the pion state vector.
The pion form factor in the bare model over predicted the experimental data for
both high and low momentum transfers, however the data can be fit with
the same wave functions using smaller quark masses (48\%).
Since the quark mass in this model appears in the
combination $m_q^2 +V_0$, it could be reduced by changing $V_0$ to keep
this combination constant.  This change leaves all of the dynamical
operators unchanged, so would not affect the bound state,
scattering or resonance calculations.  Further adjustments in the quark masses
are anticipated after contributions from sea quarks or two-body currents
are included.

These results suggest that a string breaking reaction mechanism is a
significant component of hadronic dynamics, and could provide a good
starting point for developing more realistic models of hadronic
dynamics.  This type of model could also be extended to treat baryons
modeled as diquarks, which would provide more stringent experimental
constraints.  This model is also directly applicable to treating
exotic mesons as confined diquark-anti-diquark systems, or as
molecules of confined quark and anti-quarks.  Realistic applications require
generalizing the model to include flavor dependent interactions.

This research supported by the US Department of Energy, Office
of Science, grant number DE-SC0016457 who supported this research effort.

\section{Appendix I: String breaking matrix elements}

In this and the next appendix the analytic expression (\ref{sb:22}) for the
matrix elements of
spin-independent part of the string breaking vertex the oscillator
basis is derived.  This involves a 9 dimensional integral that can be computed
analytically.

In order to calculate matrix elements of this vertex with the harmonic
oscillator states the expressions for the  smeared
delta functions are inserted in the expression for the vertex.
The resulting expression for the kernel of the string-breaking vertex becomes
\[
\langle \mathbf{r}_1,\mathbf{r}_2,\mathbf{r}_{12} \vert v_{2:1} \vert \mathbf{r}
\rangle  :=
\]
\beq
{g \over \sqrt{\lambda}}   \delta (\mathbf{r}-2\mathbf{r}_{12})
({\lambda \over 4 \pi})^3 
\int_0^1 d\eta 
e^{-{\lambda \over 4}(
(\mathbf{r}_{1}-2 \eta \mathbf{r}_{12})^2 + 
(\mathbf{r}_{2}-2 (1-\eta) \mathbf{r}_{12})^2)}
\label{sb:5}
\eeq
where the Gaussians delta functions are expressed using 
\beq
e^{-{\lambda \over 4}(\mathbf{r}_i-2 \alpha \mathbf{r}_{12})^2} = 
e^{- {\lambda \over 4}\mathbf{r}_i^2 -\lambda \alpha^2 \mathbf{r}_{12}^2 
+ \alpha \lambda \mathbf{r}_i\cdot \mathbf{r}_{12}} .
\label{sb:6}
\eeq
The angle dependent part of (\ref{sb:6})
can be represented using the partial-wave expansion of
$e^{\mathbf{k}\cdot \mathbf{x}}$ \cite{abramowitz:1964} (see section 10.2.36
page 445 - which is the analytic continuation of the usual expansion of
plane waves in spherical harmonics:)
\beq
e^{i \mathbf{k}\cdot \mathbf{r}} =
4 \pi \sum l^l j_l(kr) Y^*_{lm}(\hat{\mathbf{k}}) Y_{lm}(\hat{\mathbf{r}})
\label{sb:7}
\eeq
which becomes
\beq
e^{z \cos (\theta)}= \sum_{n=0}^\infty (2n+1) \sqrt{\pi \over 2z}
I_{n+1/2} (z) P_n (\cos (\theta))
\label{sb:8}
\eeq
or
\beq
e^{ \alpha \lambda \mathbf{r}_i \cdot \mathbf{r}_{12}} 
=4 \pi \sqrt{{\pi \over 2 \lambda \alpha r_i r_{12}}}
\sum_{l=0}^\infty \sum_{m=-l}^l I_{l+{1\over 2}}(\lambda \alpha r_i r_{12})  
Y^*_{lm}(\hat{\mathbf{r}_i}) Y_{lm}(\hat{\mathbf{r}_{12}}).
\label{sb:9}
\eeq
The Bessel function $I_{l+{1\over 2}} (2ab)$ in (\ref{sb:9}) 
is a generating function for the associated Laguerre polynomials 
(see \cite{abramowitz:1964}, equations 22.9.16 and 9.6.3 or
\cite{schwinger:2001} problem 7.12).
Abramowitz and Stegun give the following expressions
\beq
(xz)^{-\alpha/2}e^z J_\alpha (2(xz)^{1/2}) = \sum_{n=0}^\infty L^\alpha_n (x){z^n \over
\Gamma (n+a+1)}  
\label{sb:10}
\eeq
\beq
I_\nu (z) = e^{- {1\over 2}\nu \pi i }J_\nu (ze^{{1\over 2}\pi i})
\qquad -\pi < \phi < \pi/2
\label{sb:11}
\eeq
\beq
I_\nu (z) = e^{ {3\over 2}\nu \pi i }J_\nu (ze^{-{3\over 2}\pi i})
\qquad \pi/2 < \phi < \pi
\label{sb:12}
\eeq
and Schwinger gives (problem 7.12 p. 302) the equivalent but more useful   
expression
\beq
e^{-\lambda}I_\alpha (2\sqrt{\lambda x})=
(\lambda x)^{\alpha \over 2} \sum_{n=0}^\infty
(-)^n {\lambda^n \over (n+\alpha)!} L^\alpha_n (x)
\label{sb:13}
\eeq
which in this case is 
\beq
I_{l+{1\over 2}} (2ab) = e^{a^2} (ab)^{l+{1 \over 2}}
\sum_{n=0}^\infty (-)^n 
{a^{2n} \over \Gamma (n+l+{3 \over 2})}L_n^{l+{1\over 2}}(b^2).
\label{sb:14}
\eeq
In what follows this expression is used with $a=\sqrt{{\lambda \over 2}}
\alpha r_{12}$ and $b=\sqrt{{\lambda \over 2}}r_{i}$.

Using (\ref{sb:14}) in (\ref{sb:9}) along with the 
expression (\ref{cq:11_r}) for the harmonic oscillator 
wave function gives 
\[
e^{-{\lambda \over 4}(\mathbf{r}_i-2 \alpha \mathbf{r}_{12})^2} = 
\]
\[
2 \pi^{3/2} 
\sum_{n,l,m}
(\sqrt{{\lambda \over 2}}\alpha r_{12})^{2n+l}
e^{-{\lambda \over 2} \alpha^2 r_{12}^2}
(-)^n
{(\sqrt{{\lambda \over 2}}r_i)^l 
L^{l+{1 \over 2}}_n({\lambda \over 2}r_i^2 ) e^{-{\lambda \over 4}r_i^2}
\over \Gamma (n+l+{3 \over 2})}
Y_{lm}(\hat{\mathbf{r}}_i) Y^*_{lm}(\hat{\mathbf{r}_{12}}) =
\]
\beq
2 \pi^{3/2} ({2 \over \lambda})^{3/4} 
e^{-{\lambda \over 2} \alpha^2 r_{12}^2}
\sum_{n=0}^\infty
(\sqrt{{\lambda \over 2}}\alpha r_{12})^{2n+l}
{
R_{nl}({r}_i)
\over \sqrt{2 n! 
\Gamma (n+l+{3 \over 2})}}
Y_{lm}(\hat{\mathbf{r}_i})
Y_{lm}^*(\hat{\mathbf{r}_{12}})
\label{sb:15}
\eeq
where this has been expressed in terms of
the radial harmonic oscillator wave function
$R_{nl}(r)$. 

Letting
$\alpha =\eta$ and $\alpha = 1-\eta$ gives the following expression 
for the kernel of the vertex as an expansion in harmonic oscillator states:
\[
\langle \mathbf{r}_1,\mathbf{r}_2,\mathbf{r}_{12} \vert v_{2:1} \vert \mathbf{r}
\rangle =
\]
\[
{g \over \sqrt{\lambda}}  \delta (\mathbf{r}-2\mathbf{r}_{12})
({\lambda \over 4 \pi})^3 
4 \pi^3 ({2 \over \lambda })^{3/2}
\sum_{n_1l_1m_1n_2l_2m_2} \int_0^1 d\eta 
e^{-{\lambda \over 2}{r}_{12}^2(\eta^2 +(1-\eta)^2)}
\eta^{l_1+ 2n_1}(1-\eta)^{l_2+ 2n_2} 
\times
\]
\beq
(\sqrt{{\lambda \over 2}}{r}_{12})^{2n_1+l_1+2n_2+l_2} 
{\psi_{n_1l_1m_1}({\mathbf{r}}_1)
Y^*_{l_1m_1}(\hat{\mathbf{r}}_{12})
\psi_{n_2 l_2 m_2}({\mathbf{r}}_2)
Y^*_{l_2 m_2}(\hat{\mathbf{r}}_{12})  
\over \sqrt{2 n_1! \Gamma (n_1+l_1+{3 \over 2})}
\sqrt{2 n_2! \Gamma (n_2+l_2+{3 \over 2})}}
\label{sb:16}
\eeq
where 
\beq
\psi_{nlm}(\mathbf{r}) =
R_{nl} (\mathbf{r})Y_{lm}(\hat{\mathbf{r}}) .
\label{sb:17}
\eeq
Integrating this against three oscillator eigenstates
in $\mathbf{r}_{1}$, $\mathbf{r}_{2}$ and $\mathbf{r}$
reduces the vertex in coordinate space to an expression 
in terms of a single one-dimensional integral
\[
\langle n_1,l_1,m_1,n_2,l_2,m_2,\mathbf{r}_{12} \vert v_{2:1} \vert
n,l,m \rangle 
=\int 
\psi_{n_1l_1m_1}^*(\mathbf{r}_1)
\psi_{n_2l_2m_2}^*(\mathbf{r}_2)
v(\mathbf{r}_1,\mathbf{r}_2,\mathbf{r}_{12};\mathbf{r})
\psi_{nlm}(\mathbf{r})
d\mathbf{r}_1 d \mathbf{r}_2 d \mathbf{r} 
\]
\[
={g \over \sqrt{\lambda}}
 R_{nl}(2 {r}_{12})
({2\lambda})^{3/2}
{(\sqrt{{\lambda \over 2}}{r}_{12})^{2n_1+l_1+2n_2+l_2} 
\over
(\sqrt{2 n_1! \Gamma (n_1+l_1+{3 \over 2}})
(\sqrt{2 n_2! \Gamma (n_2+l_2+{3 \over 2}})
}\times
\]
\beq
\int_0^1 d\eta
e^{-{\lambda \over 2}{r}_{12}^2(\eta^2 +(1-\eta)^2)}
\eta^{l_1+2n_1}(1-\eta)^{l_2+ 2n_2} 
Y_{lm} (\hat{\mathbf{r}}_{12})
Y_{l_1m_1}^*(\hat{\mathbf{r}}_{12})
Y_{l_2m_2}^*(\hat{\mathbf{r}}_{12}) .
\label{sb:18}
\eeq
The one-dimensional integral 
\beq 
\int_0^1 d\eta e^{-{\lambda \over 2}r_{12}^2(\eta^2 +(1-\eta)^2)}
\eta^{l_1+ 2n_1}(1-\eta)^{l_2 + 2n_2}
\label{sb:19}
\eeq
can be performed analytically, although
a numerical calculation may be more efficient. 
The analytic calculation is given in appendix II.
The result of this calculation is 
\[
\int_0^1 d\eta e^{-{\lambda \over 2}r_{12}^2(\eta^2 +(1-\eta)^2)}
\eta^{l_1+ 2n_1}(1-\eta)^{l_2 + 2n_2}
=e^{-{\lambda \over 4}{r}_{12}^2} \int_0^1 d\eta 
e^{-\lambda r_{12}^2 (\eta-{1\over 2})^2}
\eta^{2n_1+l_1}(1-\eta)^{2n_2+l_2} 
\]
\[
= e^{-{\lambda \over 4} r_{12}^2} \sum_{k_1+k_2=2r} 
{(l_1+2n_1)! (l_2+2n_2)! \over
k_1! k_2! (l_1+2n_1-k_1)! (l_2+2 n_2 -k_2)!}(-)^{k_2}
({1 \over 2})^{l_1+2n_1+l_2 +2n_2 } 
\times
\]
\beq
{1 \over 2r+1}
M({1 \over 2}+r ,{3 \over 2}+r, -{\lambda r^2_{12} \over 4})
\label{sb:20}
\eeq
where the sum is over $k_1$, and $k_2$ keeping only terms with
$k_1+k_2=2r$ even, $0 \leq k_i \leq 2n_i + l_i$, and
$M(\alpha,\beta,\gamma)$ is the confluent hypergeometric function.

The kernel of the spin-independent part of the string breaking vertex
is 
\[
\langle n_1,l_1,m_1,n_2,l_2,m_2,\mathbf{r}_{12} \vert v_{2:1} \vert
n,l,m \rangle =
\]
\[
\int 
\psi_{n_1l_1m_1}^*(\mathbf{r}_1)
\psi_{n_2l_2m_2}^*(\mathbf{r}_2)
v(\mathbf{r}_1,\mathbf{r}_2,\mathbf{r}_{12};\mathbf{r})
\psi_{nlm}(\mathbf{r})
d\mathbf{r}_1 d \mathbf{r}_2 d \mathbf{r} =
\]
\[
{g \over \sqrt{\lambda}}
R_{nl}(2 {r}_{12})
({2\lambda})^{3/2}
{(\sqrt{{\lambda \over 2}}{r}_{12})^{2n_1+l_1+2n_2+l_2} 
\over
\sqrt{2 n_1! \Gamma (n_1+l_1+{3 \over 2}})
\sqrt{2 n_2! \Gamma (n_2+l_2+{3 \over 2}})
}\times
\]
\[
e^{-{\lambda \over 4} r_{12}^2} \sum_{k_1+k_2=2r} 
{(l_1+2n_1)! (l_2+2n_2)! \over
k_1! k_2! (l_1+2n_1-k_1)! (l_2+2 n_2 -k_2)!}(-)^{k_2}
({1 \over 2})^{l_1+2n_1+l_2 +2n_2 } 
\times
\]
\beq
{1 \over 2r+1}
M({1 \over 2}+r ,{3 \over 2}+r, -{\lambda r^2_{12} \over 4})
Y_{lm} (\hat{\mathbf{r}}_{12})
Y_{l_1m_1}^*(\hat{\mathbf{r}}_{12})
Y_{l_2m_2}^*(\hat{\mathbf{r}}_{12}).
\label{sb:22b}
\eeq
The radial part of this expression is the part 
of (\ref{sb:22b}) that multiplies the three
spherical harmonics:
\[
\langle 
n_1,l_1,n_2,l_2
,{r}_{12} \vert v_{2:1} \vert
n,l \rangle =
\]
\[
{g \over \sqrt{\lambda}}
R_{nl}(2 {r}_{12})
({2\lambda})^{3/2}
\times
\]
\[
{(\sqrt{{\lambda \over 2}}{r}_{12})^{2n_1+l_1+2n_2+l_2} 
\over
\sqrt{2 n_1! \Gamma (n_1+l_1+{3 \over 2}})
\sqrt{2 n_2! \Gamma (n_2+l_2+{3 \over 2}})
}\times
\]
\[
e^{-{\lambda \over 4} r_{12}^2} \sum_{k_1+k_2=2r} 
{(l_1+2n_1)! (l_2+2n_2)! \over
k_1! k_2! (l_1+2n_1-k_1)! (l_2+2 n_2 -k_2)!}(-)^{k_2}
({1 \over 2})^{l_1+2n_1+l_2 +2n_2 } 
\times
\]
\beq
{1 \over 2r+1}
M({1 \over 2}+r ,{3 \over 2}+r, -{\lambda r^2_{12} \over 4}).
\label{sb:24}
\eeq

\section{Appendix II}

The integral (\ref{sb:19}) that must be evaluated in Appendix I to get an analytic expression
for the meson form factors in the coordinate representation
has the form
\[
I_{n_1.l_1,n_2,l_2} (r_{12}) =
\]
\[
\int_0^1 e^{-{\lambda \over 2}r_{12}^2 (\eta^2 + \eta^2 - 2 \eta +1}
\eta^{l_1+2n_1} (1-\eta)^{l_2+2n_2} d\eta =
\]
\beq
e^{-{\lambda \over 4} r_{12}^2} 
\int_0^1 e^{-{\lambda}r_{12}^2 (\eta- {1 \over 2})^2}
\eta^{l_1+2n_1} (1-\eta)^{l_2+2n_2} d\eta .
\label{ap:1}
\eeq
Let $\nu= \eta-{1 \over 2}$ so this becomes
\beq
I_{n_1.l_1,n_2,l_2} (r_{12}) 
= e^{-{\lambda \over 4} r_{12}^2} 
\int_{-{1 \over 2}}^{{1 \over 2}} e^{-{\lambda}r_{12}^2 \nu^2}
({1 \over 2}+ \nu)^{l_1+2n_1} ({1 \over 2}- \nu)^{l_2+2n_2} d\nu .
\label{ap:1a}
\eeq
Using the binomial theorem this becomes
\[
I_{n_1.l_1,n_2,l_2} (r_{12}) =
\]
\[
= e^{-{\lambda \over 4} r_{12}^2} \sum_{k_1+k_2=2r} 
{(l_1+2n_1)! (l_2+2n_2)! \over
k_1! k_2! (l_1+2n_1-k_1)! (l_2+2 n_2 -k_2)!}(-)^{k_2}
({1 \over 2})^{l_1+2n_1+l_2 +2n_2 - 2r} 
\int_{-{1 \over 2}}^{{1 \over 2}} e^{-{\lambda}r_{12}^2 \nu^2}
\nu^{2r} =
\]
\[
= e^{-{\lambda \over 4} r_{12}^2} \sum_{k_1+k_2=2r} 
{(l_1+2n_1)! (l_2+2n_2)! \over
k_1! k_2! (l_1+2n_1-k_1)! (l_2+2 n_2 -k_2)!}(-)^{k_2}
({1 \over 2})^{l_1+2n_1+l_2 +2n_2 - 2r} 
\times
\]
\[
(-{1 \over r_{12}^2})^r{d^{r} \over d\lambda^{r}}
\int_{-{1 \over 2}}^{{1 \over 2}} e^{-{\lambda}r_{12}^2 \nu^2}
\]
\[
= e^{-{\lambda \over 4} r_{12}^2} \sum_{k_1+k_2=2r} 
{(l_1+2n_1)! (l_2+2n_2)! \over
k_1! k_2! (l_1+2n_1-k_1)! (l_2+2 n_2 -k_2)!}(-)^{k_2}
({1 \over 2})^{l_1+2n_1+l_2 +2n_2 - 2r} 
\times
\]
\beq
(-{1 \over r_{12}^2})^r{d^{r} \over d\lambda^{r}}
{1 \over \sqrt{\lambda}r_{12}} 
\int_{-{\sqrt{\lambda} r_{12} \over 2}}^{{\sqrt{\lambda}r_{12} \over 2}} 
e^{-\rho^2}d\rho .
\label{ap:2}
\eeq
Next express the integral in terms of the error function
\beq
\mbox{erf} (z) = {2 \over \sqrt{\pi}} \int_0^z e^{-t^2} dt
\label{ap:3}
\eeq
to get
\[
I_{n_1.l_1,n_2,l_2} (r_{12}) =
\]
\[
= e^{-{\lambda \over 4} r_{12}^2} \sum_{k_1+k_2=2r} 
{(l_1+2n_1)! (l_2+2n_2)! \over
k_1! k_2! (l_1+2n_1-k_1)! (l_2+2 n_2 -k_2)!}(-)^{k_2}
({1 \over 2})^{l_1+2n_1+l_2 +2n_2 - 2r} 
\times
\]
\beq
{\sqrt{\pi} \over 4} (-{1 \over r_{12}^2})^r{d^{r} \over d\lambda^{r}}
{2 \over \sqrt{\lambda}r_{12}} 
\left (
{2 \over \sqrt{\lambda}r_{12}} \mbox{erf}({\sqrt{\lambda}r_{12} \over 2})
-
{2 \over \sqrt{\lambda}r_{12}} \mbox{erf}(-{\sqrt{\lambda}r_{12} \over 2})
\right )
\label{ap:4}
\eeq
then express the error functions in terms of 
confluent hypergeometric function (see AS 7.1.21)
\beq
{\mbox{erf} (z) \over z} = {2 \over \sqrt{\pi}}M({1 \over 2} ,{3 \over 2},-z^2)
\label{ap:5}
\eeq
\[
I_{n_1.l_1,n_2,l_2} (r_{12}) =
\]
\[
= e^{-{\lambda \over 4} r_{12}^2} \sum_{k_1+k_2=2r} 
{(l_1+2n_1)! (l_2+2n_2)! \over
k_1! k_2! (l_1+2n_1-k_1)! (l_2+2 n_2 -k_2)!}(-)^{k_2}
({1 \over 2})^{l_1+2n_1+l_2 +2n_2 - 2r} 
\times
\]
\[
{\sqrt{\pi} \over 4} (-{1 \over r_{12}^2})^r{d^{r} \over d\lambda^{r}}
{2 \over \sqrt{\lambda}r_{12}} 
\left (
{2 \over \sqrt{\pi}} M({1 \over 2} ,{3 \over 2}, -{\lambda r^2_{12} \over 4})
+
{2 \over \sqrt{\pi}} M({1 \over 2} ,{3 \over 2}, -{\lambda r^2_{12} \over 4})
\right )
\]

\[
= e^{-{\lambda \over 4} r_{12}^2} \sum_{k_1+k_2=2r} 
{(l_1+2n_1)! (l_2+2n_2)! \over
k_1! k_2! (l_1+2n_1-k_1)! (l_2+2 n_2 -k_2)!}(-)^{k_2}
({1 \over 2})^{l_1+2n_1+l_2 +2n_2 - 2r} 
\times
\]
\beq
{1 \over 4^r} {d^{r} \over d(-\lambda r_{12}^2/4)^{r}}
{2 \over \sqrt{\lambda}r_{12}} 
M({1 \over 2} ,{3 \over 2}, -{\lambda r^2_{12} \over 4}).
\label{ap:6}
\eeq
Next use AS 13.4.9 to express derivatives of the confluent
hypergeometric functions in terms of the confluent
function.
\[
I_{n_1.l_1,n_2,l_2} (r_{12}) =
\]
\[
= e^{-{\lambda \over 4} r_{12}^2} \sum_{k_1+k_2=2r} 
{(l_1+2n_1)! (l_2+2n_2)! \over
k_1! k_2! (l_1+2n_1-k_1)! (l_2+2 n_2 -k_2)!}(-)^{k_2}
({1 \over 2})^{l_1+2n_1+l_2 +2n_2 - 2r} 
\times
\]
\[
{1 \over 4^r} {\Gamma ({1 \over 2}+r  ) \Gamma ( {3 \over 2})
\over  \Gamma ({3 \over 2}+r  ) \Gamma ( {1 \over 2})}
M({1 \over 2}+r ,{3 \over 2}+r, -{\lambda r^2_{12} \over 4})
\]
which gives (\ref{sb:20}) 
\[
= e^{-{\lambda \over 4} r_{12}^2} \sum_{k_1+k_2=2r} 
{(l_1+2n_1)! (l_2+2n_2)! \over
k_1! k_2! (l_1+2n_1-k_1)! (l_2+2 n_2 -k_2)!}(-)^{k_2}
({1 \over 2})^{l_1+2n_1+l_2 +2n_2 } 
\times
\]
\beq
{1 \over 2r+1}
M({1 \over 2}+r ,{3 \over 2}+r, -{\lambda r^2_{12} \over 4}).
\label{ap:7}
\eeq

\section{Appendix III: Cook Condition}

If the vertex is evaluated by projecting on a truncated basis the
resulting interaction is separable.  While the separable approximation
is a short-range compact interaction which has a non-trivial scattering
operator,  it is not automatically true that the sum of an infinite number
of separable terms remains short ranged.
For this reason it is desirable to prove that the limit of the
$S$ matrix with truncated numbers of channels converges in the
infinite number of channel limit to the exact $S$-matrix.  Here exact
means in the context of the $2+1$ singlet model.

On way to avoid directly dealing with infinite number of channels
is to show the existence of the channel wave operators that are used
to make the $S$ operator.
The Cook condition \cite{Cook:1957} is a sufficient condition for the
strong convergence of the time limits that define scattering
wave operators
\beq
\Omega_{\pm n_1 n_2} := \lim_{\tau \to \pm \infty} e^{i M \tau} \Phi_{n_1 n_2}
e^{-i M_{n_1n_2} \tau}
\vert f_{1} f_2 \rangle 
\label{apc:1}
\eeq
for and initial or final state in the $n_1, n_2$ two meson channel.

For the model of section VI the Cook condition is 
\beq
\int_c^{\pm\infty} 
\Vert  v_{1:2} \Phi_{n_1 n_1}  
 e^{-iM_{n_1 n_1} \tau} \vert  f_{1}f_{2} \rangle \Vert d\tau < \infty .  
\label{apc:2}
\eeq

To show that this inequality is satisfied it is necessary to show that
\beq
\Vert  v_{1:2} \Phi_{n_1 n_1}  
e^{-iM_{n_1 n_1} \tau} \vert  f_{1}f_{2} \rangle \Vert
\label{apc:3}
\eeq
is an integrable function of $\tau$ for a dense set of initial or
final wave packets.

The Cook condition follows provided 
that the square of (\ref{apc:3}) has the form
\beq
\int d\mathbf{r}_{12}
f^*(\mathbf{r}_{12},\tau)K^*(\mathbf{r}_{12})K(\mathbf{r}_{12})
f(\mathbf{r}_{12},\tau) d\mathbf{r}_{12}
\label{apc:4}
\eeq
where $\int \vert K(\mathbf{r}_{12})\vert^2 d \mathbf{r}_{12}$ is finite and
$f(\mathbf{r}_{12},\tau)$ satisfies
\beq
\vert f(\mathbf{r},2\tau)
\vert < {c_1 \over c_2 + \tau^{3/2}}
\label{apc:5}
\eeq
for constants $c_1$ and $c_2$.  This ensures that the integrand in (\ref{apc:2})
is bounded by
\beq
\sqrt{\int \vert K(\mathbf{r}_{12})\vert^2 d\mathbf{r}_{12}}
\vert < {c_1 \over c_2 + \tau^{3/2}}
\label{apc:6}
\eeq
which is integrable.

In what follows only the simplest case of identical particle scattering
is discussed.  In this case
$M_{n_1n_2}$ has the form $2 \sqrt{m_n^2 + \mathbf{q}^2}$ and
after factoring out the conserved light-front 3-momentum,  the term
$ 
\langle \mathbf{r}_{12} \vert 
e^{-i M_{n_1n_2} \tau}
\vert f_{1} f_2 \rangle $ becomes
\beq
\langle \mathbf{r}_{12} \vert 
e^{-i M_{n_1n_2} \tau}
\vert f_{1} f_2 \rangle :=
{1\over (2 \pi)^{3/2}}\int 
e^{-i 2 \sqrt{\mathbf{q}^2 +m_n^2} \tau+ i\mathbf{q}\cdot \mathbf{r}} 
d\mathbf{q}  f (\mathbf{q}).
\label{apc:7}
\eeq
This has the form of a positive energy solution of the Klein-Gordon
equation with a wave packet $f (\mathbf{q})$ in the relative momentum.
When $f(\mathbf{q})$ a Schwartz test function these solutions are known to
satisfy the bound 
(\ref{apc:5}) \cite{Ruelle:1962}\cite{jost}\cite{bogoliubov} .

What remains is to
verify $\int \vert K(\mathbf{r}_{12})\vert^2 d \mathbf{r}_{12} <  \infty$ 

To construct the  kernel $K(\mathbf{r}_{12})$ note that 
\[
K^* (\mathbf{r}_{12}) K (\mathbf{r}_{12}) =
\gamma^2
\int  
\delta (\mathbf{r} - 2 \mathbf{r}_{12}) 
\delta (\mathbf{r} - 2 \mathbf{r}_{12}') 
d\mathbf{r}_1
d\mathbf{r}_2
d\mathbf{r}_1'
d\mathbf{r}_2'
d\mathbf{r}
d\mathbf{r}_{12}'
d\eta d\eta' \times
\]
\beq
g(\mathbf{r}_1-\eta \mathbf{r}) 
g(\mathbf{r}_2-(1-\eta) \mathbf{r}) 
\langle \phi_n \vert \mathbf{r}_1  \rangle
\langle \phi_n \vert \mathbf{r}_2  \rangle
\langle \mathbf{r}'_2 \vert\phi_n \rangle
\langle \mathbf{r}'_{1} \vert \phi_n \rangle
g(\mathbf{r}'_1-\eta' \mathbf{r}) 
g(\mathbf{r}'_2-(1-\eta') \mathbf{r}) 
\label{apc:8}
\eeq
The integrals over the delta functions give a factor of
${1 \over 2}$ and replace $\mathbf{r}/2= \mathbf{r}_{12}'$
by $\mathbf{r}_{12}$.  This gives
\[
K^* (\mathbf{r}_{12}) K (\mathbf{r}_{12}) =
{\gamma^2 \over 2} \int_0^1 d\eta \int_0^1 d\eta'
\]
\[
\int
d\mathbf{r}_1
g(\mathbf{r}_1-2\eta \mathbf{r}_{12})
\langle \phi_n \vert \mathbf{r}_1  \rangle
\int
d\mathbf{r}_2
g(\mathbf{r}_2-2 (1-\eta) \mathbf{r}_{12}) 
\langle \phi_n \vert \mathbf{r}_2  \rangle
\]
\beq
\int
d\mathbf{r}_1'
\langle \mathbf{r}'_2 \vert\phi_n \rangle
g(\mathbf{r}'_1-2\eta' \mathbf{r}_{12})
\int
d\mathbf{r}_2'
g(\mathbf{r}'_2-2(1-\eta') \mathbf{r}_{12})
\langle \mathbf{r}'_{1} \vert \phi_n \rangle
\label{apc:9}
\eeq
Ignoring the $\eta$ and $\eta'$ which are over the compact interval
$[0,1]$,
this expression is a product of four integrals of identical structure.
It is enough to consider one of them; for example
\beq
\int
d\mathbf{r}_1
g(\mathbf{r}_1-2\eta \mathbf{r}_{12})
\langle  \mathbf{r}_1 \vert \phi_n   \rangle .
\label{apc:10}
\eeq
In this case
$g(\mathbf{r}_1-2\eta \mathbf{r}_{12})$ is a Gaussian and the oscillator
function,
$\langle \phi_n \vert \mathbf{r}_1  \rangle$, is a Gaussian
multiplied by a polynomial.  In what follows it will be shown that the result
of the integration over $\mathbf{r}_i$  is a
Gaussian multiplied by a polynomial in $\mathbf{r}_{12}$.
The product of these four factors are Gaussians multiplied by polynomials.
This ensures that the bound $\int \vert K(\mathbf{r}_{12})\vert^2 d \mathbf{r}_{12} <  \infty$ is satisfied, which along with
(\ref{apc:5}) verifies that the interaction satisfies the Cook condition.
It also implies that truncations of
the channel sums to finite numbers of one and two-body channels is an
approximation with controlled errors rather than an uncontrolled
truncation, however this is only in the context of the trunctation to the
two-meson sector.  The case of mesons with unequal masses should have
the same qualitative behavior, but the proof is not as clean as in
the equal mass case.

Note that the $\tau^{-3/2}$ dependence is exactly the same falloff that
is seen the non-relativistic case. The origin of this behavior is the
phase space factor $q^2 dq$.

\section{Appendix IV}

The numerical methods used to compute
$M_{mn} (e+i0^+)$ in equation (\ref{sc:36}) are discussed in
this appendix.

The calculations below evaluate this matrix
using an orthonormal set of box functions
\cite{Rubtsova:2017kvn}, that are constant
on a finite interval, $[q_{n-1},q_n]$ and are normalized to unity:
\beq
1= N_n^2 \int_{\kappa_{n-1}}^{\kappa_n} \kappa^2 d\kappa =
{N_n^2 \over 3}(\kappa_n^3 - \kappa_{n-1}^3) =
{N_n^2  \over 3}(\kappa_n - \kappa_{n-1})(\kappa_n^2 + \kappa_n \kappa_{n-1}
+ \kappa_{n-1}^2). 
\label{sc:39}
\eeq
The resulting normalization coefficient is
\beq
N_n = \sqrt{{3 \over (\kappa_n - \kappa_{n-1})(\kappa_n^2 + \kappa_n
\kappa_{n-1} + \kappa_{n-1}^2)}}.
\label{sc:40}
\eeq
This is equivalent to approximating the integral in the integral equation
by a limit of Riemann sums.
For a given grid of momentum variables $\{ \kappa_0 , \cdots , \kappa_N\}$
the normalized box functions are
\beq
\chi_n(\kappa) :=
\left \{
\begin{array}{cc}
0 & \kappa \notin [\kappa_{n-1},\kappa_n)\\
N_n & \kappa \in [\kappa_{n-1},\kappa_n)\\
\end{array}
\right . .
\label{sc:41}
\eeq
They satisfy the orthonormality condition 
\beq
\int_0^\infty \chi_n (\kappa) \chi_m(\kappa)\kappa^2 d\kappa = \delta_{mn}.
\label{sc:42}
\eeq
Using these functions, $M_{mn}(e+i0^+)$ is approximated by
\beq
M_{nm} (e+i0^+) \approx
\sum_{k,m_1,m_2} 
\langle n \vert v_{1:2} \vert m_{1}, m_{2},\chi_k \rangle
\langle \chi_k \vert
{1 \over  e-M_{2m_1,m_2}(\kappa) +i 0^+} \vert \chi_k \rangle 
\langle  m_{1}, m_{2}, \chi_k  \vert
v_{2:1} \vert m \rangle .
\label{sc:43}
\eeq
The advantage of this basis is that matrix elements of the propagator
are diagonal \cite{Rubtsova:2017kvn}.  In addition to the channel
truncations the calculations also use a momentum cutoff at $\kappa_N$.
The box functions are chosen so the  on-shell point always sits at the
midpoint of an interval.

The quantities that appear in the sum are the elementary integrals
over the vertex in momentum space
\beq
\langle n \vert v_{1:2} \vert m_{1}, m_{2},\chi_n \rangle =
N_n \int_{\kappa_{n-1}}^{\kappa_n} 
\langle n \vert v_{1:2} \vert m_{1}, m_{2},\kappa \rangle \kappa^2 d\kappa
\label{sc:44}
\eeq
\beq
\langle  m_{1}, m_{2}, \chi_n  \vert
v_{2:1} \vert m \rangle =
N_n \int_{\kappa_{n-1}}^{\kappa_n}
\langle  m_{1}, m_{2}, \kappa  \vert
v_{2:1} \vert m \rangle \kappa^2 d\kappa
\label{sc:45}
\eeq
and
\beq
g_n(e+i0^+) = 
\langle \chi_n \vert
{1 \over  e-M_2(m_1,m_2,\kappa) +i 0^+} \vert \chi_n \rangle =
N_n^2 \int_{\kappa_{n-1}}^{\kappa_n}         
{\kappa^2 d\kappa \over e-M_2(m_1,m_2,\kappa) +i 0^+}.
\label{sc:46}
\eeq
The first two integrals are computed using a Gauss Legendre quadrature.
For reasonably narrow intervals $[\kappa_{n-1},\kappa_n]$ only a
small number of quadrature points are sufficient for convergence.
For the Green's function integrals there are three cases to consider.  The first step is to check to see if
\beq
M_2(m_1,m_2,0) > e = m_{os} .
\label{sc:47}
\eeq
where $m_{os}$ is the on-shell invariant mass.
If this is true then there is no singularity in the denominator and
the $i0^+$ can be removed.  The integral can be computed
numerically using a Fejer or Gauss Legendre quadrature rule 
with a small number of quadrature points.

If
\beq
M_2(m_1,m_2,0) \leq  e = m_{os}
\label{sc:48}
\eeq
then there is a real $\kappa_s$ satisfying
\beq
M_2(m_1,m_2,\kappa_s) = e = m_{os}.
\label{sc:49}
\eeq
In this case $\kappa_s$ is given by
\beq
\kappa^2_s ={e^4 + m_1^4 + m_2^4 - 2e^2 m_1^2 - 2 e^2 m_2^2 -2 m_1^2 m_2^2
\over 4 e^2}.
\label{sc:50}
\eeq
If this condition is satisfied then there are two possibilities.
If $\kappa_s \notin [\kappa_{n-1},\kappa_n]$ there is no singularity in the
domain of integration and the integral can again be 
performed numerically, setting 
$\epsilon \to 0$, using a small number of quadrature points.  In this case
\[
g_n(e) := \langle \chi_n \vert
{1 \over e-M_2(m_1,m_2,\kappa) } \vert \chi_n \rangle =
N_n^2 \int_{\kappa_{n-1}}^{\kappa_n}         
{\kappa^2 d\kappa \over e-M_2(m_1,m_2,\kappa)}
\]
where $\epsilon$ can be set to 0. 
If $\kappa_s \in [\kappa_{n-1},\kappa_n]$ then there is a singularity in the
domain of integration.  In this case the singular part
can be isolated as follows:
\[
g_n(e) := \langle \chi_n \vert
{1 \over e-M_2(m_1,m_2,\kappa) +i 0^+} \vert \chi_n \rangle =
N_n^2 \int_{\kappa_{n-1}}^{\kappa_n}         
{\kappa^2 d\kappa \over e-M_2(m_1,m_2,\kappa) +i 0^+}=
\]
\[
N_n^2  \int_{\kappa_{n-1}}^{\kappa_n}
{\kappa^2 d\kappa \over
M_2(m_1,m_2,\kappa_s) -M_2(m_1,m_2,\kappa) +i 0^+} =
\]
\[
N_n^2 \int_{\kappa_{n-1}}^{\kappa_n}
{\kappa^2 d\kappa \over
\sqrt{m_1^2 +\kappa_s^2}+ \sqrt{m_2^2 +\kappa_s^2}-
\sqrt{m_1^2 +\kappa^2}- \sqrt{m_2^2 +\kappa^2}
+i 0^+} =
\]
\[
N_n^2 \int_{\kappa_{n-1}}^{\kappa_n}
{\kappa^2 d\kappa \over
\sqrt{m_1^2 +\kappa_s^2}- \sqrt{m_1^2 +\kappa^2}+
\sqrt{m_2^2 +\kappa_s^2}- \sqrt{m_2^2 +\kappa^2}
+i 0^+} =
\]
\[
N_n^2 \int_{\kappa_{n-1}}^{\kappa_n}
{\kappa^2 d\kappa \over
(\kappa_s^2 - \kappa^2)  
({1 \over \sqrt{m_1^2 +\kappa_s^2}+ \sqrt{m_1^2 +\kappa^2}}+
{1 \over \sqrt{m_2^2 +\kappa_s^2}+ \sqrt{m_2^2 +\kappa^2}}
) +i 0^+}=
\]
\[
N_n^2 \int_{\kappa_{n-1}}^{\kappa_n}
{\kappa^2 d\kappa \over
(\kappa_s^2 - \kappa^2)  
({1 \over \sqrt{m_1^2 +\kappa_s^2}+ \sqrt{m_1^2 +\kappa^2}}+
{1 \over \sqrt{m_2^2 +\kappa_s^2}+ \sqrt{m_2^2 +\kappa^2}}
) +i 0^+}=
\]
\[
N_n^2 \int_{\kappa_{n-1}}^{\kappa_n}
{( \sqrt{m_1^2 +\kappa_s^2}+ \sqrt{m_1^2 +\kappa^2})
(\sqrt{m_2^2 +\kappa_s^2}+ \sqrt{m_2^2 +\kappa^2})\over
\sqrt{m_1^2 +\kappa_s^2}+ \sqrt{m_1^2 +\kappa^2}+
\sqrt{m_2^2 +\kappa_s^2}+ \sqrt{m_2^2 +\kappa^2}
}
{\kappa^2 d\kappa \over 
\kappa_s^2 - \kappa^2 + i0^+} .
\]
In this form the singularity is at $\kappa=\kappa_s$.
Defining
\[
f (\kappa) :=
-
N_n^2 
{( \sqrt{m_1^2 +\kappa_s^2}+ \sqrt{m_1^2 +\kappa^2})
(\sqrt{m_2^2 +\kappa_s^2}+ \sqrt{m_2^2 +\kappa^2})\over
\sqrt{m_1^2 +\kappa_s^2}+ \sqrt{m_1^2 +\kappa^2}+
\sqrt{m_2^2 +\kappa_s^2}+ \sqrt{m_2^2 +\kappa^2}}
{\kappa^2\over \kappa+\kappa_s}  
\]
gives
\[
g_n(e) = \int_{\kappa_{n-1}}^{\kappa_n} {f(\kappa) - f(\kappa_s)\over \kappa-\kappa_s} d\kappa
+ f(\kappa_s)\int_{\kappa_{n-1}}^{\kappa_n} {d\kappa \over \kappa-\kappa_s -i0^{+}}  =
\]
\beq
\int_{\kappa_{n-1}}^{\kappa_n} {f(\kappa) - f(\kappa_s)\over \kappa-\kappa_s} d\kappa
+ f(\kappa_s)
(
\ln ({\kappa_n - \kappa_s \over \kappa_s -\kappa_{n-1}})
+ i \pi
) 
\eeq
where 
\beq
f(\kappa_s)=
-N_n^2 { \kappa_s \sqrt{m_1^2 +\kappa_s^2}\sqrt{m_2^2 +\kappa_s^2}
\over  \sqrt{m_1^2 +\kappa_s^2}+\sqrt{m_2^2 +\kappa_s^2}}.
\eeq 
These singular terms contain all of the imaginary parts of $g_n(e)$.

\bibliography{duality.bib}
\end{document}